\documentclass[final,5p,times,twocolumn,authoryear]{elsarticle}

\usepackage{nccmath, 
            amssymb}

\usepackage{xspace} 
\usepackage{hyperref} 
\hypersetup{colorlinks=true}
\usepackage{caption} 
\usepackage{subcaption} 
\usepackage{setspace} 

\usepackage{booktabs} 
\usepackage{multirow, bigdelim} 
\usepackage{lscape} 
\usepackage{longtable} 

\graphicspath{{figures/}}

\newcommand{\integral}{\textsl{INTEGRAL}\xspace}
\newcommand{\xmm}{\textsl{XMM-Newton}\xspace}

\newcommand{\einstein}{\textsl{Einstein}\xspace}
\newcommand{\exosat}{\textsl{EXOSAT}\xspace}
\newcommand{\arielv}{\textsl{Ariel V}\xspace}
\newcommand{\swift}{\textsl{Swift}\xspace}

\newcommand{\asca}{\textsl{ASCA}\xspace}

\newcommand{\rosat}{\textsl{ROSAT}\xspace}

\newcommand{\ginga}{\textsl{GINGA}\xspace}
\newcommand{\heaoI}{\textsl{HEAO-1}\xspace}
\newcommand{\uhuru}{\textsl{Uhuru}\xspace}
\newcommand{\velaVb}{\textsl{Vela 5B}\xspace}

\newcommand{\erosita}{\textsl{eROSITA}\xspace}

\newcommand{\eupper}{\textsc{eupper}\xspace}
\newcommand{\hiligt}{HILIGT\xspace}

\newcommand{\postgres}{\textsc{PostgreSQL}\xspace}
\newcommand{\sql}{\textsc{SQL}\xspace}

\newcommand{\nasa}{NASA\xspace}
\newcommand{\heasarc}{HEASARC\xspace}
\newcommand{\esa}{ESA\xspace}
\newcommand{\mpe}{Max-Planck Institute for Extraterrestrial Physics\xspace}

\newcommand{\kev}{\ensuremath{\text{keV}}\xspace}

\newcommand{\cps}{\ensuremath{\mathrm{cts}\,\mathrm{sec}^{-1}}\xspace}

\newcommand{\nh}{\ensuremath{{N}_\mathrm{H}}\xspace}

\renewcommand{\deg}{\ensuremath{^\circ}\xspace}
\newcommand{\amin}{\ensuremath{'}\xspace}
\newcommand{\asec}{\ensuremath{''}\xspace}
\newcommand{\ergcms}{\ensuremath{\text{erg\,cm}^{-2}\text{s}^{-1}}\xspace}

\journal{Astronomy and Computing}

\begin{document}

\begin{frontmatter}



\title{HILIGT, Upper Limit Servers II - Implementing the data servers}


\author[1]{Ole~K{\"o}nig}
\ead{ole.koenig@fau.de}
\author[2]{Richard~D.~Saxton}
\author[3]{Peter~Kretschmar}
\author[4]{Lorella~Angelini}
\author[3]{Guillaume~Belanger}
\author[5]{Phil~A.~Evans}
\author[6]{Michael~J.~Freyberg}
\author[7]{Volodymyr~Savchenko}
\author[8]{Iris~Traulsen}
\author[1]{J\"{o}rn~Wilms}

\address[1]{Dr.~Karl-Remeis-Sternwarte and Erlangen Centre for Astroparticle Physics, Friedrich-Alexander-Universit\"at Erlangen-N\"urnberg, Sternwartstr. 7, 96049 Bamberg, Germany} 
\address[2]{Telespazio-Vega UK for the European Space Agency (ESA), XMM-Newton SOC, European Space Astronomy Centre (ESAC), Camino Bajo del Castillo s/n, 28692 Villanueva de la Ca{\~{n}}ada, Madrid, Spain}
\address[3]{European Space Agency (ESA), European Space Astronomy Centre (ESAC), Camino Bajo del Castillo s/n, 28692 Villanueva de la Ca{\~{n}}ada, Madrid, Spain}
\address[4]{NASA, Goddard Space Flight Center, Greenbelt, Maryland, United States}
\address[5]{School of Physics and Astronomy, University of Leicester, Leicester, LE1 7RH, UK}
\address[6]{Max-Planck-Institut f\"ur extraterrestrische Physik, Giessenbachstra{\ss}e, 85748 Garching, Germany}
\address[7]{ISDC, Department of Astronomy, University of Geneva, Chemin d'\'Ecogia 16, 1290 Versoix, Switzerland}
\address[8]{Leibniz-Institut f\"ur Astrophysik Potsdam (AIP), An der Sternwarte 16, 14482 Potsdam, Germany}

\begin{abstract}
  The \emph{High-Energy Lightcurve Generator} (\hiligt) is a new
  web-based\tnoteref{label1} tool which allows the user to generate
  long-term lightcurves of X-ray sources. It provides historical data
  and calculates upper limits from image data in real-time. 
  \hiligt utilizes data from twelve satellites, both modern missions such as \xmm and \swift, and earlier facilities such as \rosat, \exosat, \einstein or
  \arielv. Together, this enables the user to query 50 years of X-ray
  data and, for instance, study outburst behavior of transient
  sources. In this paper we focus on the individual back-end servers
  for each satellite, detailing the software layout, database design,
  catalog calls, and image footprints. 
  We compile all relevant calibration information of these missions and provide an in-depth summary of the details of X-ray astronomical instrumentation and data.
\end{abstract}
\tnotetext[label1]{\url{http://xmmuls.esac.esa.int/hiligt/}}


\end{frontmatter}


\section{Introduction}
\label{sec:introduction}

X-ray sources can be extremely variable on times scales of fractions of a second (for instance during quasi-periodic oscillations in black hole binaries, see, e.g., \citealt{Strohmayer01a}, or millisecond neutron star pulsations, see, e.g,  \citealt{Hartman08a}) to decades \citep[e.g.,][]{Lin17a}, with the flux sometimes ranging over several orders of magnitude (for instance during thermonuclear X-ray bursts of neutron stars, see, e.g., \citealt{Grindlay76a} or \citealt{Kuulkers09a}). Some sources, like pulsars with a Be companion star (e.g., EXO~2030$+$375, see \citealt{Wilson02a}) exhibit regular outbursts, others, like tidal disruption events \citep[e.g.,][]{Komossa99a} typically show one bright flare followed by a slow decay.
The lifetime of a space-based observatory is very short compared to
astrophysical timescales, and usually of the order of 5--10 years. In
order to resolve variability patterns which are decades-long, it is
therefore necessary to combine data of several instruments. With the
increasing number of X-ray missions, however, it has become difficult
to get a comprehensive overview of all available data.
Furthermore, old data formats, missing analysis pipelines, and incomplete documentation make the extraction of data from many missions sometimes cumbersome. 
Due to these complexities and the time-consuming process of compiling flux data over multiple decades, detailed studies of source variability on these time scales are rare. 
Although there are services available that combine data of current and past missions (e.g., HEAVENS\footnote{\url{http://astroh.unige.ch/heavens/}}), so far no generalized tool is available which automatically computes flux upper limits for multiple missions.
For this reason, the \emph{High-Energy Lightcurve Generator} (\hiligt)
was developed. It is a web-based tool which enables quick access to
the data of past and current X-ray missions, without the need to
browse through several catalogs in person, or install software (see Saxton et al.\,2021; hereafter Paper~I, for an overview of the front-end). 

\begin{figure}
    \centering
    \includegraphics[width=1.\linewidth]{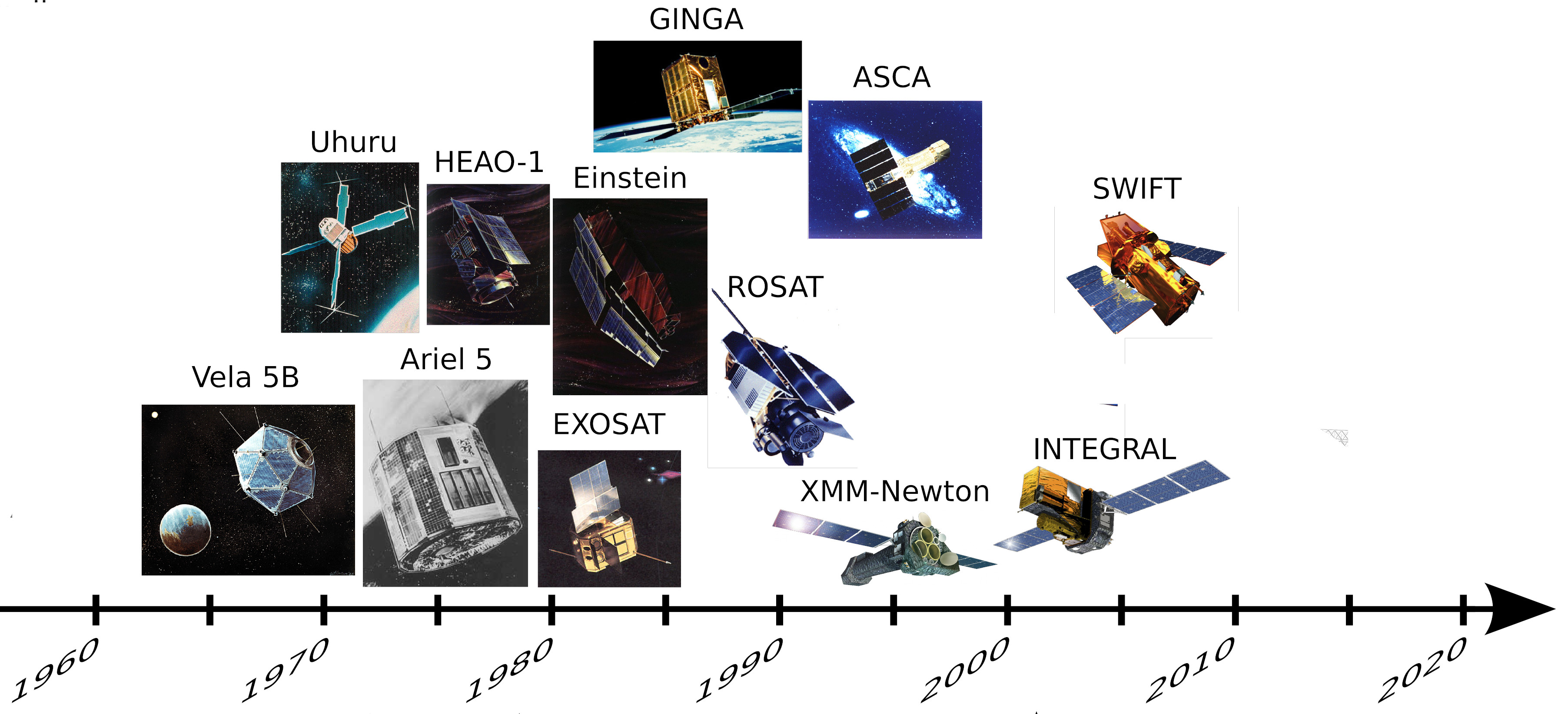}
    \caption{An overview of X-ray satellites implemented in \hiligt. The dates are approximately from the launch.}
    \label{fig:history}
\end{figure}
Figure~\ref{fig:history} compiles an overview of the X-ray satellites implemented in \hiligt, which span a time period of over 50 years.
In order to fully exploit all the available data from these satellites, \hiligt
provides a framework which either returns a catalog flux or calculates
a flux upper limit at any celestial position. Adding upper limits to
the lightcurve augments its diagnostic capability, for instance, in
studying long-term flux variation of sources.
In quiescent states (e.g., \citealt{Menou99a}), 
sources are often observed serendipitously, and upper limits provide
the possibility to constrain their flux in these periods. 
A tool like \hiligt is especially important for sources typically studied in the X-rays, because they often show dramatic changes in luminosity (see, e.g., \citealt{Remillard06a}).

Therefore, a further major application is research into outbursts and state changes of transient sources (see, e.g., \citealt{Belloni11a} or \citealt{Hasinger89a}), as performed for instance in recent nova studies \citep{Sokolovsky21a}, or the ongoing \erosita all-sky survey (e.g., \citealt{GokusATel20a} or \citealt{MalyaliATel20a}).
As it can be used to easily query the history of a source in the X-rays, \hiligt also bridges a connection to other multi-wavelength efforts such as ASSA-SN\footnote{\url{http://www.astronomy.ohio-state.edu/asassn/}}, which often observe new transient phenomena. 

In Paper~I we give an overview of the design of \hiligt and the details of the front-end clients. In this paper we concentrate on the back-end. Section~\ref{sec:eupper} describes the upper limit calculation and Sect.~\ref{sec:convfactor} shows how \hiligt performs the count rate to flux conversion. 
Catalog calls are described in Sect.~\ref{sec:cat}, followed by the footprint
computation and database implementation (Sect.~\ref{sec:database}).
Individual mission servers are described in
Sect.~\ref{sec:description_of_missions}. An overview of all mission
parameters can be found in Table~\ref{tab:params}. We summarize and
outline future plans in Sect.~\ref{sec:summary}.

\section{Design}
\label{sec:hiligt}

In order to achieve the goal of providing flux upper limits and
catalog values of multiple missions we
\begin{itemize}
\item download the available data (images, lightcurves, exposure and background maps) and store the image footprints in databases,
\item design a framework which queries existing catalogs and matches
  them against the locally stored data,
\item write a tool which calculates upper limits from image data,
\item provide a set of spectral models and output energy bands to
  perform the count rate to flux conversion, and
\item design a web interface for the end user.
\end{itemize}

\begin{figure}
  \centering
  \includegraphics[width=1\linewidth]{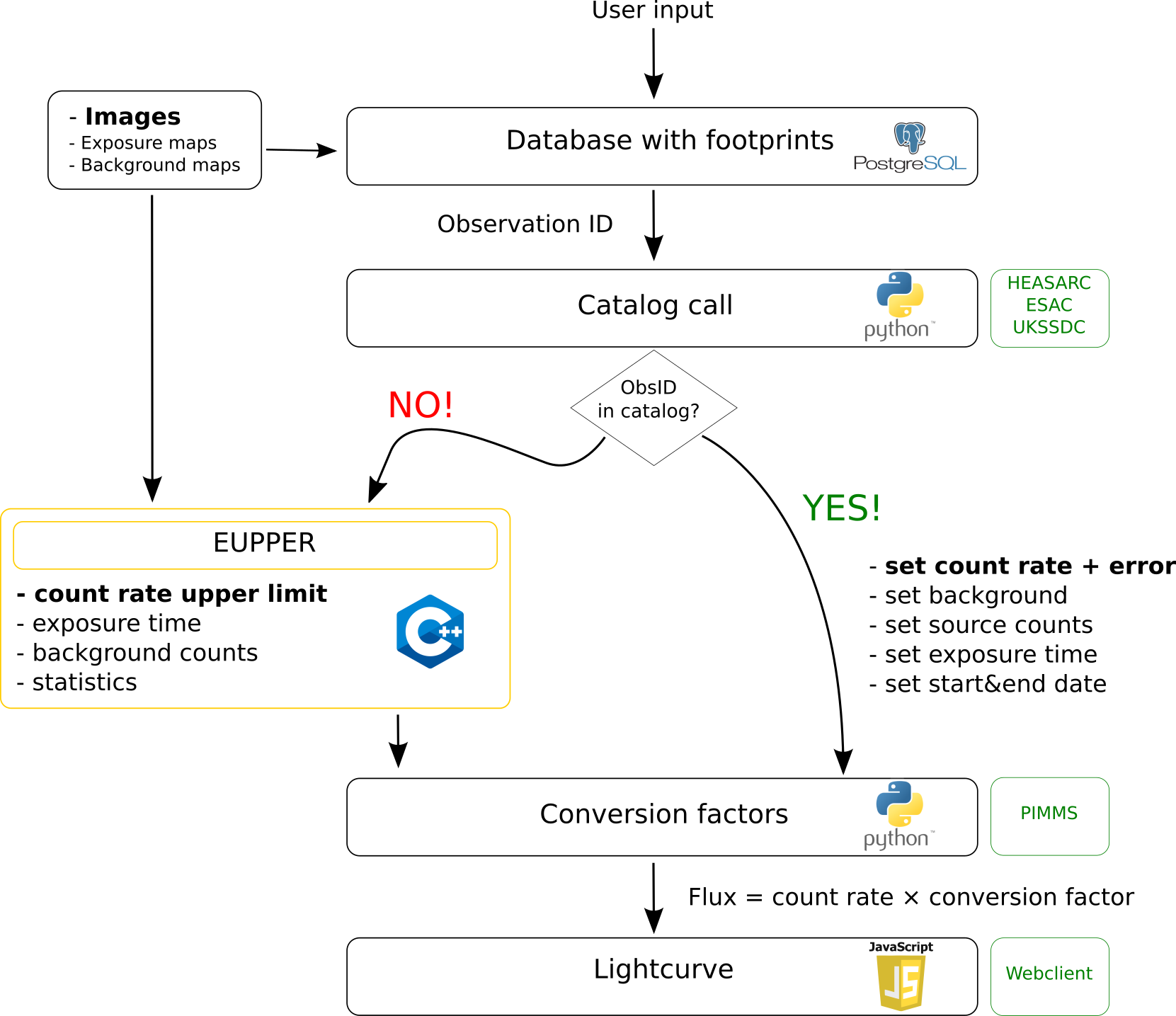}
  \caption{Basic scheme of \hiligt:
    The lightcurves exhibit data from catalogs in combination
    with upper limits, which are computed from the images.}
  \label{fig:diagramn}
\end{figure}

We describe the top level design and server interaction of \hiligt in Paper~I. 
Briefly, a user provides a set of celestial coordinates and a spectral model
appropriate to the source's spectral shape. For each mission, a local
database query finds all images in which the coordinates match the
image footprint. In other words, the query finds all images which
contain data at the input position and then returns a list with the
corresponding observation IDs. This list of observation IDs is
reconciled with the existing catalogs: If the observation ID in the
database contains a source corresponding to the input location, the
count rate of that source from the mission catalog is returned.
Otherwise, the image is used to compute a count rate upper limit
(taking exposure, background, vignetting effects, and the point spread
function into account). Finally, a conversion factor dependent on the
input spectral model is used to convert the count rate into a flux.
These data can be either queried from a web-interface in the browser,
or using a URL query from the command line. A scheme of how \hiligt functions can be
seen in Fig.~\ref{fig:diagramn}.
We refer to Paper~I for the details of the front-end implementation, run-time, and a sample of possible scientific uses.

\subsection{Calculation of upper limits}
\label{sec:eupper}

In the following we describe how \hiligt uses the tool ``\eupper'' to calculate upper limits according to \citet{1991ApJ...374..344K}. 
Although the tool is part of the \xmm SAS \citep[available since SAS 16]{2004ASPC..314..759G}, it is designed to be mission-independent. Firstly, one needs to define an appropriate source region for the extraction of counts.
Let N be the total number of counts in this source region $\mathrm{A}_\mathrm{S}$. $\mathrm{A}_\mathrm{S}$ is assumed to be circular and depends on the point spread function (PSF), angular resolution and pointing accuracy of the mission.

In order to obtain the source counts S=N-B, the background counts B in
background area $\mathrm{A}_\mathrm{B}$ of a given position on the sky
is needed. Ideally, this is done by analyzing a background map,
whereby the counts of the background map in the source area
($\mathrm{A}_\mathrm{B}=\mathrm{A}_\mathrm{S}$) are summed up. If no
background map is available the background counts are taken from an
annulus of radius $2.5r$ and $3r$ around the source position. $r$ is a
mission-dependent source radius defined from the size of the PSF (see
Table~\ref{tab:params}), such that the PSF does not overlap with the
annulus. The background counts are scaled to the source area by
multiplying with the area ratio
$\mathrm{A}_\mathrm{S}/\mathrm{A}_\mathrm{B}$. We note that this
simplistic background estimation can cause errors if positions close
to bright sources or at the edge of an image are evaluated. Upper
limits in crowded fields or for extended sources can thus be estimated
too low due to the high background counts. However, we emphasize that
the annulus background estimation is only applied in the case where no
background maps are available (only for \einstein, \exosat, and \rosat
PSPC, see also ``bkg'' column in Table~\ref{tab:params}).

To calculate the count rate, one has to divide the background
subtracted counts by the exposure time. The exposure is taken from the
exposure map at the specified position (which includes also the
vignetting). If no exposure map is available, the exposure is taken
from the header of the Flexible Image Transport System (FITS, \citealt{Wells81a}) image
and multiplied by an off-axis dependent vignetting correction factor.
In order to correct the count rate for the PSF, one also has to divide the source counts by the encircled energy fraction (EEF), given by the radius-integrated PSF normalized to the full PSF:
\begin{equation}
    \mathrm{EEF}(E,r,\theta) = \int_0^r \mathrm{PSF}(E,r,\theta)~dr \Biggm/ \int_0^\infty \mathrm{PSF}(E,r,\theta)~dr
    \label{eq:eef}
\end{equation}
\hiligt takes into account that the EEF changes as a function of source radius
and off-axis angle $\theta$, but does not account for the energy dependence of
the PSF. The background subtracted count rate is therefore computed by
\begin{equation}
  \mathrm{CR} = \frac{\mathrm{N}-\mathrm{B}\cdot \mathrm{A}_\mathrm{S}/\mathrm{A}_\mathrm{B}}{\text{EEF} \cdot \text{exposure}} \quad .
\end{equation}
The error propagation and thus the final upper limit depends on the number of counts. \citet{Gehrels86a} shows that the application of classical Gaussian statistics is unsuitable in the case of low counts and Poisson or, best, Bayesian statistics \citep[e.g.,][]{1991ApJ...374..344K} must be used.
Because such estimates are computationally more demanding, and the distribution asymptotically becomes Gaussian, we set a threshold of 80 counts to divide the two regimes.

For N$>$80 classical Gaussian statistics are applied and the
uncertainty is computed by
\begin{equation}
    \label{eq:GaussUpplim}
    \text{Upper limit (\cps)} \:=\: \frac{max(\mathrm{S},~0) + f \cdot
      \sqrt{\mathrm{N} + \mathrm{B}}}{\text{EEF} \cdot \text{exposure}}
\end{equation}
with $\sigma_S=\sqrt{\mathrm{N} + \mathrm{B}}$ being the standard deviation estimated from the source counts based on Poisson statistics. 
These are obtained by linear Gaussian error propagation under the
assumption of a Poissonian distribution on N and B. The factor $f$
can be 1, 2, or 3 depending on the desired 1, 2 or 3-$\sigma$
confidence level for a single parameter of interest (68.2\%, 95.5\% or 99.7\%). In the case of low
counts, Bayesian statistics are applied. We refer to
\citet{1991ApJ...374..344K} for a description of the relevant
statistics.

We want to stress that \eupper calculates upper limits (i.e., single-sided confidence limits) in most cases. Only if the value is $\geq2$ times the error it returns a data point (i.e., a two-sided confidence limit). At this stage, \hiligt has computed count rates from the images, exposure and background maps. However, count rates cannot be directly compared between instruments. Therefore, \hiligt converts the count rate into a flux by multiplying it by an instrument- and model-specific conversion factor, as described in the next section.

\subsection{Count rate to flux conversion}
\label{sec:convfactor}
Photon counting detectors, such as CCDs, have counts as fundamental
measurement property. In the linear regime, which is valid for low
count rates, the number of source counts, $S$, in detector channel $h$
within the exposure $T$ is given by
\begin{equation}
  \label{eq:count_rate}
  S(h) = T \int_0^\infty \mathrm{RMF}(h,E)\cdot \mathrm{ARF}(E)\cdot
  F(E)\,dE \qquad ,
\end{equation}
where $\mathrm{ARF}(E)$ describes the effective area of the optics,
including the detector efficiency. The detector response
$\mathrm{RMF}(h,E)$, often called the response matrix, gives the
probability of detecting a photon of energy $E$ in channel $h$. More
details can be found, for instance, in \citet{2011hxra.book.....A}.
The physically important parameter in Eq.~\ref{eq:count_rate} is the
flux $F(E)$ of the observed source.
This quantity is, however, difficult to obtain because generally
Eq.~\ref{eq:count_rate} cannot be inverted \citep[e.g.,][]{Lampton76a}.
In spectral analyses one therefore uses some kind of a forward folding process to estimate the
predicted counts for a given spectral model, and then minimizes the
deviation between the model predicted counts and the data using an
appropriate statistical model (e.g., $\chi^2$-statistic).

Extracting spectra and model based fitting of the data for each observation is
beyond the scope of this tool. Therefore, the user selects a spectral
model and \hiligt produces the flux \emph{under the assumption} of
this model. The currently available models are a power-law with photon
index $\Gamma=0.5$, 1.0, 1.5, 1.7, 2.0, 2.5, 3.0, or 3.5, or a black
body with temperature $kT=60$, 100, 300, or 1000\,eV, respectively,
and equivalent hydrogen column densities of
$N_\mathrm{H}=(1,3,10)\times 10^{20}\,\mathrm{cm}^{-2}$. We use the
PIMMS software \citep[version 4.11a]{Mukai93PIMMS}, which solves
Eq.~\ref{eq:count_rate} and calculates the count rate to flux
conversion factors. We extract conversion factors for each filter,
spectral model, and spectral index/temperature configuration (for
efficiency reasons the conversion factors are hard-coded into the
\hiligt code). They can be found in Table~\ref{tab:cf} of the
appendix.

The data of the different satellites can only be compared if they are
given for the same flux range. Therefore, we interpolate (or
extrapolate if the mission band is slightly smaller than our reference
band) the energy bands of the different satellites onto three
predefined energy bands. These reference bands are called
\textit{soft} for 0.2--2.0\,\kev, \textit{hard} for 2.0--12.0\,\kev,
and \textit{total} for 0.2--12.0\,\kev.

It must be noted that an accurate selection of the spectral shape is
of crucial importance for the correct flux determination. As an
example we show the effect of a wrongly modeled spectrum for the
blazar PKS~0537$-$286, which is known for having a hard power-law
spectrum. Using \xmm data, \citet{Reeves01a} determine a 0.2--10\,keV
flux of $3.1\times 10^{-12}$\,\ergcms, assuming Galactic foreground
absorption and $\Gamma=1.3$. If we use \hiligt to reconstruct the flux
for this pointed \xmm observation\footnote{ObsID 0114090101,
  $\Gamma=1.5$, $\nh=3\times 10^{20}\,\mathrm{cm}^{-2}$}, we obtain
$(3.262\pm 0.021)\times 10^{-12}$\,\ergcms in the 0.2--12\,keV band.
Given the slightly broader energy band, this is in line with the value
determined by \citet{Reeves01a}. Contrarily, using a very soft
spectral model of $\Gamma=3.5$ yields a flux of
$(1.092\pm0.007)\times 10^{-12}$\,\ergcms, a factor 3 off the correct
flux value.

\subsection{Reliability of upper limits}
We address the problem of nearby bright sources by choosing the source
radius according to the PSF. However, an upper limit close to a bright
source (where the source region intersects with the wings of the PSF
of the bright nearby source) can be estimated too high, and can
possibly even cause false detections. Also straylight can result in
biased upper limits. The number of bright sources is, however,
relatively small, and we estimate that source confusion is relatively
unlikely for the majority of observations (see
Sect.~\ref{subsubsec:xmmp}). Furthermore, we note that for crowded
fields or fields with extended emission the upper limits can be
biased. This is less of a problem for older missions where the source
density is lower due to lower sensitivity. We do not make an attempt
to correct for bright pixels or columns and rely on the calibration
data with which the images were created. Finally, we note that also
chip gaps (e.g., \xmm) or telescope features (e.g., \rosat PSPC) can
cause errors in the upper limit computation.

\subsection{Catalog calls}
\label{sec:cat}
In order to complement the upper limits with archival data, \hiligt
performs cone searches around the specified coordinates in the
available mission catalogs. If an entry is found, the catalog data is
provided and no upper limit calculation is performed. This can be
changed by setting a flag to ignore all catalog data. \nasa's
\textit{Goddard Space Flight Center} provides most catalogs and for
most missions, \hiligt uses the High Energy Astrophysics Science
Archive Research Center (\heasarc) to access the
data\footnote{\url{https://heasarc.gsfc.nasa.gov/}}. The \esa mission
\xmm is accessed via a Table Access Protocol (TAP) to the \xmm Science
Archive \citep{2017RMxAC..49..146L}. \swift is accessed through a
custom-built \textsc{https} interface to the 2SXPS upper limit server
\citep{Evans20a}. We note that the back-end calls these servers
on-the-fly, which implies issues with \hiligt in case of malfunction.
The search radii for the different missions can be found in
Sect.~\ref{sec:description_of_missions} and Table~\ref{tab:params}.
The search radius is chosen based on the spatial resolution of the
telescope and typically follows the recommendation by the
\heasarc\footnote{This value is shown above the observation table when
  searching for a source in the W3Browse interface and selecting the
  relevant catalog in the ``Query Results'' tab. The catalog sizes can be found at \url{https://heasarc.gsfc.nasa.gov/cgi-bin/W3Browse/w3catindex.pl}}. However, source
confusion in the catalog is a problem in crowded fields and for old
missions (large search radii), which can result in too high flux
values close to bright sources.

\subsection{Footprints and database implementation}
\label{sec:database}
Finally, we describe how \hiligt queries its databases to find the images, which match the source coordinates.
The so-called footprint delineates the sky region where the image
contains useful data and for which \hiligt is able to compute upper
limits. \hiligt queries the database with the given world coordinates
(right-ascension, declination for the epoch J2000) and looks for all
images where the coordinates lie within the footprint shape. Unless a
catalog count rate for this location is available, an upper limit is
calculated. We note that the footprints of \xmm slew, \xmm stacked, \einstein
HRI/IPC, \exosat LE, \rosat HRI are computed with a custom-built
algorithm \citep[see][]{KoenigMastersThesis19}, however, the detailed information about the calculation can be found in the following Sect.~\ref{sec:description_of_missions}.

\hiligt uses \postgres to handle the image meta data (mostly footprint
shapes)\footnote{\url{https://www.postgresql.org/}}. In order to
determine whether the input coordinates match the footprint,
\textsc{pgSphere} is used\footnote{\url{https://pgsphere.github.io/}}.
It provides a fast search of spherical coordinates in a \sql database.
\hiligt uses \textsc{spoly} for the representation of polygons and
\textsc{scircle} for circular footprints.  The database has the following
fields:
\begin{itemize}
    \item \texttt{OBSID}: The observation ID is used to link the
      images to the mission catalogs
    \item \texttt{FILENAME}: By using the file name the actual image
      can be found to parse its location to the \eupper tool
    \item \texttt{FOV}: The polygon or circular information of
      the footprint in radians (for \textsc{spoly/scircle})
    \item \texttt{FILT/INSTRUME}: If necessary, a filter or instrument
      column is appended, to apply the correct conversion factors
      (e.g. for \einstein)
\end{itemize}
\einstein, launched in 1978, and \exosat, launched in 1983, used the B1950 epoch (in FK4). 
For consistency, all footprints in \hiligt are computed in epoch J2000 (FK5). This means that a coordinate transformation \citep{2013AA...558A..33A} of the source position has to be done to analyze the actual B1950 FK4 images for the upper limit.

\section{Description of the mission servers}
\label{sec:description_of_missions}

In the following, we outline the implementation of each mission server.

\subsection{\velaVb (1969--1979)}
\label{subsec:vela}
\velaVb's instruments \citep{Whitlock92b,Whitlock92} had an energy range of 3--750\,keV. 
We concentrate only on the scintillation X-ray detector (XC), which was
an all-sky monitor and had an energy range of 3--12\,keV. The collimator aperture was $\sim$6\,\deg and the satellite had an orbital period of 56\,hours.
\heasarc provides lightcurves of 99 sources\footnote{\url{https://heasarc.gsfc.nasa.gov/FTP/vela5b/data/}}, out of which 35 sources have 56\,h binned lightcurves and 64 sources have 112\,h binned lightcurves (see \citealt{Whitlock92} for more information about the deconvolution of crowded fields). 
We use a circular footprint of 30\,\amin around the source position (converted from FK4 to FK5) to query the lightcurves.
To convert count rates into fluxes, \hiligt uses the conversion factor
from the \texttt{COUFLU} keyword, which is set to $4.5\times
10^{-10}\,\text{erg\,cm}^{-2}\text{cnt}^{-1}$ for a Crab-like spectrum
in the 3--12\,keV band. Because this conversion factor is applied to
all sources regardless of their spectral shape, we add a systematic uncertainty of $1.5\times 10^{-10}\,\text{erg\,cm}^{-2}\text{cnt}^{-1}$. To make the passband comparable to the other missions within \hiligt, we scale the 3--12\,keV flux to 2--12\,keV according to\footnote{The missions \velaVb, \uhuru, \arielv and \heaoI are not included in PIMMS. For these missions, only conversion
  factors for a power-law spectrum in the missions energy band are available ($\text{ConvFac}_{\text{Mission band}}$). Generally, this range is different from the \hiligt bands and the calculated flux cannot be directly compared. We therefore estimate a new conversion factor $\text{ConvFac}_\text{\hiligt}$ for the \hiligt energy range through multiplying $\text{ConvFac}_{\text{Mission band}}$ by the ratio of the fluxes in the two energy bands. This factor is hard-coded. The integrated energy flux in the \hiligt ($\text{Flux}_\text{\hiligt band}$) and mission ($\text{Flux}_{\text{Mission band}}$) band, respectively, is computed with the Interactive
  Spectral Interpretation System \citep[ISIS version 1.6.2-47]{2002hrxs.confE..17H} with the model
  \texttt{tbabs*powerlaw}. We note that since we only have knowledge of this one conversion factor for each mission, the output spectral
  model will always be a power-law of fixed slope, indifferent of the
  user choice in \hiligt. To account for this uncertainty, we add a systematic error on the conversion factors.}
\begin{equation}
  \label{eq:CV_old}
  \text{ConvFac}_\text{\hiligt} = \text{ConvFac}_{\text{Mission band}} \cdot
  \frac{\text{Flux}_\text{\hiligt band}}{\text{Flux}_{\text{Mission band}}}
\end{equation}


\subsection{\uhuru (1970--1973)}
\label{subsec:uhuru}
\uhuru \citep{1971ApJ...165L..27G} was an all-sky survey mission which monitored the sky from 1970 December 12 until 1973 March 18.
A total of 339 unique sources were identified and published in the Fourth Uhuru (\citealt{1978ApJS...38..357F}) catalog. \hiligt accesses the \textsc{UHURU4} catalog\footnote{\url{https://heasarc.gsfc.nasa.gov/W3Browse/uhuru/uhuru4.html}} with a search radius of 1\,\deg.
\hiligt does not access the raw lightcurves but only to the time-averaged catalog data. The catalog count rate is transformed to a 2--6\,keV flux using the conversion factor $1.7\times 10^{-11}\,\text{erg\,cm}^{-2}\text{cnt}^{-1}$, given by \cite{1978ApJS...38..357F} for a Crab-like spectrum.
A systematic uncertainty of 20\% is added to the flux and the energy range extrapolated to 2--12\,keV using Eq.~\ref{eq:CV_old}.
For transient and highly variable sources (see Table~5 and 8 of \citealt{1978ApJS...38..357F}) the count rate error field is not populated in the catalog. We adopt this approach and set the error to zero in these cases. The resulting error on the flux will then solely originate from the 20\% flux conversion uncertainty.


\subsection{\arielv (1974--1980)}
\label{subsec:arielv}
\begin{figure}
    \centering
    \includegraphics[width=1\linewidth]{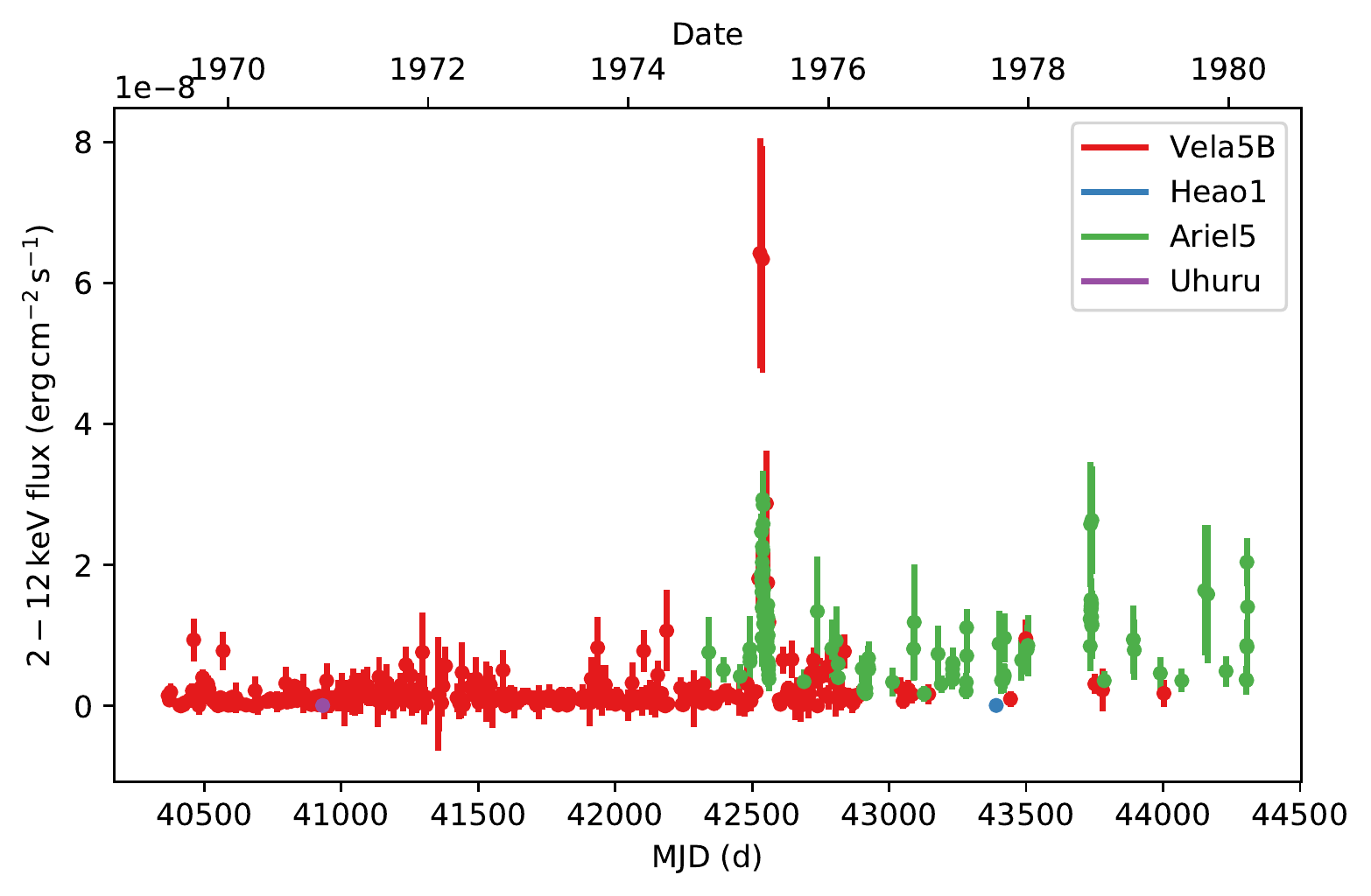}
    \caption{The historical 1975 outburst of 1A~0535+262 covered by the all-sky monitors on \velaVb and \arielv \citep{Rosenberg75a}.}
    \label{fig:a0535}
\end{figure}

As \uhuru, \arielv was an all-sky survey mission which scanned the sky for about 5.5 years with a spin period of $\sim$6\,s. 
\hiligt provides the lightcurves of the All Sky Monitor (ASM, \citealt{Holt76a}).
In order to select the lightcurve corresponding to the query position, we define a circular footprint of 30\,\amin. 
The files\footnote{\url{https://heasarc.gsfc.nasa.gov/FTP/ariel5/asm/data/}} provide flux data in units of photons\,cm$^{-2}$\,s$^{-1}$ in the 3--6\,keV range. 
As outlined in \citet[p.~106, Table~A.1]{Kaluzienski77a} we first convert the photon flux to a \uhuru count rate and afterwards to ``Uhuru Flux Units'' (UFU):
\begin{equation}
    \text{\arielv (3--6\,keV photons\,cm$^{-2}$\,s$^{-1}$)} = \kappa \cdot \text{U (\uhuru \cps)} ,
\end{equation}
where 1\,\uhuru \cps equals $1.7\times 10^{-11}$\,\ergcms in 2--6\,keV. \citet{Kaluzienski77a} state that a value of $\kappa = (1.5\pm 0.5) \times 10^{-3}$ is reasonable for a relatively wide range of incident spectra.
As for \velaVb and \uhuru we estimate the 2--12\,keV flux by Eq.~\ref{eq:CV_old}, assuming a Crab-like spectrum.
Figure~\ref{fig:a0535} shows an example lightcurve of the 1975 outburst of 1A~0535+262 with \velaVb and \arielv all-sky monitor data combined.

In addition, \hiligt accesses the catalog of the Sky Survey Instrument (SSI, \citealt{Villa76a}) which were two pairs of two proportional counters (LE and HE). One LE detector failed shortly after launch.
The \textsc{ARIEL3A} catalog\footnote{\url{https://heasarc.gsfc.nasa.gov/W3Browse/ariel-v/ariel3a.html}} contains 109 X-ray sources at low Galactic latitudes
($|b|<10\,\deg$, \citealt{1981MNRAS.197..865W}), and 142 sources at high Galactic latitudes \citep{1981MNRAS.197..893M}. 
\hiligt searches this catalog with a radius of 0.5\,\deg.
We assume a Crab-like conversion factor of $(5.3\pm 0.8)\times
10^{-11}\,\text{erg\,cm}^{-2}\text{cnt}^{-1}$, as outlined by \citealt[p.~880]{1981MNRAS.197..865W}) and extrapolate the resulting 2--10\,keV flux to \hiligt's \textit{hard} band (2--12\,keV) using Eq.~\ref{eq:CV_old}.

\subsection{\heaoI (1977-1979)}
\label{subsec:heaoI}
The payload of \heaoI \citep{1979SSI.....4..269R} consisted of four major instruments (A1--A4). 
\hiligt provides fluxes from the \textsc{A1} catalog\footnote{\url{https://heasarc.gsfc.nasa.gov/W3Browse/heao1/a1.html}} \citep{Wood84a}, which we access with a 1\,\deg search radius. This catalog contains data of 842 sources from the first six months of the \heaoI mission. We use a Crab-like conversion factor of $4.78\times 10^{-9}$\,erg\,cm$^{-2}$\,cnt$^{-1}$ to convert the count rate to a 2--10\,keV flux \citep{Wood84a}, and extrapolate to 2--12\,keV using Eq.~\ref{eq:CV_old}.

Furthermore, \hiligt gives access to data of the A2 experiment, which was divided into
six proportional counters: two low energy detectors (LED,
0.15--3\,keV), the medium energy detector (MED, 1.5--20\,keV) and
three high energy detectors (HED, 2.5--60\,keV).
MED and HED data is accessed through the the \citet{1982ApJ...253..485P} catalog \textsc{A2PIC}, which contains 68 extra-galactic sources\footnote{\url{https://heasarc.gsfc.nasa.gov/W3Browse/heao1/a2pic.html}}. 
The data originates from two six month long scans and is confined to Galactic latitudes $|b|>20\,\deg$, totaling 65.5\% of the entire sky. 
The catalog also provides conversion factors depending on the spectral shape of the source. For sources without sufficient spectral information, the catalog assumes a 6\,keV thermal Bremsstrahlung spectrum for clusters and a slope 1.65 powerlaw spectrum for AGN. For unidentified sources, an average conversion factor of $2.5\times 10^{-11}$\,erg\,cm$^{-2}$\,cnt$^{-1}$ is assumed. 
We extrapolate this to \hiligt's 2--12\,keV \textit{hard} band (Eq.~\ref{eq:CV_old}). 
The catalog, which is queried with a 1\,\deg search radius, provides two count rates from the first and second scan, which we treat as two individual data points.

\heaoI surveyed over 95\% of the sky to a sensitivity limit of $\sim 3\times 10^{-11}$\,\ergcms in the 0.44--2.8\,keV band. \hiligt utilizes the LED data through the \textsc{A2LED} catalog \citep{Nugent83a}, which contains 114 sources\footnote{\url{https://heasarc.gsfc.nasa.gov/W3Browse/heao1/a2led.html}}. 32 of the 114 sources have a count rate of zero and \hiligt returns the count rate error as an upper limit in these cases. Conversion factors for a black body and powerlaw can be found in \citet[Fig.~5-6]{Nugent83a}. These are calculated for the ``1\,keV band'' (0.44--2.8\,keV). Again, \hiligt uses Eq.~\ref{eq:CV_old} to map this onto \hiligt's soft (0.2--2\,keV) band. 
Due to the differences in energy bands of all three catalogs with respect to the \hiligt bands, a systematic uncertainty of 20\% is added on the flux.

\heasarc does not provide a catalog with count rate data of the A3 experiment, which is therefore not accessed. The all-sky monitoring data of the A4 experiment in the 13--180\,keV range \citep{Levine84a} will be included in a future release of \hiligt.


\subsection{\einstein (1978--1981)}
A mission overview of \einstein is given by
\citet{1979ApJ...230..540G} and the user manual can be found in
\citet{RUM}.
We concentrate on the Wolter Type I telescope with the Imaging
Proportional Counter (IPC, \citealt{1981ITNS...28..869G,1984SAOSR.393.....H})
and the High Resolution Imager (HRI, \citealt{1977SPIE..106..196H}). 
The data of both instruments are interpolated to \hiligt's \textit{soft} band
(0.2--2.0\,keV).

\subsubsection{High Resolution Imager (HRI)}
The HRI was the first high-resolution X-ray camera on-board a
spacecraft. It had a high spatial resolution of 3\,\asec over the
central 25\,\amin of the focal plane, and even 2\,\asec within 5\,\amin of
the optical axis axis.
Two catalogs are available through \heasarc, \textsc{HRIIMAGE} \footnote{\url{https://heasarc.gsfc.nasa.gov/W3Browse/einstein/hriimage.html}} and \textsc{HRICFA} \footnote{\url{https://heasarc.gsfc.nasa.gov/W3Browse/einstein/hricfa.html}}.
The count rates of the \textsc{HRICFA} catalog do not have a time stamp or exposure time but can be linked to the \textsc{HRIIMAGE} catalog, which provides this information, using the sequence number field. The sequence number is further used as key to link the catalog entries to the images of the database (where we usually use the observation ID, see Sect.~\ref{sec:hiligt}).
As there is no direct background information in the catalogs, \hiligt provides background counts from a constant background estimate of $5\times 10^{-3}\,\cps\mathrm{arcmin}^{-2}$ \citep{1979ApJ...230..540G} with a circular source radius of 3.98\,\asec \citep[Ch.~4, p.~10]{RUM}. 
The column \verb|Net_Time| in \textsc{HRIIMAGE} is used as exposure time. 
Furthermore, we adopt the catalog search radii of 1\,\amin for \textsc{HRICFA}, and 15\,\amin for \textsc{HRIIMAGE} from the \heasarc.

\begin{figure}
\centering
\begin{subfigure}[t]{0.35\linewidth}
    \includegraphics[width=1\linewidth]{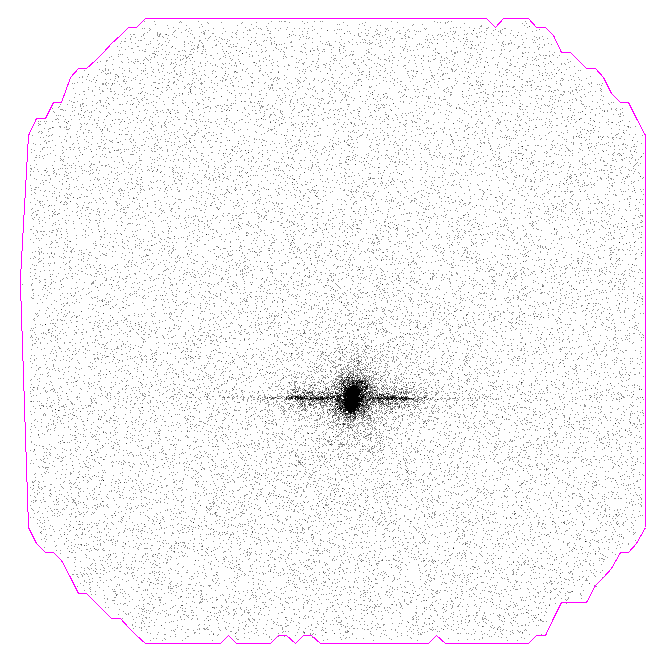}
    \subcaption{LMXB
      \href{http://simbad.u-strasbg.fr/simbad/sim-id?Ident=3U+1636\%E2\%88\%9253++&NbIdent=1&Radius=2&Radius.unit=arcmin&submit=submit+id}{3U~1636$-$53}
    (Seq. No. \texttt{6769})}
    \label{fig:hri_sample:a}
\end{subfigure}
\hfil
\begin{subfigure}[t]{0.35\linewidth}
    \includegraphics[width=1\linewidth]{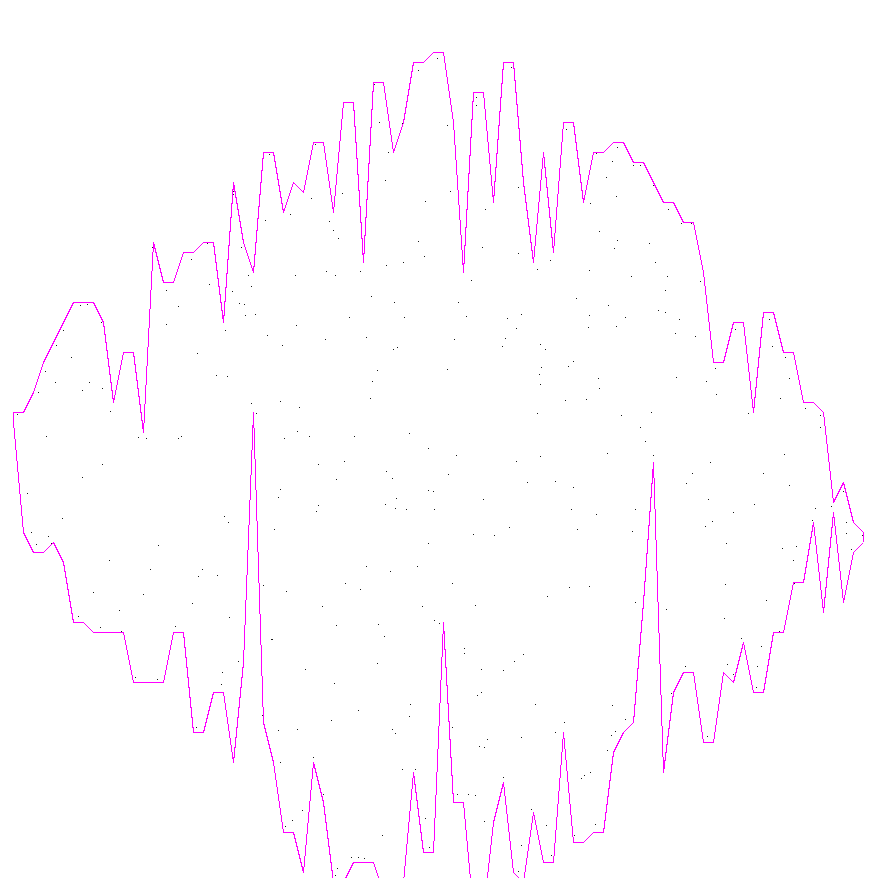}
    \subcaption{\href{http://simbad.u-strasbg.fr/simbad/sim-id?Ident=4U+0813-38&NbIdent=1&Radius=2&Radius.unit=arcmin&submit=submit+id}{4U~0813$-$38}
    (Seq. No. \texttt{981})}
    \label{fig:hri_sample:b}
\end{subfigure}
\vfil
\begin{subfigure}[t]{0.35\linewidth}
    \includegraphics[width=1\linewidth]{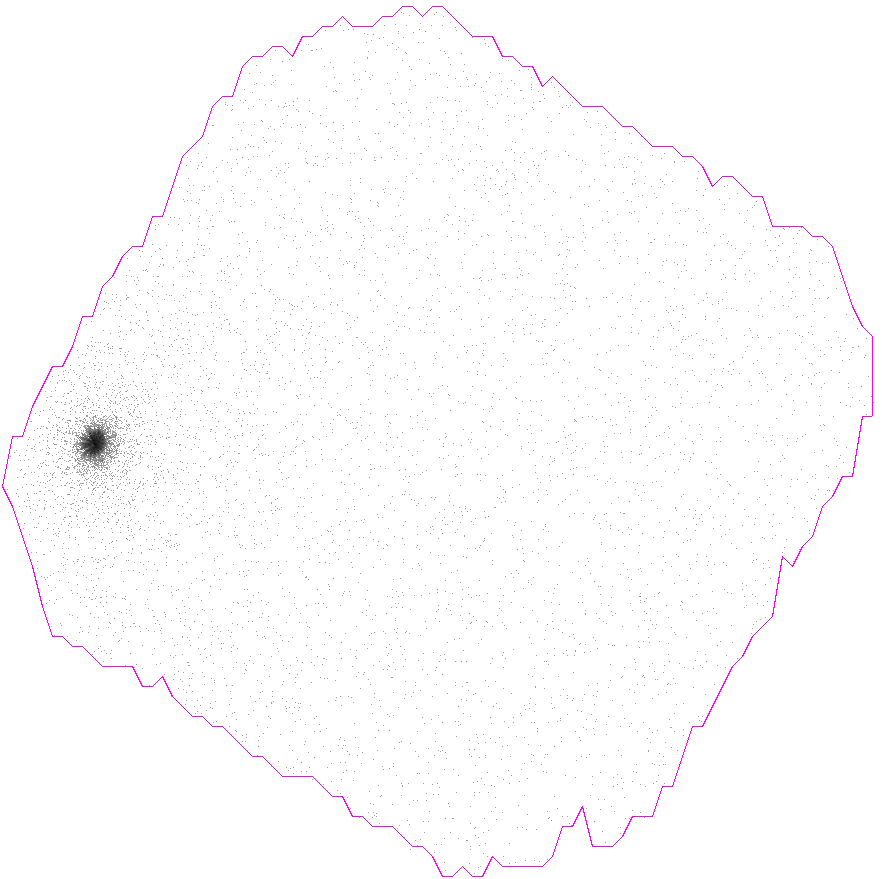}
    \subcaption{HMXB
      \href{http://simbad.u-strasbg.fr/simbad/sim-id?Ident=Cyg+X-1&NbIdent=1&Radius=2&Radius.unit=arcmin&submit=submit+id}{Cygnus~X-1}
    (Seq. No. \texttt{3})}
    \label{fig:hri_sample:c}
\end{subfigure}
\hfil
\begin{subfigure}[t]{0.35\linewidth}
    \includegraphics[width=1\linewidth]{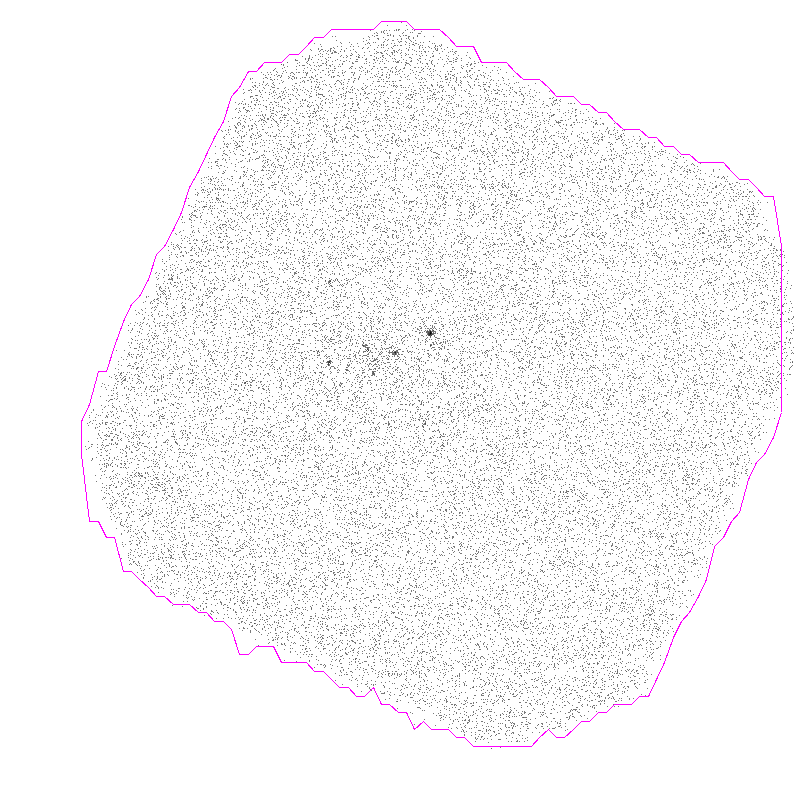}
    \subcaption{\href{http://simbad.u-strasbg.fr/simbad/sim-id?Ident=M31&submit=submit+id}{M31}
    (Seq. No. \texttt{7066})}
    \label{fig:hri_sample:d}
\end{subfigure}
\caption{Example of the footprints calculated with our custom-built
  footprint algorithm for the \einstein HRI images.}
\label{fig:hri_sample}
\end{figure}

In case no catalog entry is found for the requested position, we use 870 HRI images, produced by the Harvard Center for Astrophysics, to calculate upper limits. 
The images were taken with a 24\,\amin FOV in the range
0.15--3.5\,keV. The footprints are calculated with our custom-built
algorithm (see an example in Fig.~\ref{fig:hri_sample}).
About 5\% of the images exhibit regions with low count rates, making an exact border distinction difficult and resulting in frayed footprints. Typically, this reduction in footprint area is acceptable (see Fig.~\ref{fig:hri_sample}\subref{fig:hri_sample:b}) and the image can be used for upper limit calculations. However, for 34 images the footprint becomes too distorted, or other file problems arise. These are excluded from the database, which contains a total of 836 footprints.

\begin{figure}
  \centering
  \includegraphics[width=.8\linewidth]{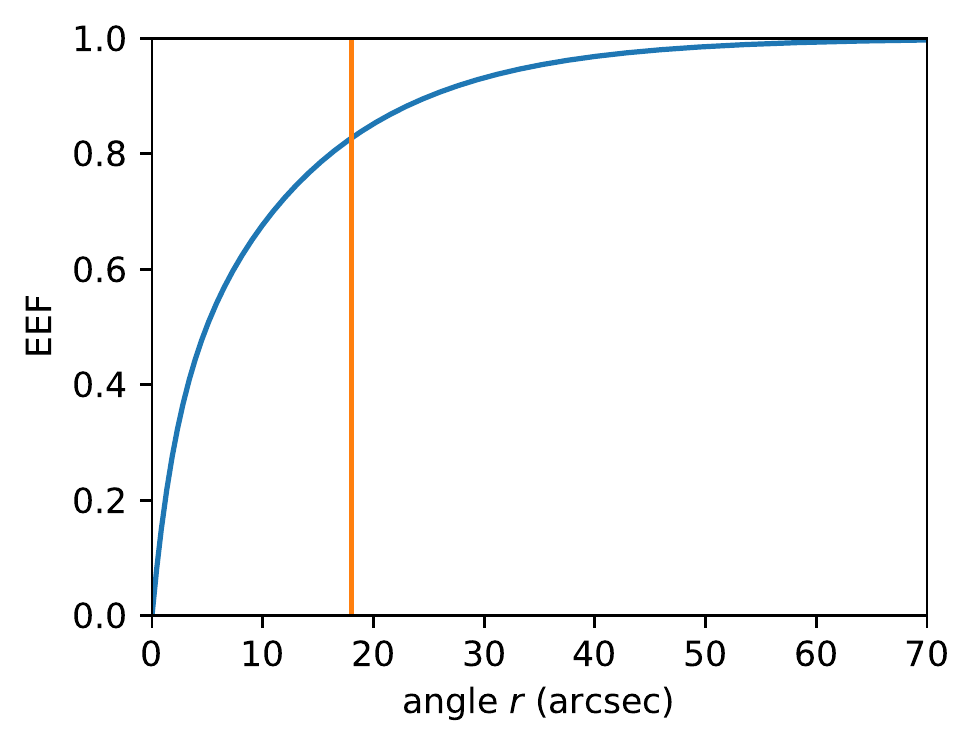}
  \caption{\einstein HRI encircled energy fraction, computed by integrating Eq.~\ref{eq:einstein_psf} using Eq.~\ref{eq:eef}. The orange line shows the 18\,\asec source radius.}
  \label{fig:einstein_hri_eef}
\end{figure}

The PSF \citep[Ch.~4,~p.~10ff]{RUM} can be approximated by
\begin{equation}
\label{eq:einstein_psf}
\begin{aligned}
    \mathrm{PSF}(r)~=~
    & 2.885\cdot10^{-2} \cdot \exp\left (-\frac{r}{1.96\,\asec}\right ) \\
    & + 0.01 \cdot \exp \left (-\frac{r}{12.94\,\asec} \right )\,\mathrm{arcsec}^{-2}
\end{aligned}
\end{equation}
This approximation is accurate for a 5\,\amin circle around the field
center (on-axis) at 1.5\,keV and $r\le 1\,\amin$.  We were unable to  find quantitative information about the off-axis behavior of the PSF outside of 5\,\amin.
The source radius for upper limit calculations is set to 18\,\asec
(\einstein's ``standard circle''), corresponding to an EEF of 0.83
for this empirical model (see Fig.~\ref{fig:einstein_hri_eef}).
No vignetting correction is available to our knowledge. This may overestimate upper limits at large off-axis angles because the vignetting effectively decreases the exposure time, which increases the count rate. Therefore the upper limits for sources detected at the edges may be too stringent.

\subsubsection{Imaging Proportional Counter (IPC)}
The IPC had lower 1\,\amin spatial and spectral resolution than the HRI but a full focal plane coverage.
Two identical IPC detectors (except for the entrance material) were mounted on \einstein.
The background count rate was $\sim 10^{-2}\,\cps$ and the instruments
sensitivity $1\,\cps$ per $4\times 10^{-11}\,\ergcms$ \citep{1979ApJ...230..540G}.

There are two catalogs for the IPC containing 4132 images in the
0.2--3.5\,keV range, \textsc{IPCIMAGE} \footnote{\url{https://heasarc.gsfc.nasa.gov/W3Browse/einstein/ipcimage.html}}
and \textsc{IPC} \footnote{\url{https://heasarc.gsfc.nasa.gov/W3Browse/einstein/ipc.html}}.
We use cone search radii of 2\,\amin for the \textsc{IPC} catalog and
15\,\amin for \textsc{IPCIMAGE}, respectively. To infer all necessary
information, we merge these two catalogs by the sequence number in
\textsc{IPC} and the object column in \textsc{IPCIMAGE}.  We also use
this sequence number as key to link the catalog entries to the images
in the \einstein database.  The following fields are queried from the
\textsc{IPCIMAGE} catalog:
\begin{itemize}
    \item \verb|Object|: Same as sequence number in
      \textsc{IPC} (excluding the first letter); links the entries
      to the \textsc{IPC} catalog and to the images in \hiligt's \einstein database
    \item \verb|Live_Time|: Exposure time, equaling the keyword
      \verb|TIME_LIV| in the images
    \item \verb|Time| and \verb|End_Time|: Start and end time of the
      observation
\end{itemize}
The \textsc{IPC} catalog gives the source and background count rates of the
observations \citep{1990eoci.book.....H},
\begin{itemize}
\item \verb|Sequence_Num|: A two to five digit number which uniquely
  identifies an \einstein observation, and which we use as link to
  the \textsc{IPCIMAGE} entry
\item \verb|Count_Rate| and \verb|Count_Rate_Error|: Corrected count
  rate in the 0.2--3.5\,keV band
\item \verb|Background_Count|: The total number of background counts
  in the $2.4\,\amin\times 2.4\,\amin$ detection cell
\end{itemize}

\begin{figure}
\centering
\begin{subfigure}[t]{0.41\linewidth}
    \includegraphics[width=1\linewidth]{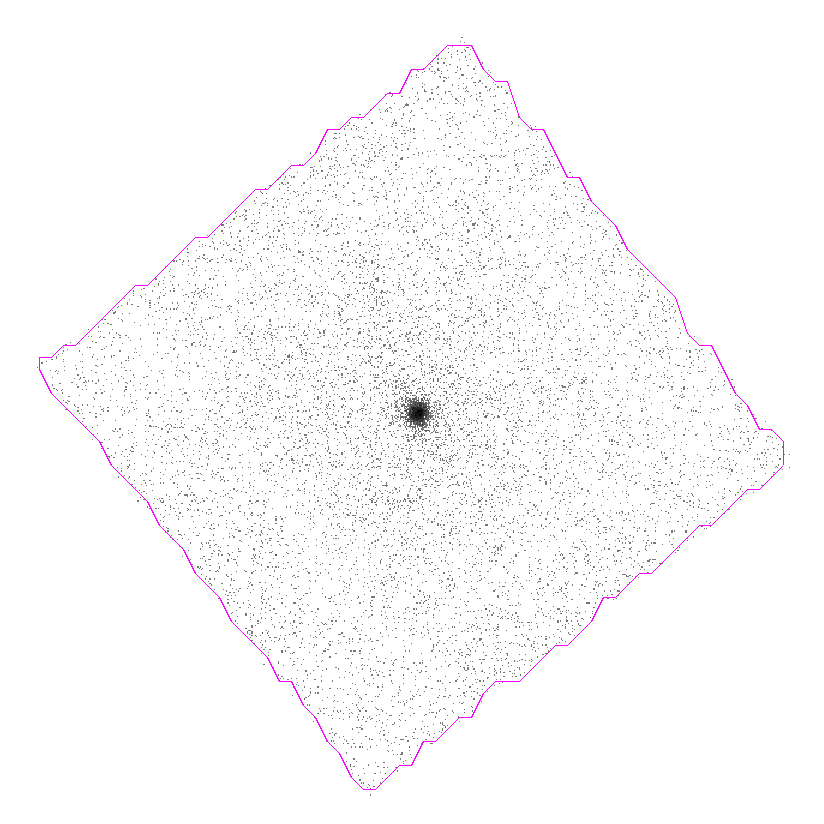}
    \subcaption{HMXB
      \href{http://simbad.u-strasbg.fr/simbad/sim-id?Ident=Cen+X-3&submit=submit+id}{Cen~X-3}
    (Seq. No. \texttt{817})}
    \label{fig:ipc_sample:a}
\end{subfigure}
\hfil
\begin{subfigure}[t]{0.41\linewidth}
    \includegraphics[width=1\linewidth]{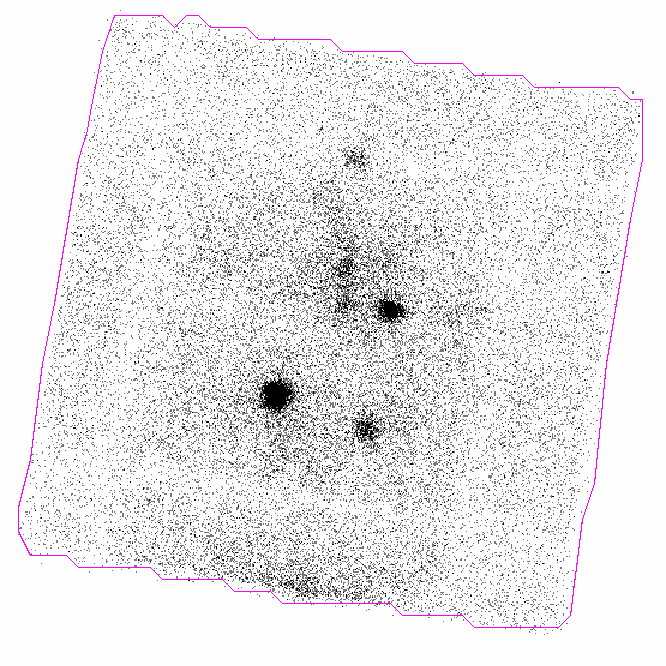}
    \subcaption{\href{http://simbad.u-strasbg.fr/simbad/sim-id?Ident=30+Dor&NbIdent=1&Radius=2&Radius.unit=arcmin&submit=submit+id}{30~Dor},
      a cluster of stars in the LMC (Seq. No. \texttt{4559})}
    \label{fig:ipc_sample:d}
\end{subfigure}
\vfil
\begin{subfigure}[t]{0.41\linewidth}
    \includegraphics[width=1\linewidth]{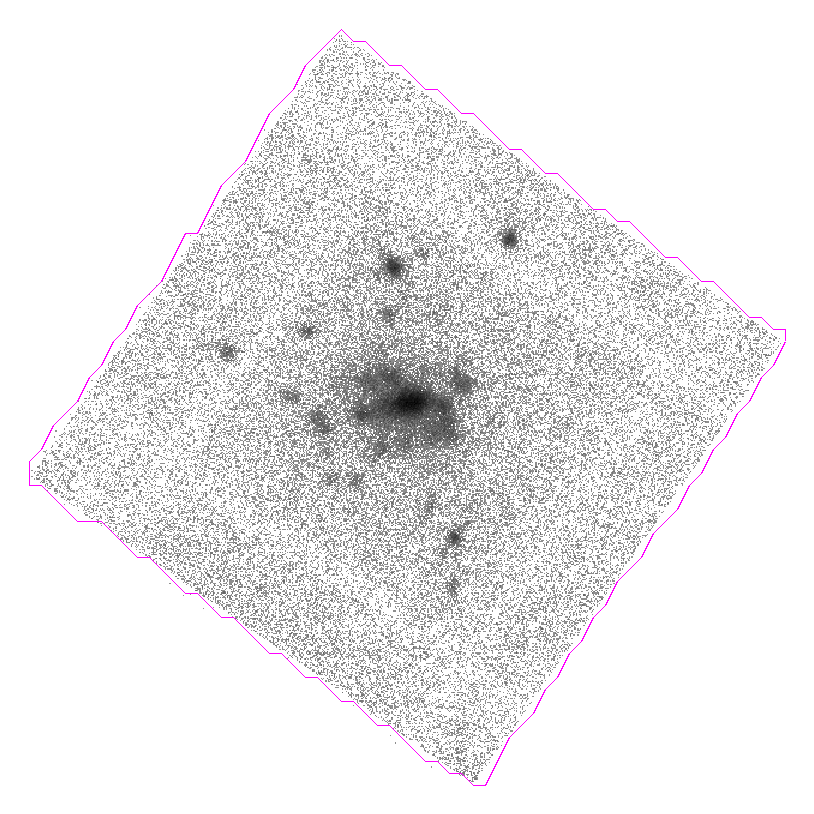}
    \subcaption{\href{http://simbad.u-strasbg.fr/simbad/sim-id?Ident=M31&submit=submit+id}{M31}
    (Seq. No. \texttt{574})} 
    \label{fig:ipc_sample:e}
\end{subfigure}
\hfil
\begin{subfigure}[t]{0.41\linewidth}
    \includegraphics[width=1\linewidth]{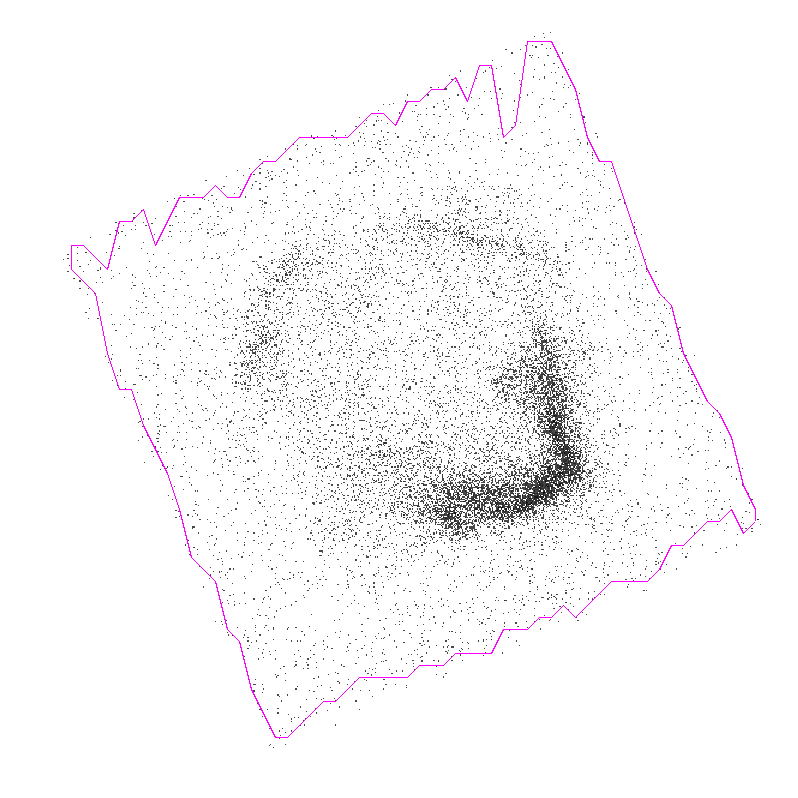}
    \subcaption{SNR
      \href{http://simbad.u-strasbg.fr/simbad/sim-id?Ident=MSH+14-63&NbIdent=1&Radius=2&Radius.unit=arcmin&submit=submit+id}{MSH~14$-$63}
    (Seq. No. \texttt{2164})}
    \label{fig:ipc_sample:g}
\end{subfigure}
\caption{Example of \einstein IPC footprints.}
\label{fig:einstein_ipc_footprint}
\end{figure}

The images available at the \heasarc have been smoothed with a
$\sigma=32\,\asec$ Gaussian and were background-subtracted, which makes
the processing with \eupper difficult because it works on
uncorrected images. We therefore use the event files and
create images with the SAOImageDS9 software \citep[Beta version
  8.0rc4]{ds9Joye03}. Examples for the footprint calculated with our
custom-built algorithm can be found in
Fig.~\ref{fig:einstein_ipc_footprint}. In total the \einstein
database contains 3923 footprints with the corresponding images.

\begin{figure}
  \centering
  \includegraphics[width=.8\linewidth]{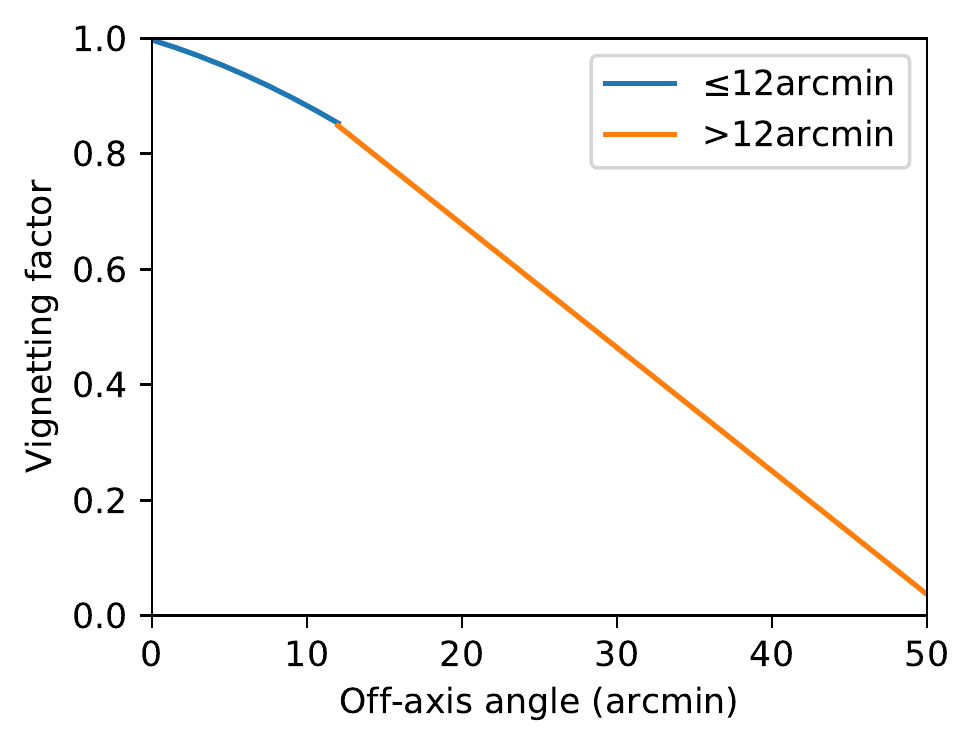}
  \caption{\einstein IPC vignetting function
    (Eq.~\ref{eq:einstein_ipc_vign1}--\ref{eq:einstein_ipc_vign2}).}
  \label{fig:einstein_ipc_vign}
\end{figure}

The vignetting (see \citealt[Ch.~5, p.~18]{RUM}, and plotted in
Fig.~\ref{fig:einstein_ipc_vign}) as a function of off-axis angle $\theta$
(the difference between the optical axis and source position in arcmin) is
\begin{eqnarray}
\label{eq:einstein_ipc_vign1}
 \mathrm{Vign}(\theta\leq 12\,\amin) &= (-0.0003125 \cdot \theta - 0.00825) \cdot \theta + 0.997\\
\label{eq:einstein_ipc_vign2}
 \mathrm{Vign}(\theta>12\,\amin) &= 1.1049 - 0.02136 \cdot \theta \qquad .
\end{eqnarray}

\begin{figure}
  \centering
  \includegraphics[width=.8\linewidth]{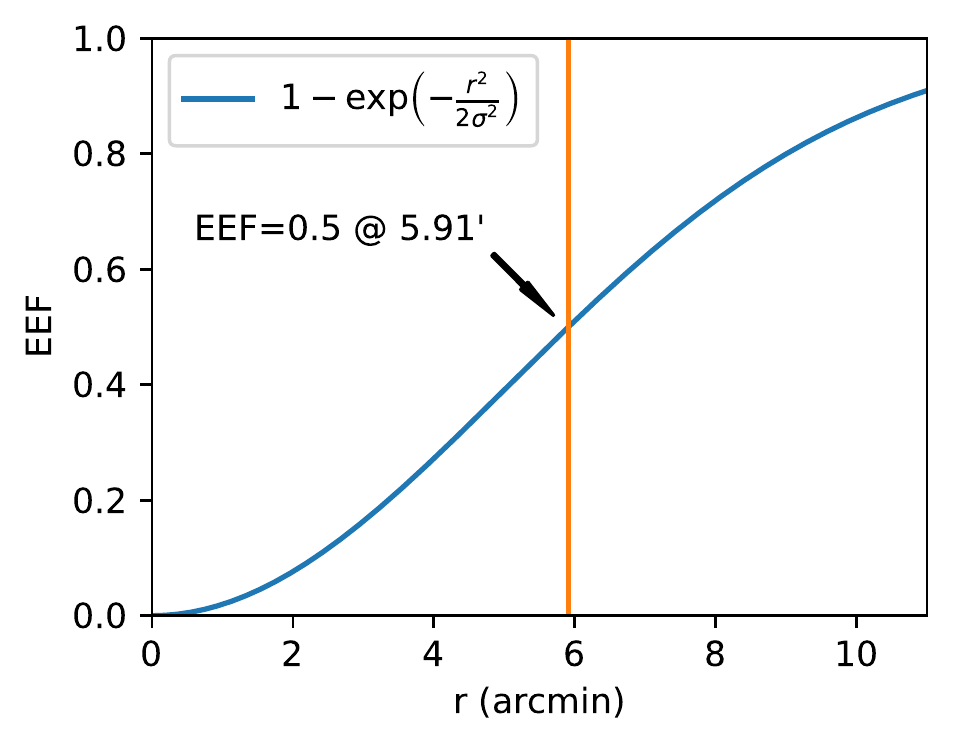}
  \caption{\einstein IPC EEF function (Eq.~\ref{eq:einstein_ipc_eef})
    with source radius at 5.91\,\amin.}
  \label{fig:einstein_ipc_eef}
\end{figure}

Under the assumption of a circular Gaussian response with width $\sigma$, \citet[p.~24]{1984SAOSR.393.....H} show that the fraction of total power enclosed within radius $r$ (see Fig.~\ref{fig:einstein_ipc_eef}) is given by
\begin{equation}
    \label{eq:einstein_ipc_eef}
    \mathrm{EEF}(r) = 1-\exp \left ( -\frac{1}{2} \frac{r^2}{\sigma^2} \right )
\end{equation}
\hiligt uses a source radius of 5.911\,\amin, corresponding to an EEF
of 0.5. This is a very large source radius, originating from the low 1\,\amin
spatial resolution of the IPC.


\subsection{\exosat (1983--1986)}
\label{sec:exosat}

The payload of \exosat consisted of two low energy imaging telescopes
(LE, \citealt{1981SSRv...30..495D}), a medium energy proportional
counter (ME, \citealt{1981SSRv...30..513T}) and a gas scintillation
proportional counter \citep{1981SSRv...30..525P}. A full mission
overview is given by \citet{1988MmSAI..59....7W}. The \exosat LE
instrument is of great importance for \hiligt, since it produced 3677
images which provide the possibility of upper limit estimates.
Additionally, the \exosat ME slew and pointed catalogs are included in \hiligt.

\subsubsection{\exosat LE}
The Low Energy (LE) instrument consisted of two identical Wolter I telescopes with a focal length of 1.1\,m. Each instrument had a \textit{channel multiplier array} (CMA1 and CMA2) assembled in the focal plane.
The energy band of the instrument was determined by the effective area of the applied filter (see Fig.~4 of \citealt{1981SSRv...30..495D}). Overall, the LE instrument had an energy range of $\sim$0.05--2.0\,keV. The CMA2 instrument failed on 1983-10-28, only five months into the three year mission.
Therefore, most images were detected by CMA1. 
The CMA detectors were sensitive to UV radiation and bright O- and B-stars could contaminate the image quality. \exosat LE therefore used optical blocking filters. From the nine filters on the filter wheel (FW), the most attendant to the \exosat LE images used by \hiligt are: Polypropylene (``PPL'' at FW Pos.~2), thick 400\,nm Lexan (``4Lx'' at FW Pos.~3), Aluminium-parylene (``Al/P'' at FW Pos.~6), thin 300\,nm Lexan (``3Lx'' at FW Pos.~7), and a Boron (``Bor'' at FW Pos.  8) filter. Usually, 3Lx, Bor and Al/P were used.
By applying filter dependent conversion factors, \hiligt gives the flux in the \textit{soft} 0.2--2\,keV band.

\hiligt accesses the \textsc{LE} catalog\footnote{\url{https://heasarc.gsfc.nasa.gov/W3Browse/exosat/le.html}} with a catalog search radius of 1\,\amin.
The catalog gives the background in the field \texttt{background\_per\_sqpix} in units of cts pixel$^{-2}$.
One pixel equals 4\,\asec and the on-axis HEW of the PSF is 24\,\asec \citep{1988MmSAI..59....7W}.
Thus, we use a circular background region of area $\pi(12\,\mathrm{arcsec})^2$ and determine the background counts by
\begin{equation}
  \texttt{bkg\_cts} = \texttt{background\_per\_sqpix}\cdot \pi (12\,\asec)^2 / (4\,\asec)^2
\end{equation}

\begin{figure}
\centering
\begin{subfigure}[t]{0.45\linewidth}
  \centering
  \includegraphics[width=1\linewidth]{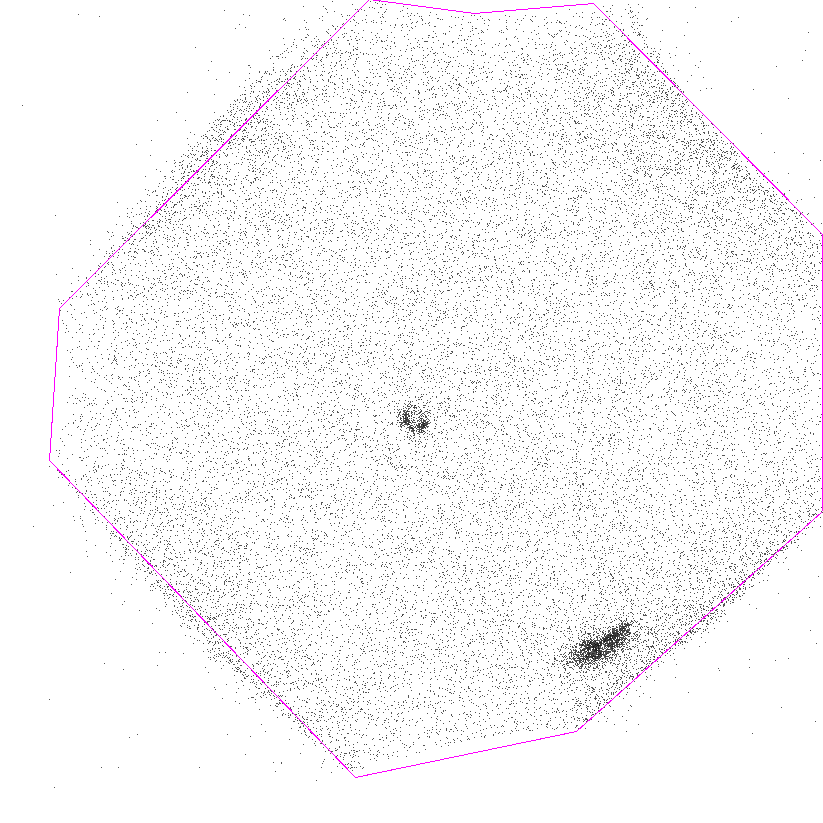}
  \subcaption{CMA1 PPL obs. of SNR
    \href{https://heasarc.gsfc.nasa.gov/FTP/exosat/data/cma1/im_gif/a01979.gif}{Cas~A} (Obs.~ID \texttt{01979})}
  \label{fig:exosat_le_footprint:a}
\end{subfigure}
\hfil
\begin{subfigure}[t]{0.45\linewidth}
  \centering
  \includegraphics[width=1\linewidth]{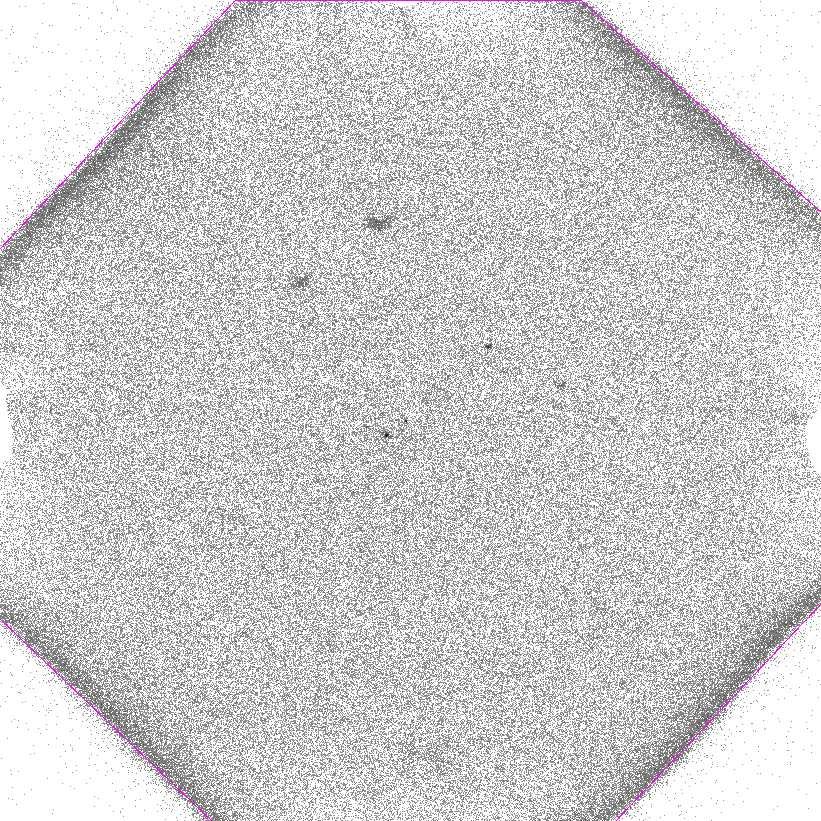}
  \subcaption{CMA2 3Lx obs. of HMXB
    \href{https://heasarc.gsfc.nasa.gov/FTP/exosat/data/cma2/im_gif/a02788.gif}{Cyg~X-3} (Obs.~ID \texttt{02788})}
  \label{fig:exosat_le_footprint:b}
\end{subfigure}
\vfil
\begin{subfigure}[t]{0.45\linewidth}
  \centering
  \includegraphics[width=1\linewidth]{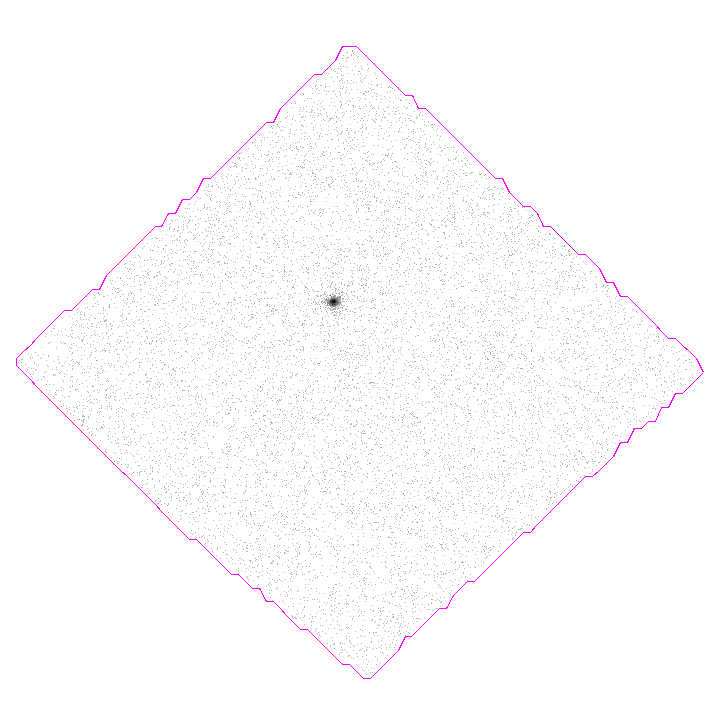}
  \subcaption{CMA1 3Lx obs. of the pulsar
    \href{https://heasarc.gsfc.nasa.gov/FTP/exosat/data/cma1/im_gif/a27503.gif}{Her~X-1} (Obs.~ID \texttt{27503})}
  \label{fig:exosat_le_footprint:c}
\end{subfigure}
\hfil
\begin{subfigure}[t]{0.45\linewidth}
  \centering
  \includegraphics[width=1\linewidth]{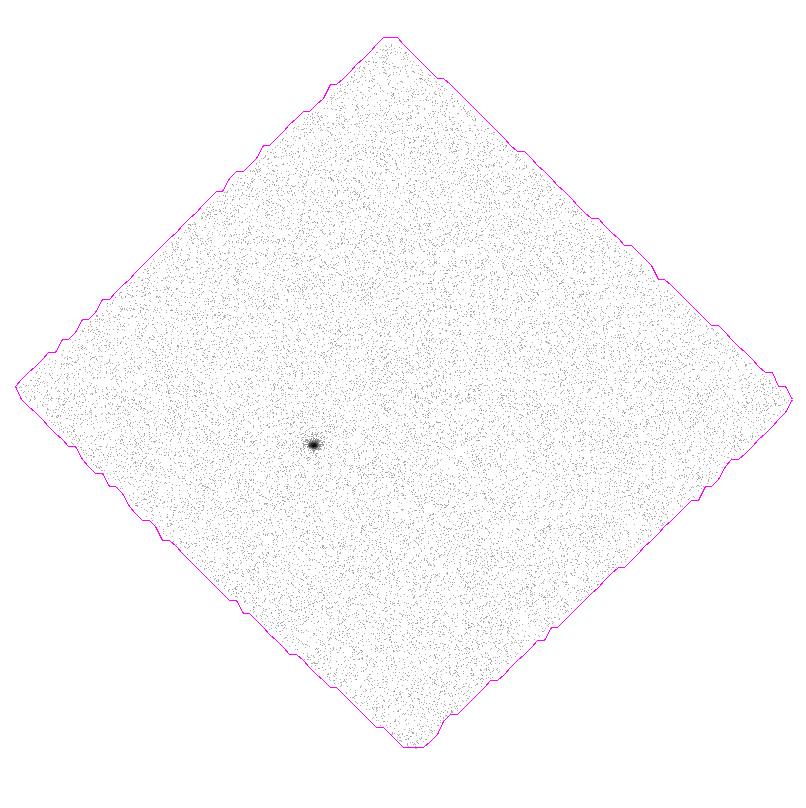}
  \subcaption{CMA1 3Lx obs. of the CV
    \href{https://heasarc.gsfc.nasa.gov/FTP/exosat/data/cma1/im_gif/a18315.gif}{VV~Pup} (Obs.~ID \texttt{18315})}
  \label{fig:exosat_le_footprint:d}
\end{subfigure}
\caption{An example of \exosat LE images with the calculated
  footprints. Fig.~(\subref{fig:exosat_le_footprint:a}) shows the most
  abundant ($\sim$81\%) octagonal CMA1 footprint shape with nine defined
  footprint points. In Fig.~(\subref{fig:exosat_le_footprint:b}) one
  can see an example of the CMA2 octagonal footprint shape
  ($\sim$11\%) with eight defined points (see
  Table~\ref{tab:exosat_pixelpos}). The illuminated edges are most
  likely stray light effects.
  Fig.~(\subref{fig:exosat_le_footprint:c}--\subref{fig:exosat_le_footprint:d})
  show the two remaining rectangular and smaller shapes of CMA1 3Lx, which were
  computed with our custom-built algorithm.}
\label{fig:exosat_le_footprint}
\end{figure}

We download 3677 \exosat LE images from the \heasarc\footnote{\url{https://heasarc.gsfc.nasa.gov/FTP/exosat/data}} and identify four different image shapes in the data set: CMA1 images with the 3Lx filter have a rectangular ($\sim$8\%) shape, while the other CMA1 filters and CMA2 images have an octagonal ($\sim$92\%) shape.
The octagonal CMA1 and CMA2 images tend to have illuminated
edges -- likely due to stray light -- with noise outside of the main
shape. Some also exhibit extremely low count rates in the whole
image. The latter makes an automatized footprint calculation with our
footprint algorithm difficult. The footprint position on the detector
plane, however, is constant for all octagonal images. Therefore, we
define nine (eight for CMA2) fixed footprint points, specified in
Table~\ref{tab:exosat_pixelpos}.  An example can be seen in
Fig.~\ref{fig:exosat_le_footprint}\subref{fig:exosat_le_footprint:a},\subref{fig:exosat_le_footprint:b}.
The remaining rectangular images do not exhibit noise outside of the
FOV, which makes them ideally suited for our footprint algorithm
(Fig.~\ref{fig:exosat_le_footprint}\subref{fig:exosat_le_footprint:c},\subref{fig:exosat_le_footprint:d}).

The observation ID is five letters long and occurs as \verb|ORIGFILE|
keyword in the FITS images. It is also contained in the filename (letter
5--9) and in the catalog as \verb|File_Image| with an additional
letter in front. Removing the letter allows us to generate a key that can
be used to link the image to the corresponding catalog entry.

The point spread function of \exosat has 50\% enclosed power (HEW) at
24\,\asec on-axis, degrading to 4\,\amin at 1\,\deg off-axis. Vignetting
effects of the telescopes reduced the off-axis effective area to 45\%
of its peak value at 1\,\deg off-axis \citep[p.~11]{1988MmSAI..59....7W}.
In case the image is analyzed with \eupper (i.e., no catalog entry available),
\hiligt uses a source radius of 1\,\amin.


\subsubsection{\exosat ME}
The Medium Energy (ME) instrument of \exosat consisted of eight proportional counters, which took photometric data (lightcurves) in the 1--20\,keV (Argon filled-gas cell) and 5--50\,keV (Xenon filled) energy range. 
For more information about \exosat ME we refer to \citet{1981SSRv...30..513T} and \citet{1999AAS..134..287R}.

\hiligt accesses the \textsc{ME} catalog\footnote{\url{https://heasarc.gsfc.nasa.gov/W3Browse/exosat/me.html}} for the pointed mission phase with a search radius of 45\,\amin.
This catalog provides 2291 entries with start and end time, count rate in the 1--15\,keV band, error, and exposure.
The slew phase covered 98\% of the sky and the \textsc{EXMS} catalog\footnote{\url{https://heasarc.gsfc.nasa.gov/W3Browse/exosat/exms.html}} contains 1210 detected sources \citep{1999AAS..134..287R}. \hiligt accesses this catalog with a search radius of 60\,\amin. Next to the count
rate (1--8\,keV), error, and detection time, this catalog provides no processable
information for \hiligt.
Using PIMMS, we calculate two sets of conversion factors with an input range of 1--15\,keV and 1--8\,keV for the ME and EXMS catalogs, respectively. To compare the data to the other missions, both catalogs are mapped onto \hiligt's \textit{hard} (2--12\,keV) band.


\subsection{\ginga (1987-1991)}

\ginga \citep{1987ApL....25..223M} was the third Japanese X-ray mission and had a very large energy range from 1--500\,keV. The main instrument was the Large Area Proportional Counter (LAC, 1.5--37\,keV, \citealt{1989PASJ...41..345T}). \ginga LAC had no imaging capability, thus upper limit calculations from images is not possible in this case. 
However, we include the \textsc{GINGALAC} catalog\footnote{\url{https://heasarc.gsfc.nasa.gov/W3Browse/ginga/gingalac.html}}.
The 1.5--37\,keV range is interpolated to the \textit{hard} band
(2--12\,keV). \ginga was in a low-earth orbit and regularly crossed
Earth's radiation belts and a further background source is the diffuse
cosmic X-ray background. Count rates of the catalog are background subtracted \citep[see][]{1989PASJ...41..373H}.
\ginga's pointing stability was about 6\,\amin \citep{1989PASJ...41..345T} and we use a conservative 10\,\amin as catalog search radius. The catalog contains 419 entries.
We plan to include the ASM lightcurve data in a future \hiligt release. 


\subsection{\rosat (1990--1998)}

A position sensitive proportional counter (PSPC) and a High Resolution Imager (HRI) were located in the focal plane of \rosat \citep{1987SPIE..733..519P}.
A second telescope with a wide-field camera \citep{1993MNRAS.260...77P}, operating in the hard UV range (0.06--0.2\,keV), is not included in \hiligt (also note that the UV instrument was very insensitive due to the Sun). Due to its long operational period, large sky coverage and high exposure times, \rosat is a key mission for \hiligt, and we include both images and available catalogs. For a mission overview we refer to the ROSAT User's Handbook by \citet{ROSAT_User_Handbook}.

\subsubsection{Position Sensitive Proportional Counter (PSPC)}

The PSPC consisted of multi-wire proportional counters and had modest
energy and high spatial resolution (25\,\asec at 1\,keV) with a
circular 2\,\deg diameter FOV. Two redundant units were assembled on a
carousel, PSPC-B was used for the pointed phase while detector PSPC-C
was used for the survey. Shadows of the wires in the detector could
be suppressed by dithering, however, shadows originating from the
mirror mount (``spider'') remain in the images (see
Fig.~\ref{fig:rosat_pspc_ptn_footprint}). We note that if the source
or background position coincides with the spider, the resulting upper
limit may be biased. We use a source radius of
100\,\asec for the PSPC instrument and interpolate its 0.1--2.4\,keV passband to
\hiligt's \textit{soft} band (0.2--2\,keV). The same conversion factors are used for the RASS and pointed data.

\subsubsection{PSPC Pointed}
\label{sec:rosat_pspc}

\begin{figure}
\centering
\begin{subfigure}[t]{0.49\linewidth}
    \includegraphics[width=1\linewidth]{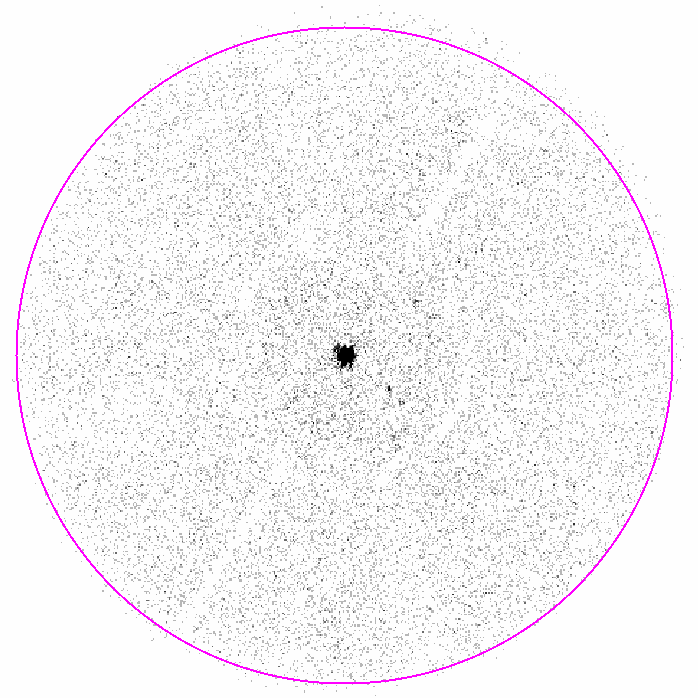}
    \subcaption{Quasar
      \href{http://simbad.u-strasbg.fr/simbad/sim-basic?Ident=3C273&submit=SIMBAD+search}{3C~273}
    (Obs.~ID \texttt{UK701265P.N1})}
    \label{fig:rosat_pspc_ptn_footprint:a}
\end{subfigure}
\hfil
\begin{subfigure}[t]{0.49\linewidth}
    \includegraphics[width=1\linewidth]{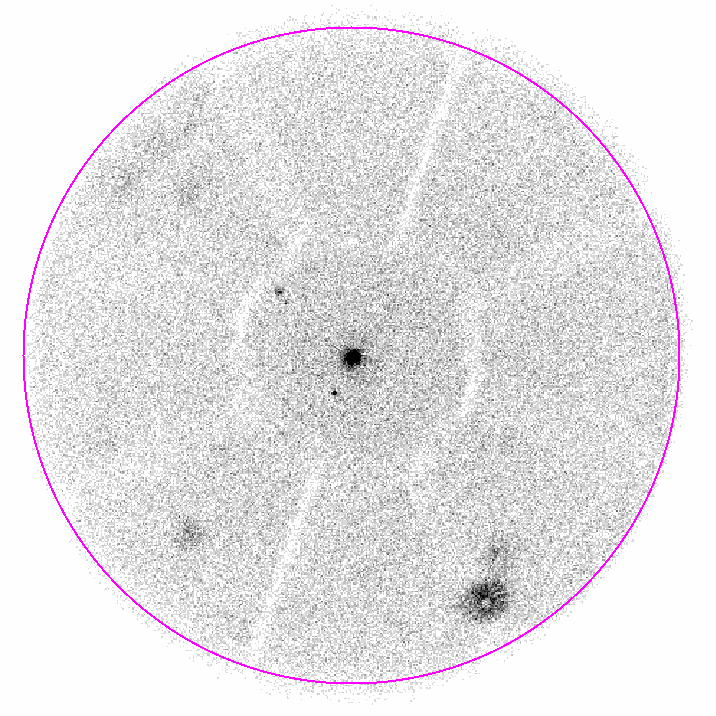}
    \subcaption{HMXB
      \href{http://simbad.u-strasbg.fr/simbad/sim-id?Ident=Vela+X-1&NbIdent=1&Radius=2&Radius.unit=arcmin&submit=submit+id}{Vela~X-1}
    (Obs.~ID \texttt{WG400313P.N1})}
    \label{fig:rosat_pspc_ptn_footprint:b}
\end{subfigure}
\caption{Examples of \rosat PSPC pointed images with the circular
  footprint. The purple circle has a radius of 0.95\,\deg. The radial
  spikes are shadows of the so-called spider (the mount of the
  nested shells). Point sources become extended and
  smeared out at large off-axis angles due to the off-axis aberration that smears out the PSF.}
\label{fig:rosat_pspc_ptn_footprint}
\end{figure}

The second \rosat source catalog of pointed observations (ROSPSPC/2RXP) contains 100\,048 source detections with arcsecond positions from 17.3\% of the sky, including 54\,133 high confidence detections \footnote{\url{https://heasarc.gsfc.nasa.gov/mail_archive/rosnews/msg00131.html}}.
\hiligt accesses the \textsc{ROSPSPC} catalog\footnote{\url{https://heasarc.gsfc.nasa.gov/W3Browse/rosat/rospspc.html}} with a search radius of 30\,\asec. 
Furthermore, we use the 0.1--2.5\,keV images\footnote{\url{https://heasarc.gsfc.nasa.gov/FTP/rosat/data/pspc/processed\_data/}} for the calculation of upper limits. 
As footprint we choose a circle of radius 0.95\,\deg around the center. Examples of the resulting footprint are shown in Fig.~\ref{fig:rosat_pspc_ptn_footprint}. 
In total, our \rosat database contains 5490 footprints. Additionally, exposure and background maps are provided by the \heasarc. While the exposure maps are included into the upper limit calculation, the background maps are currently not included due to compatibility issues with the SAS.

\begin{figure}
  \centering
  \includegraphics[width=1\linewidth]{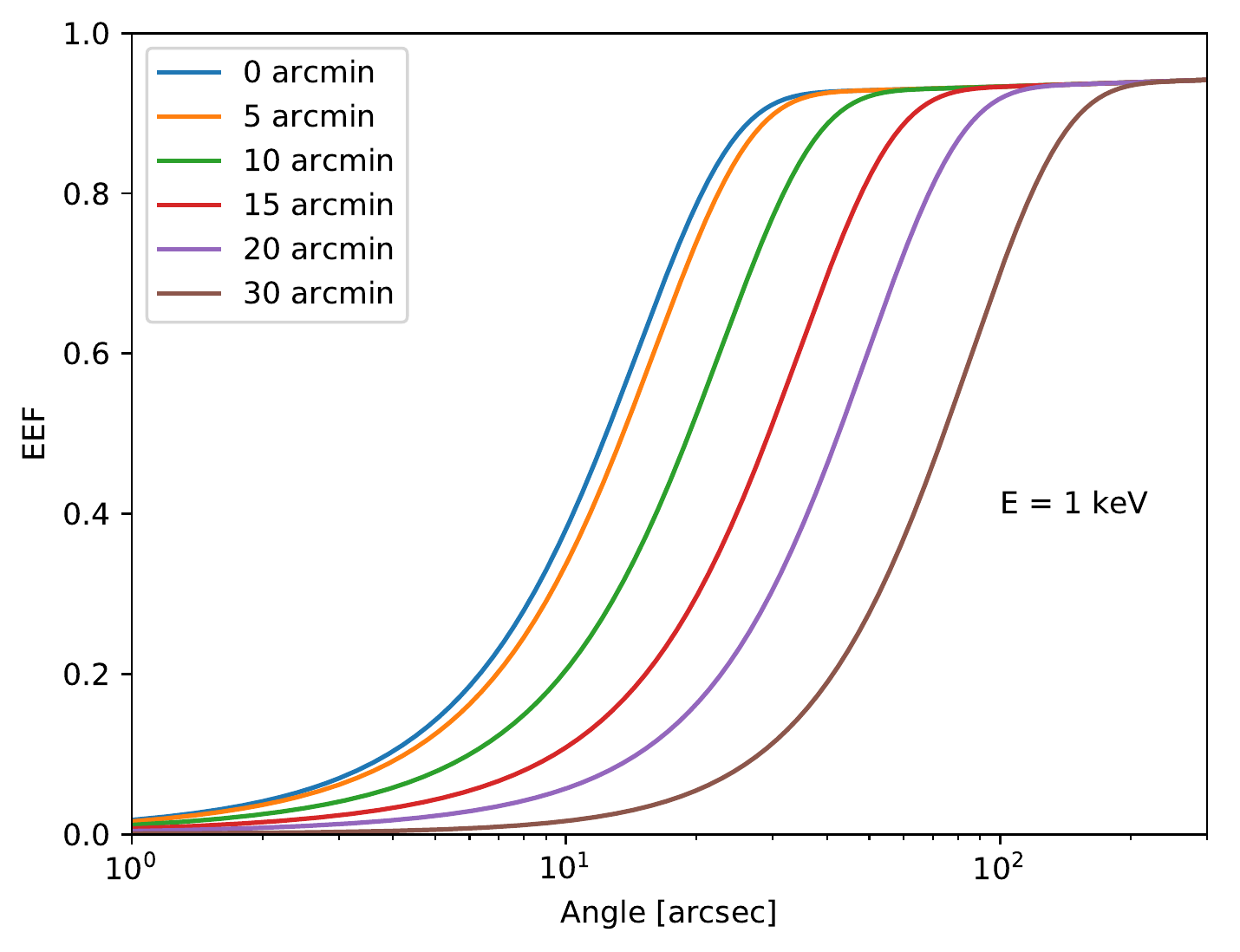}
  \caption{\rosat PSPC cumulative point spread function ($\equiv$
    EEF) as a function of angle with off-axis angle dependence
    at energy E=1\,keV \citep[Eq.~5.13]{EXSAS}.}
  \label{fig:rosat_pspc_ptn_eef}
\end{figure}

We implement the radius and off-axis angle dependent encircled energy
fraction according to \citep[Eq.~5.13]{EXSAS}.
The 1\,keV representation can be seen in Fig.~\ref{fig:rosat_pspc_ptn_eef}.
Note that the formula is energy dependent. This energy dependence is
impossible to include into the \hiligt code because the information about photon energy is irretrievably lost in the image unless one has access to the corresponding event files. 
Therefore, we assume\footnote{This approximation has a
  large uncertainty of about 50\% at low off-axis angles
  \citep[Fig.~5.17]{EXSAS}. 
  The energy variation decreases at larger off-axis angles and is in the
  order of 5\% at 30\,\amin off-axis \citep[Fig.~5.18]{EXSAS}.} a
constant energy of 1\,keV.

\begin{figure}
  \centering
  \includegraphics[width=1\linewidth]{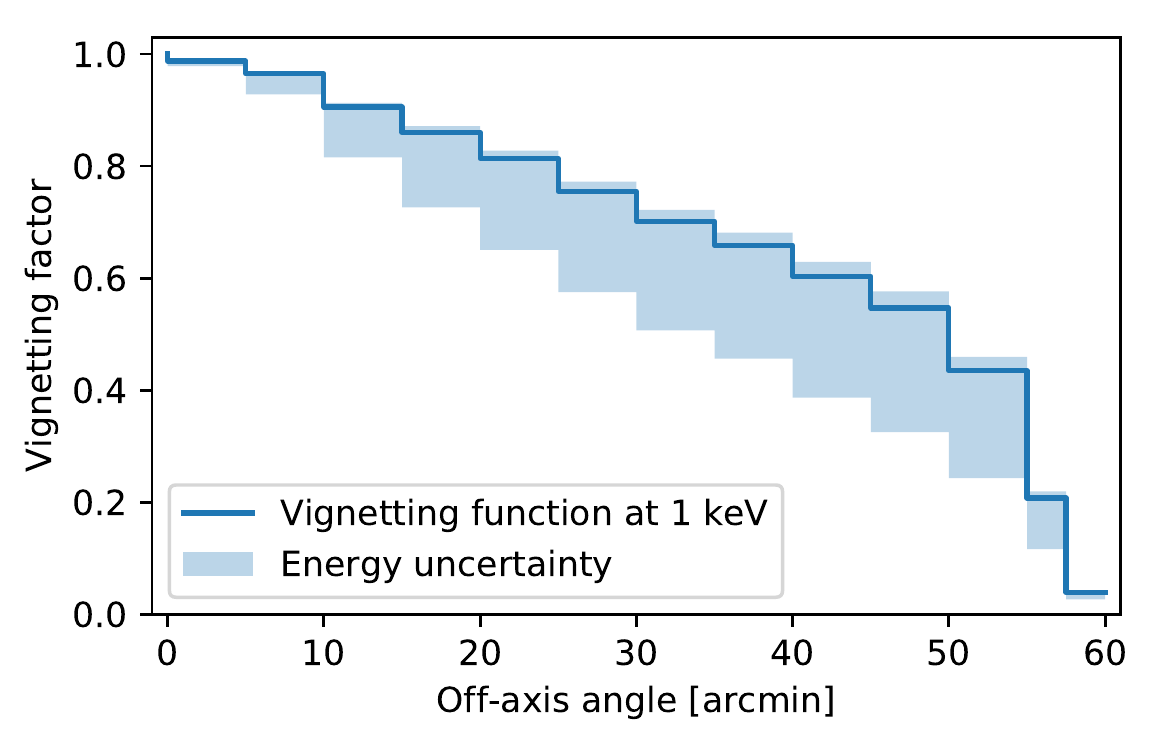}
  \caption{\rosat PSPC vignetting function. The blue line represents
    the implemented 1\,keV approximation. The shaded region shows the
    uncertainty originating from the energy dependence.}
  \label{fig:rosat_pspc_ptn_vign}
\end{figure}

We infer the vignetting correction factor from table
\texttt{vignet\_pspc.fits} which is available at the \mpe (MPE, see also Fig.~5.29 in \citealt{EXSAS}).
Due to the reasons outlined above, we only use the 1\,keV values and
obtain the vignetting correction function in
Fig.~\ref{fig:rosat_pspc_ptn_vign}.

\subsubsection{PSPC Survey}

\begin{figure}
\centering
\begin{subfigure}[t]{0.49\linewidth}
    \includegraphics[width=1\linewidth]{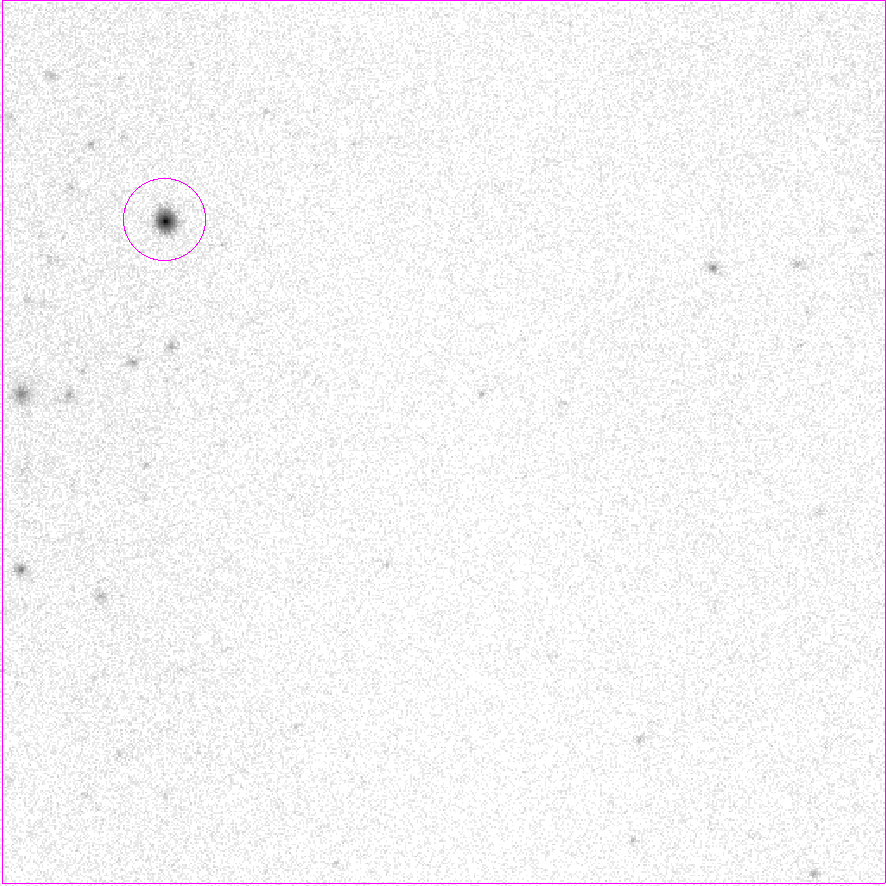}
    \subcaption{LMXB
      \href{http://simbad.u-strasbg.fr/simbad/sim-id?Ident=Her+X-1&submit=submit+id}{Her~X-1}
    (Obs.~ID \texttt{WG931139P\_N1})}
    \label{fig:rosat_pspc_rass_footprint:a}
\end{subfigure}
\hfil
\begin{subfigure}[t]{0.49\linewidth}
    \includegraphics[width=1\linewidth]{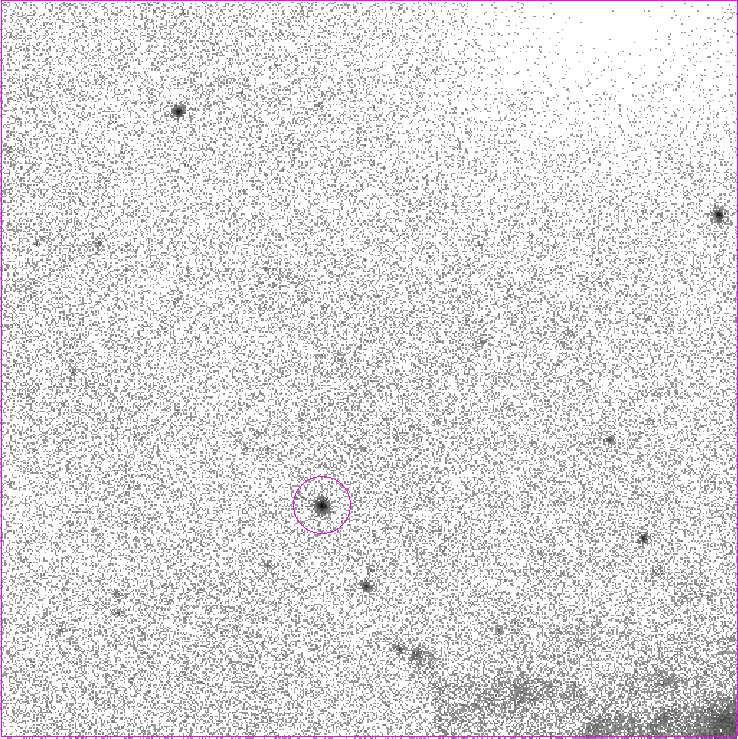}
    \subcaption{HMXB
      \href{http://simbad.u-strasbg.fr/simbad/sim-id?Ident=Vela+X-1&submit=submit+id}{Vela~X-1}
    (Obs.~ID \texttt{WG932420P\_N1})}
    \label{fig:rosat_pspc_rass_footprint:b}
\end{subfigure}
\caption{An example of \rosat PSPC survey images from the RASS catalog
  with rectangular footprints. The targets are indicated by the 100\,\asec source radius circle.}
\label{fig:rosat_pspc_rass_footprint}
\end{figure}

The Second \rosat All-Sky Survey (RASS) Point Source Catalog
(\textsc{RASS2RXS} \footnote{\url{https://heasarc.gsfc.nasa.gov/W3Browse/rosat/rass2rxs.html}})
originates from the survey phase of the mission between 1990 June and
1991 August and contains over 135\,000 sources at a likelihood
threshold of 6.5 \citep{2016AA...588A.103B}. \hiligt uses a catalog
search radius of 2\,\amin.  Since the downloaded survey images are
rectangular and represent the full FOV, the footprint is a simple
rectangle.  See Fig.~\ref{fig:rosat_pspc_rass_footprint} for two
example images. Our \rosat PSPC survey database contains 1378
footprints.

\subsubsection{High Resolution Imager}

Information about the HRI can be found in \citet{1987SPIE..733..519P},
\citet{ROSAT_User_Handbook}, and \citet{EXSAS}. The \textsc{ROSHRI}
catalog\footnote{\url{https://heasarc.gsfc.nasa.gov/W3Browse/rosat/roshri.html}} contains arcsecond positions and count rates for 56\,401
detected sources from 5393 \rosat HRI observations. In total, 1.94\%
of the sky is covered with 13\,452 high-confidence detections. The
catalog is accessed with a search radius of 1\,\amin.

\begin{figure}
  \centering
  \includegraphics[width=.9\linewidth]{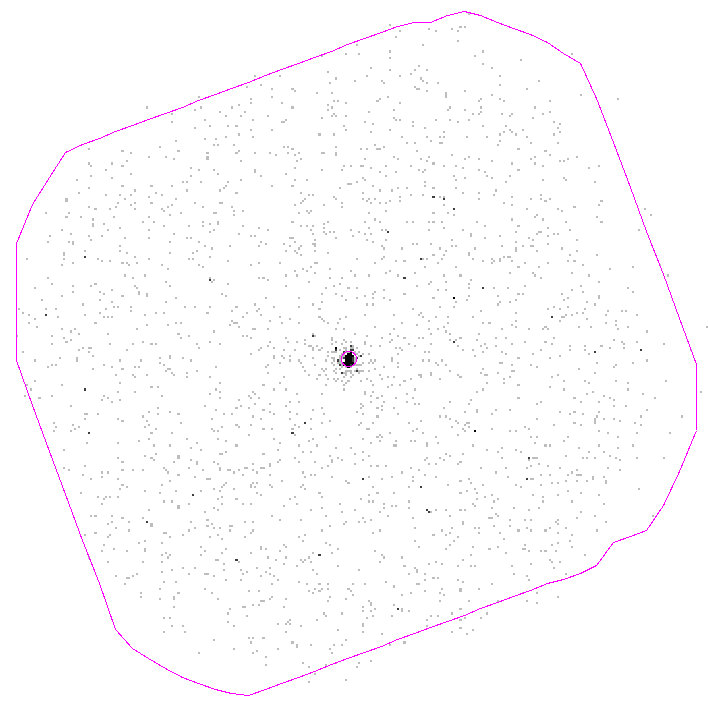}
  \caption{\rosat HRI image of the quasar 3C~273 (Obs.~ID \texttt{CA141520H.N1}). The footprint is calculated from the background image. The small circle displays the sources radius of 30\,\asec.}
  \label{fig:rosat_hri_3c273}
\end{figure}

We download 5347 images from the \heasarc and use the background
images for the footprint extraction with our footprint algorithm. The
background images are smoothed and exhibit a clear, distinct border.
We exclude negative background regions occurring in some images from
the footprint. An example footprint is shown in
Fig.~\ref{fig:rosat_hri_3c273}.
The data set includes 245 images with zero exposure and no counts, which were rejected from the database. 
The exposure time is inferred from the \texttt{EXPOSURE} keyword in the image. 
Overall, our database covers 5094 footprints.

\begin{figure}
  \centering
  \includegraphics[width=1\linewidth]{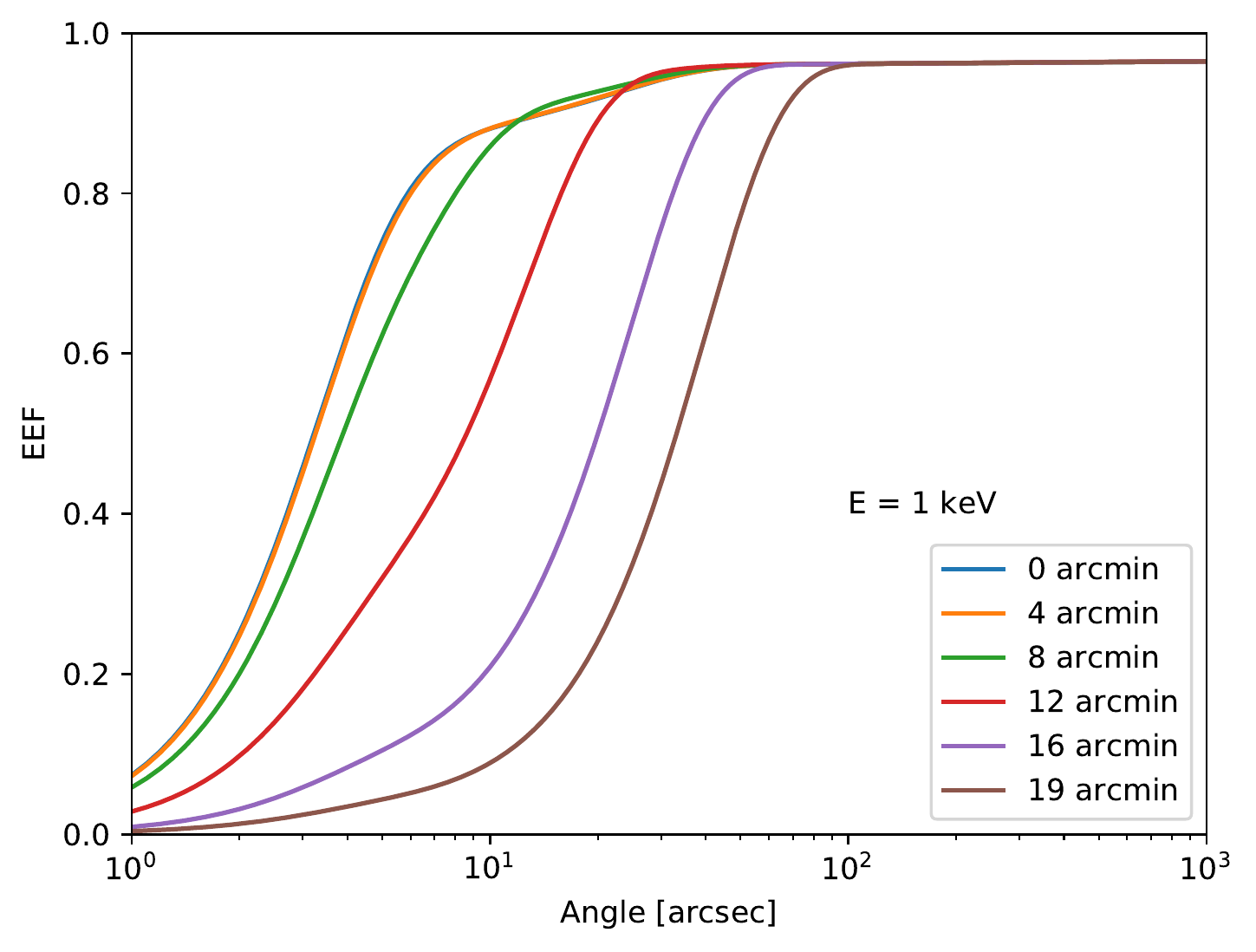}
  \caption{\rosat HRI cumulative point spread function ($\equiv$ EEF)
    as a function of angle for various off-axis
    angles and fixed energy $E=1$\,keV \citep[Eq.~5.19]{EXSAS}.}
  \label{fig:rosat_hri_ptn_eef}
\end{figure}

\begin{figure}
  \centering
  \includegraphics[width=1\linewidth]{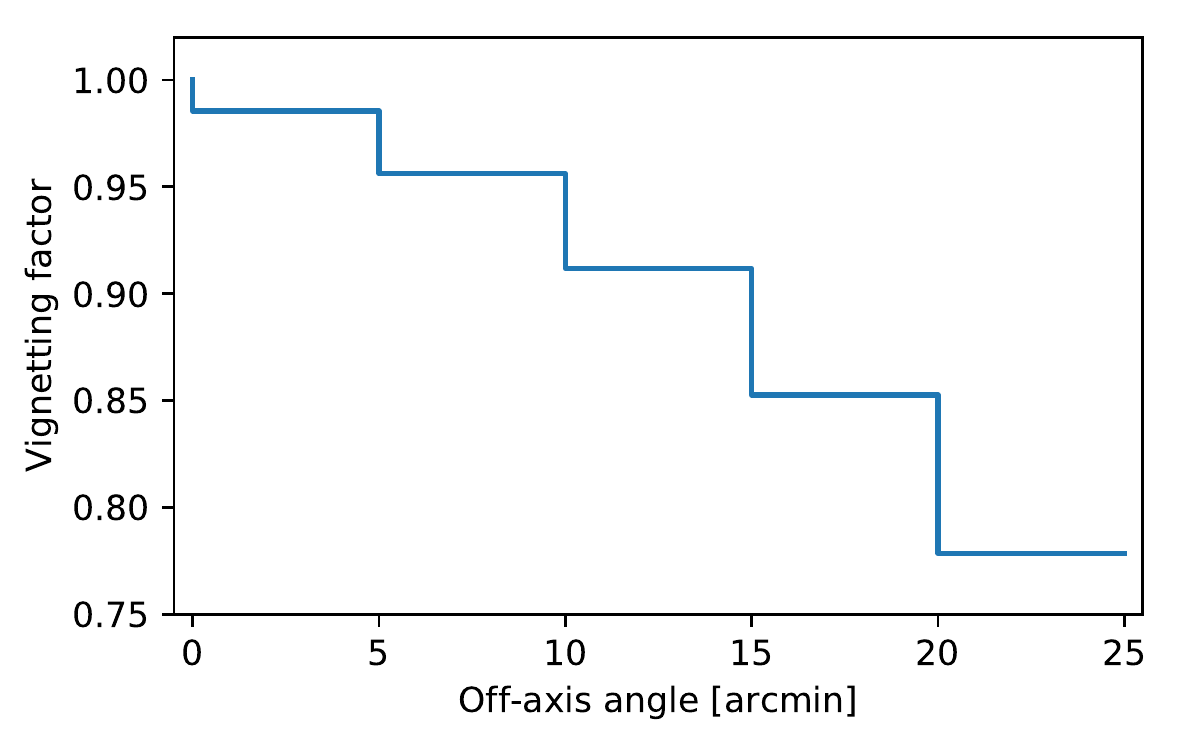}
  \caption{\rosat HRI vignetting factors as a function of off-axis
    angle.}
  \label{fig:rosat_hri_vign}
\end{figure}

The encircled energy fraction is shown in
Fig.~\ref{fig:rosat_hri_ptn_eef}. \hiligt uses a source radius of
30\,\asec for the upper limit calculation, which corresponds to an EEF of
94\%.  As for \rosat PSPC, we fix the energy to 1\,keV.  This is a
much better approximation (scattering $<10\%$) compared to the PSPC
instrument \citep[Fig.~5.23--24]{EXSAS}. The vignetting of the HRI is
small compared to the PSPC instrument. It is inferred from table
\texttt{vignet\_hri.fits}, which is available at the MPE \citep[see
  also p.~246 in][]{EXSAS}, and plotted in Fig.~\ref{fig:rosat_hri_vign}.

\subsection{\asca (1993--2000)}
 
Four X-ray telescopes \citep{1995PASJ...47..105S} with two detectors
were assembled on \asca \citep{Tanaka94A},
a Gas Imaging Spectrometer (GIS, 0.8--12\,keV,
\citealt{1996PASJ...48..157O,1996PASJ...48..171M}) and a
Solid-state Imaging Spectrometer (SIS, 0.4--12\,keV,
\citealt{1995PhDT.........7G}).
Further information can be found in
\citet{1995ApOpt..34.4848T} and \citet{AscaTechReport}.

\subsubsection{\asca GIS}
The GIS consisted of two imaging gas scintillation proportional
counters with a circular FOV of 50\,\amin and a spatial resolution of
$\sim$0.5\,\amin at 5.9\,keV. A major problem for the upper limit
calculation is the intrinsic PSF of the GIS detector. Usually, the PSF
is only constrained by the optics, however, the GIS detector induces
additional image distortion which is heavily dependent on off-axis
angle. We do not attempt to model the PSF and provide only results for
the detected sources that are contained in the \asca GIS catalog.

The \asca Medium Sensitivity Survey (AMSS) is a serendipitous source
survey for the extra-galactic sky $|b|>10\,\deg$ and described by
\cite{2001ApJS..133....1U,2005ApJS..161..185U}. The catalog lists a
total of 2533 detected sources ($5\sigma$) from an area of
278\,deg$^2$ in the bands 0.7--7\,keV (\textit{total}), 2--10\,keV
(\textit{hard}) and 0.7--2\,keV (\textit{soft} band). The \textsc{ASCAGIS}
catalog\footnote{\url{https://heasarc.gsfc.nasa.gov/W3Browse/asca/ascagis.html}} consists of two sub-catalogs: \textsc{AMSS-I} with 1343
sources detected between 1993 May and 1996 December and
\textsc{AMSS-II} with 1190 sources detected between 1997 January and
2000 May. The catalog gives the count rate in the three bands
mentioned above. We compute the uncertainty of the count rate as
\citep[p.~13]{2001ApJS..133....1U} 
\begin{equation}
  \sigma_{\text{CR}} = \frac{\text{CR}}{\sigma_{\text{D}}} \qquad ,
\end{equation}
where CR is the count rate and where $\sigma_{\text{D}}$ is the signal-to-noise
ratio in units of sigma, given in column \verb|SNR_Total/Hard/Soft| of
the \heasarc catalog. The catalog does not provide background information. Furthermore, we use a catalog
search radius of 3\,\amin for the \asca GIS catalog cone search, the same value as the \heasarc.

\subsubsection{\asca SIS}
The Solid-state Imaging Spectrometer consisted of two cameras with front-side illuminated CCD chips. It had an energy range of 0.4--10\,keV and a FOV of $22\,\amin\times 22\,\amin$ with a spatial resolution of 30\,\asec \citep{1995PhDT.........7G}. 
The \textsc{ASCASIS} catalog\footnote{\url{https://heasarc.gsfc.nasa.gov/W3Browse/asca/ascasis.html}} is populated with target and serendipitous sources in the SIS FOV.  
The catalog was published by \citet{1997xisc.conf...31G}, resulting from a search for point-like sources in the public \asca data archive.
The 0.5--12\,keV count rate is extrapolated to the \textit{total} band (0.2--12\,keV), and we use a catalog search radius of 5\,\amin for the \asca SIS catalog cone search.
The catalog does not provide background information.


\subsection{\xmm (2000--2021+)}
The European Space Agency's (ESA) X-ray Multi-Mirror Mission
(\xmm; \citealt{Jansen01a}) was launched on 1999 December 10 and is
still working nominally at the time of writing. It carries three X-ray telescopes
with high effective area and an optical monitor and orbits the Earth
in a highly-elliptical 48\,h orbit, which allows it to take long
uninterrupted exposures. 
The telescopes have been designed to give a point spread function
which is uniform as a function of photon energy and of off-axis
angle. For reference, an extraction region of radius 30\,\asec
contains 88\% of 1.5\,keV photons at an intermediate off-axis angle of
6\,\amin.
Three CCD cameras operate in parallel, the
EPIC-pn \citep{Strueder01a} back-illuminated detector and two EPIC-MOS
front-illuminated CCDs \citep{Turner01a}, each receiving light from a
nearly identical, nested Wolter I telescope. The MOS cameras
share their telescope with one  Reflection Grating Spectrometer each, and
receive roughly half of the light. The EPIC instrumentation is
sensitive to photons in the energy range 0.2--12\,keV.

EPIC-pn has an energy resolution of 100\,eV (FWHM) at 1\,keV and a
spatial resolution of 16\,\asec (half energy width) at 1.5\,keV. It
has a high quantum efficiency and has been shown to return a stable
flux throughout the mission\footnote{
"Stability of the EPIC-pn camera": 
\url{http://xmmweb.esac.esa.int/docs/documents/CAL-TN-0212-1-0.pdf}}
It has $>2$ times the throughput of the EPIC-MOS cameras and is the preferred instrument for
upper limit measurements. Performance of the telescope and detector is
well described in the \textit{XMM-Newton Users Handbook}
\citep{XMM_UHB}.

Each of the EPIC cameras carry a thin, medium and thick filter
designed to block out optical light. They produce an increasingly
important reduction in the soft X-ray ($<2$\,keV) photon flux and
factors to convert count rates to flux are computed independently for
each filter.


\subsubsection{\xmm Pointed}
\label{subsubsec:xmmp}

\begin{figure}
  \centering
  \includegraphics[width=.8\linewidth]{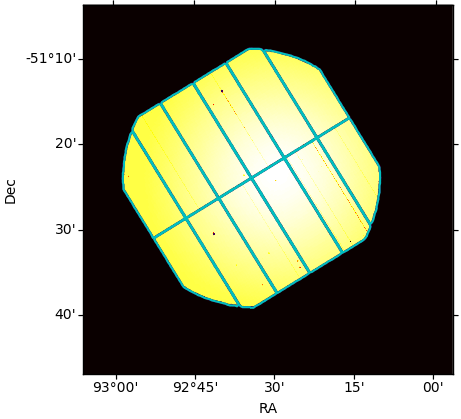}
  \caption{Example \xmm pointed exposure map with footprint.}
  \label{fig:xmmpnt_footprint}
\end{figure}

The products from pointed observations are stored in the (New)
XMM-Newton science archive (NXSA) with a unique ten-digit identifier
starting with ``0''. They have been made in relatively narrow bands
and we combine bands 1, 2 and 3 to make the soft band (0.2--2\,keV)
image and exposure map and bands 4 and 5 to make the hard band
(2--12\,keV) products. The full band image is available directly as
band 8. Lists of observations and associated images which contain a
particular sky position are obtained by making TAP queries to the
\texttt{public\_observations}, \texttt{exposure} and
\texttt{uls\_exposure\_image} tables within NXSA. The camera has a FOV
of $27.2\,\amin \times 26.2\,\amin$. Footprints for each observation
and for each slew are stored in the NXSA (see
Fig.~\ref{fig:xmmpnt_footprint}).

The vignetting function is energy-dependent, giving a throughput of
$\sim$30\% at the edge of the FOV for photons of 1.5\,keV and
$\sim$20\% for 8\,keV photons. This quantity is folded into the
exposure map and is not used directly in the calculation. 
The \xmm mirrors were designed to give a consistent encircled energy correction across the field of view. \hiligt uses the encircled energy correction for 1.5\,keV photons located 6\,\amin off-axis. The difference between this encircled energy correction and that of photons at 1\,\amin or 10\,\amin off-axis or of energy 1\,keV or 6\,keV is less than 5\%.

The 4XMM-DR11 pointed source catalog \citep{Webb20a} contains 602\,543
unique sources from 12\,210 EPIC-pn pointed observations that were public
on 2020-12-31. The sky coverage is 1239\,deg$^2$ (3\% of the sky). 117\,124 sources were observed multiple times\footnote{\url{http://xmmssc.irap.omp.eu/Catalogue/4XMM-DR11/4XMM-DR11_Catalogue_User_Guide.html} Sect.~5.1}. Sources are
located to within a median error of $\sim1$\,\asec and are flagged to
indicate issues which may potentially affect the returned count
rates. \hiligt looks for sources within a radius of 10\,\asec of the
input position, which have been detected with the EPIC-pn camera and
where the quality flag, \texttt{SUM\_FLAG}$<$2. 112\,084 detections are classified as extended with 34\,173 marked as ``clean'' (\texttt{SUM\_FLAG}$<$1).

Both pointed and slew observations have an exposure map and pointed
observations also have a background map for each of the energy
bands. These are used in the upper limit calculation. The background
is calculated in slew observations from an annulus around the source
center of inner radius 60\,\asec and outer radius 120\,\asec. We note
that \hiligt does not account for chip gaps in the EPIC-pn images and
that the upper limit may be biased if the queried position falls into such a gap.

To estimate the chance of source confusion, we note that the \xmm
pointed observations exhibit about 50 sources per FOV for a deep
observation. At an approximate (circular) FOV of 14\,\amin radius and a
cone search radius of 10\,\asec, the chance of randomly hitting a
source in an average field would be roughly $\pi 10^2 / (\pi
(60\cdot 14)^2)\cdot 50 \approx 0.1\%$. Of course in the Galactic
center or very crowded regions, the source density and confusion
chance is higher.

\subsubsection{\xmm Slew}
\xmm performs a shallow survey of the sky while slewing between
pointed observation targets \citep{Saxton08a}, reaching a sensitivity
of $6\times10^{-13}\,\ergcms$ in the soft band,
$4\times10^{-12}\,\ergcms$ in the hard band, and
$1\times10^{-12}\,\ergcms$ in the total band. Pointing accuracy is
limited by the ability to reconstruct the slew path and has a
1$\sigma$ error of 8\,\asec.

\begin{figure}
  \centering
  \includegraphics[width=.6\linewidth]{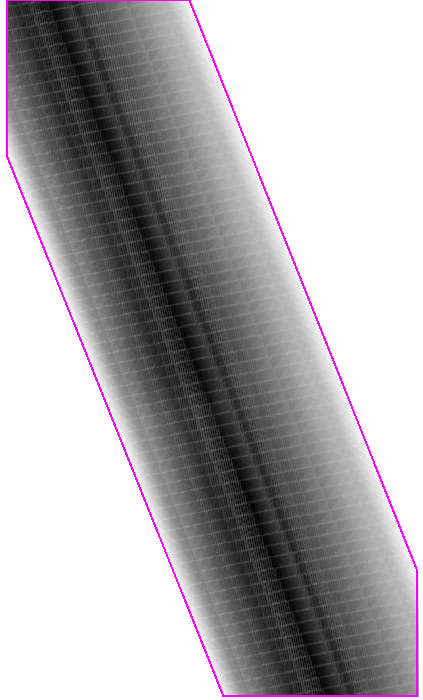}
  \caption{An example \xmm slew exposure map with footprint.}
  \label{fig:xmmslew_footprint}
\end{figure}

Slew images (Fig.~\ref{fig:xmmslew_footprint}) containing a given sky
position are found from TAP calls to the \texttt{slew\_exposure} and
\texttt{slew\_exposure\_image} tables. A limitation in some of the
early pipeline products has led to some slew images being extremely
large and impossible to process with the \eupper task. These are
excluded from the upper limit calculations. It is expected that the
entire slew archive will be reprocessed in the near future to remove
this limitation.

The slew catalog, XMMSL2, contains over 20\,000 sources detected in
85\% of the sky. Given the low sky density of sources a radius of
30\,\asec is used in the catalog search. \hiligt dynamically links the
\xmm slew and pointed footprints and images to the NXSA, and always
uses the latest catalog release.

\subsubsection{\xmm Stacked}
The XMM-Newton Serendipitous Source Catalogue from overlapping observations (``stacked catalog'') is a source catalog made from overlapping observations \citep{Traulsen19a,Traulsen20a}.
Due to the longer effective exposure, upper limits of the stacks can provide a much deeper
constraint on the flux than the pointed observations.
However, the stacking also results in an averaging of the source count rate. 
We caution that for bright, flaring sources the averaged stacked flux can
in fact be above the flux of a pointed observation, if it is in a low
flux state\footnote{An example is 3C~273, which has a stacked
  0.2--12\,keV flux of $(1.330\pm0.001)\times 10^{-10}$\,\ergcms but
  had, on 2000-06-16 (ObsID 0126700501) a flux of $(9.84\pm0.04)\times
  10^{-11}\,\ergcms$}. The start and end date of the stacked data point or
upper limit is given by the first and last observation of the stack,
and should be interpreted as a mean over the whole time period.

\begin{figure}
\centering
\begin{subfigure}[t]{0.49\linewidth}
    \includegraphics[width=1\linewidth]{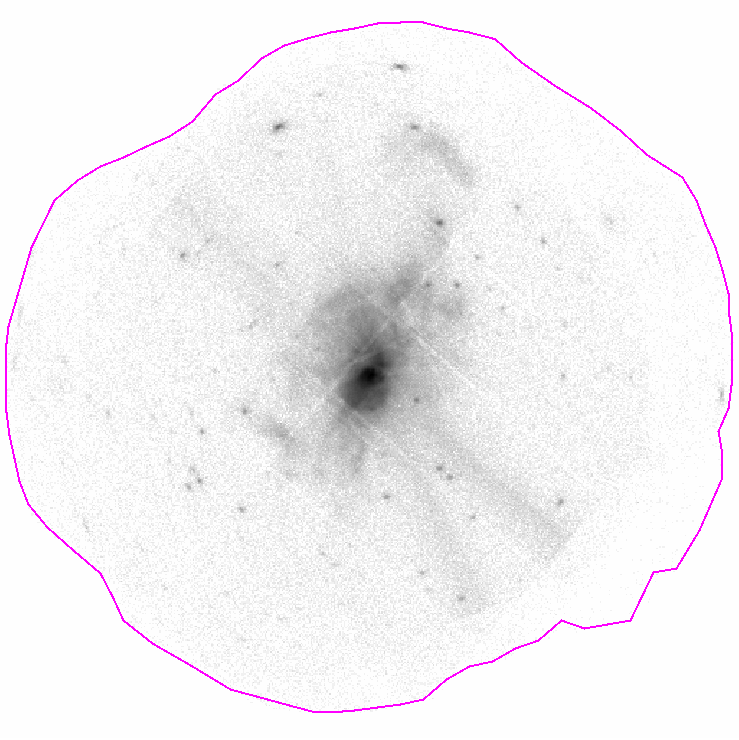}
    \subcaption{The interacting galaxies
      \href{http://simbad.u-strasbg.fr/simbad/sim-basic?Ident=M82&submit=SIMBAD+search}{M82}
    (StackID \texttt{0434\_009})}
    \label{fig:xmmstacked_footprint:a}
\end{subfigure}
\hfil
\begin{subfigure}[t]{0.49\linewidth}
    \includegraphics[width=1\linewidth]{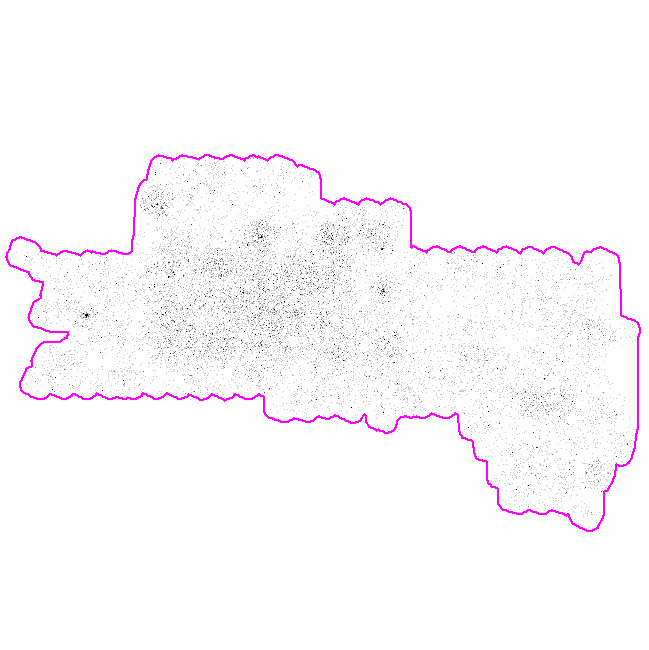}
    \subcaption{The XMM-Large Scale Structure region \citep[see
        also][]{Pierre07a} (StackID \texttt{0072\_352})}
    \label{fig:xmmstacked_footprint:d}
\end{subfigure}
\vfil
\begin{subfigure}[t]{0.49\linewidth}
    \includegraphics[width=1\linewidth]{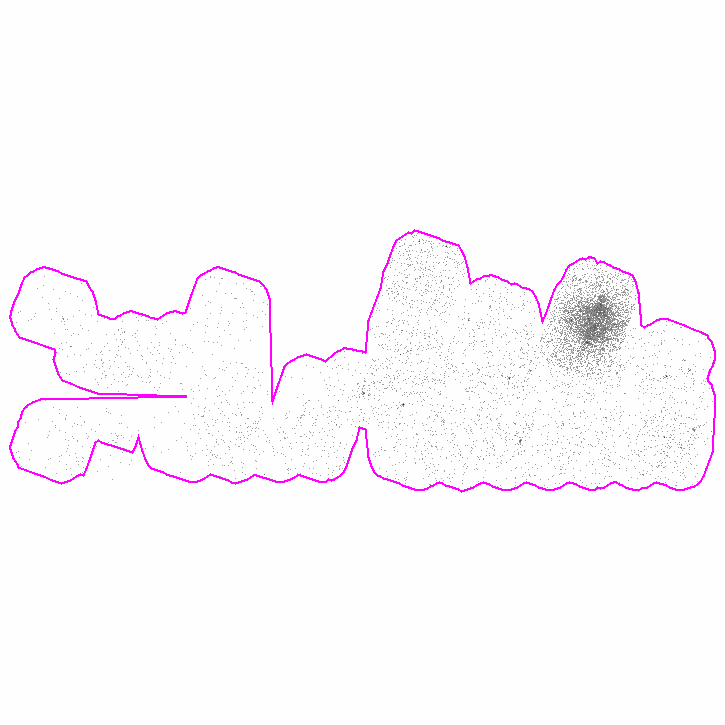}
    \subcaption{Stacked PN images around position
      01\,19\,05.14~+00\,37\,45.0 containing multiple faint X-ray
      sources (StackID \texttt{0861\_045})}
    \label{fig:xmmstacked_footprint:c}
\end{subfigure}
\hfil
\begin{subfigure}[t]{0.49\linewidth}
    \includegraphics[width=1\linewidth]{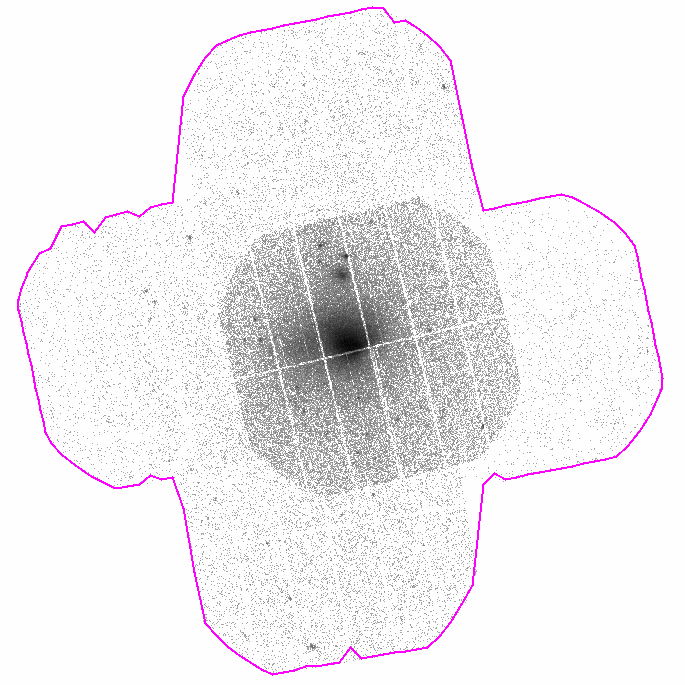}
    \subcaption{The galaxy cluster
      \href{http://simbad.u-strasbg.fr/simbad/sim-basic?Ident=A+2163&submit=SIMBAD+search}{ACO
        2163} (StackID \texttt{0424\_007})}
    \label{fig:xmmstacked_footprint:b}
\end{subfigure}
\caption{Examples of the \xmm stacked footprints. The footprints are
  calculated based on EPIC-pn coverage maps.}
\label{fig:xmmstacked_footprint}
\end{figure}

\hiligt currently accesses only EPIC-pn data of the 4XMM-DR11s stacked catalog. 
The footprints of each stack are computed based on the EPIC-pn coverage maps
(Fig.~\ref{fig:xmmstacked_footprint}). The \hiligt database contains 1328
footprints covering a total sky area of $\sim$560\,deg$^2$, from which
350\,deg$^2$ are multiply observed. The catalog
\texttt{epic\_xmm\_stack\_cat} at the NXSA is queried with the options
\texttt{pn\_rate}$>$0 (to obtain EPIC-pn data only) and
\texttt{STACK\_FLAG<=1} (to flag spurious entries). To connect the
catalog entry to the database counterpart, we use a conversion table
(Traulsen, priv. comm.) to map the \texttt{SRCID} to the
\texttt{STACKID}.
Finally, the upper limits are computed based on stacked images, exposure, and background maps in
five energy bands. The same conversion factors, encircled energy
fraction and vignetting as for \xmm pointed is used
(Sect.~\ref{subsubsec:xmmp}).

\subsection{\integral (2002--2021+)}

ESA's INTErnational Gamma-Ray Astrophysics Laboratory (\integral, \citealt{Winkler03a}) was launched on 2002 October 17 and is the only gamma-ray and coded-mask instrument included in \hiligt. 
The imager (IBIS, \citealt{Ubertini03a}) has a $8.3\,\deg\times 8.0\,\deg$ FOV with an angular resolution of 12\,\amin (FWHM) and total energy range of 15\,keV-10\,MeV. 
For \hiligt, we concentrate on the IBIS/ISGRI \citep{Lebrun03a} count rates in
three energy bands: 20--40, 40--60, and 60--100\,keV. 
\integral observations are performed as a series of pointings
called science windows (\textit{scw}) that last between 30 and 60
minutes. Each \textit{scw} is processed individually to produce
reconstructed images \citep{Goldwurm03}.

The count rates and variances in the three bands at each sky position
are extracted from these reconstructed images and stored in a
table. The values provided are based on the significance or
signal-to-noise ($\sigma$) following one of three cases:
\begin{itemize}
\item $\sigma \ge 3$: The rate is given by the weighted mean count
  rate computed from the complete set of pointings that contain the
  given sky position, and the uncertainty is given by the error on
  that weighted mean.
\item $3<\sigma\le 0$: For $\sigma$ values below 3, the rate is given
  by twice the error on the mean divided by the average detection
  efficiency, and the uncertainty is given by $-1.0$ to indicate that
  the value is an upper limit. The average detection efficiencies are
  computed from the response files, and are 0.87, 0.64,
  and 0.64, respectively, for the three bands 20--40, 40--60, and
  60--100 keV.
\item $\sigma<0$: This case exists because the reconstructed count
  rate in each sky pixel in the absence of sources is normally
  distributed, and the significance approximates a standard normal
  variable. Signal and background counts are equal and given by the
  error on the weighted mean rate multiplied by the exposure time and
  the data point is displayed as an upper limit.
\end{itemize}
In the first two cases of the estimated signal and background counts,
for any value of $\sigma$ above 0, the signal counts, $S$, are given
by the rounded value of the weighted mean rate multiplied by the
effective position-dependent exposure time.
The background counts, $B$, are derived from the standard formula $\sigma = S/\sqrt{S+B}$, and thus given by $(S/\sigma)^2 - S$.

We use \integral data up to May 2016. We note that we do not include
any catalog values (as available, e.g., in \citealt{Krivonos07a} or at
\url{www.isdc.unige.ch/integral/science/catalogue}). \integral's FOV
is $8.3\,\deg \times 8.0\,\deg$, larger than that of most X-ray missions,
and each sky position is observed a large number of times. We
currently provide one data point averaged over the 16 years of
observation. As \integral is a hard X-ray/soft $\gamma$-ray mission, we provide only a powerlaw spectral model. 
These conversion factors are time dependent due to slow ageing of the IBIS/ISGRI detectors by cosmic radiation, changing the effective area over time. 
Nevertheless, we use conversion factors which are averaged over the 16 years time period and computed with the OSA~10 software (see Table~\ref{tab:cf_integral}).

It is important to note that because IBIS/ISGRI is a coded mask
instrument, the variance map from which the errors are taken, and the
intensity map from which the rates are taken, are computed
independently: the intensity map is reconstructed based on spatial
information derived from knowledge of the mask pattern, whereas the
variance map relies only on the number of photons detected on the
detector plane.


\subsection{\swift (2005--2021+)}

\begin{figure}
  \centering
  \includegraphics[width=.8\linewidth]{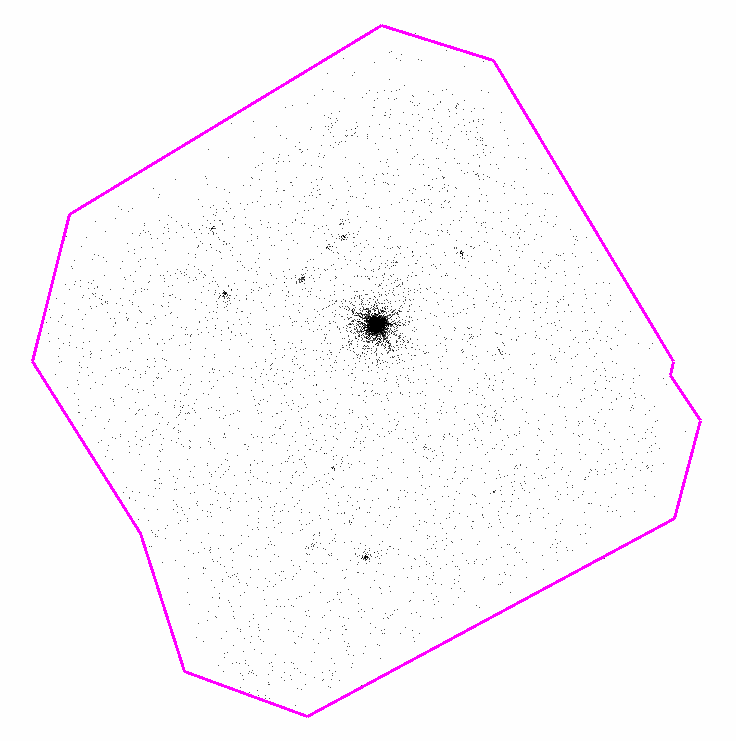}
  \caption{An example \swift XRT image of 3C~273 (ObsID 00035019004)
    with footprint.}
  \label{fig:swift_xrt_image}
\end{figure}

The X-ray Telescope (XRT; \citealt{BurrowsXRT}) on the \emph{Neil
  Gehrels Swift Observatory} \citep{GehrelsSwift} has a CCD detector
with a circular field of view radius 12.3\,\amin, and an energy range of
0.3--10\,keV.  A typical \swift XRT image is shown in
Fig.~\ref{fig:swift_xrt_image}.

\hiligt makes use of the 2SXPS catalog \citep{Evans20a}. It contains
data from the end of early operations phase 2005-01-01 to 2018-08-01,
and provides exposure and background maps (including the model PSF of
any detected sources) for each field, and in four energy bands. 
\hiligt queries the 2SXPS upper limit server provided at the
UK Swift Science Data
Centre\footnote{\url{https://www.swift.ac.uk/2SXPS/ulserv.php}}.  The
majority of fields within 2SXPS have been observed multiple times,
thus multiple upper limits or count-rates can be returned.

To identify the \swift footprint at the requested position, initially
any observation within 28\,\amin was found. Within each observation, and
for each requested energy band, a cone search radius 20\,\asec
identified any matching source and its count rate was extracted from
2SXPS. If no source was found, a circular region of radius 28\,\asec (12
image pixels) centered at the input position was created. The exposure
at this position was taken from the exposure map within 2SXPS; only
datasets where the mean exposure in this circle was at least 1\% of
the on-axis exposure were accepted. For those datasets, the number of
counts in the image, the sum of the background map and average value
of the exposure map in this region were returned to \hiligt. A
correction factor was also calculated which accounted both for the
fraction of the EEF outside the circular region and for vignetting or
bad columns within the region (see \citealt{Evans20a}, Sect.~3.3 for
details). The values from the soft (0.3--1\,keV) and medium
(1--2\,keV) bands were combined to give values in the 0.3--2\,keV
band, and then the upper limits were calculated as described in
Sect.~\ref{sec:eupper}-\ref{sec:convfactor}.


\begin{landscape}

\begin{table}[]
\centering
\caption{Overview of parameters for each mission. The energy range of the catalog count rates is in keV with the corresponding \hiligt band in parenthesis. The source radius ($r$ in Sect.~\ref{sec:eupper}) defines the size of the circular region for source counts estimation, and, if no background map is available, of the background annulus. The number behind the catalog name shows the cone search radius. The columns ``UL/Bkg/Exp/P'' indicate whether there is an upper limit calculation, and whether it makes use of background and exposure maps, and whether PIMMS is used to compute the conversion factors. In the ``Notes'' column we indicate, if available, the number of sources in the catalog, and the number of images that are included in the upper limit calculations. The reference numbers correspond to:
  \citenum{AscaTechReport}:\citet{AscaTechReport},
  \citenum{1988ApOpt..27.1404A}:\citet{1988ApOpt..27.1404A}, 
  \citenum{2016AA...588A.103B}:\citet{2016AA...588A.103B},
  \citenum{ROSAT_User_Handbook}:\citet{ROSAT_User_Handbook},
  \citenum{BurrowsXRT}:\citet{BurrowsXRT},
  \citenum{RUM}:\citet{RUM}, 
  \citenum{1981SSRv...30..495D}:\citet{1981SSRv...30..495D},
  \citenum{Evans20a}:\citet{Evans20a},
  \citenum{1978ApJS...38..357F}:\citet{1978ApJS...38..357F},
  \citenum{GehrelsSwift}:\citet{GehrelsSwift},
  \citenum{1995PhDT.........7G}:\citet{1995PhDT.........7G}, 
  \citenum{1979ApJ...230..540G}:\citet{1979ApJ...230..540G}, 
  \citenum{1971ApJ...165L..27G}:\citet{1971ApJ...165L..27G},
  \citenum{1990ApJS...72..567G}:\citet{1990ApJS...72..567G}, 
  \citenum{1981ITNS...28..869G}:\citet{1981ITNS...28..869G},
  \citenum{1997xisc.conf...31G}:\citet{1997xisc.conf...31G},
  \citenum{1984SAOSR.393.....H}:\citet{1984SAOSR.393.....H},
  \citenum{1990eoci.book.....H}:\citet{1990eoci.book.....H},
  \citenum{1989PASJ...41..373H}:\citet{1989PASJ...41..373H}, 
  \citenum{1977SPIE..106..196H}:\citet{1977SPIE..106..196H},
  \citenum{Holt76a}:\citet{Holt76a},
  \citenum{Jansen01a}:\citet{Jansen01a},
  \citenum{Kaluzienski77a}:\citet{Kaluzienski77a},
  \citenum{Lebrun03a}:\citet{Lebrun03a},
  \citenum{1987ApL....25..223M}:\citet{1987ApL....25..223M},
  \citenum{1996PASJ...48..171M}:\citet{1996PASJ...48..171M},
  \citenum{1981MNRAS.197..893M}:\citet{1981MNRAS.197..893M},
  \citenum{Nugent83a}:\citet{Nugent83a},
  \citenum{1996PASJ...48..157O}:\citet{1996PASJ...48..157O},
  \citenum{1987SPIE..733..519P}:\citet{1987SPIE..733..519P}, 
  \citenum{1982ApJ...253..485P}:\citet{1982ApJ...253..485P},
  \citenum{1999AAS..134..287R}:\citet{1999AAS..134..287R},
  \citenum{1979SSI.....4..269R}:\citet{1979SSI.....4..269R},
  \citenum{Saxton08a}:\citet{Saxton08a},
  \citenum{1995PASJ...47..105S}:\citet{1995PASJ...47..105S},
  \citenum{Strueder01a}:\citet{Strueder01a},
  \citenum{Tanaka94A}:\citet{Tanaka94A},
  \citenum{Terrell82a}:\citet{Terrell82a},
  \citenum{Traulsen19a}:\citet{Traulsen19a},
  \citenum{Traulsen20a}:\citet{Traulsen20a},
  \citenum{1982AdSpR...2..241T}:\citet{1982AdSpR...2..241T},
  \citenum{1995ApOpt..34.4848T}:\citet{1995ApOpt..34.4848T}, 
  \citenum{1981SSRv...30..513T}:\citet{1981SSRv...30..513T},
  \citenum{1989PASJ...41..345T}:\citet{1989PASJ...41..345T},
  \citenum{Ubertini03a}:\citet{Ubertini03a},
  \citenum{2001ApJS..133....1U}:\citet{2001ApJS..133....1U},
  \citenum{2005ApJS..161..185U}:\citet{2005ApJS..161..185U}, 
  \citenum{Villa76a}:\citet{Villa76a},
  \citenum{1981MNRAS.197..865W}:\citet{1981MNRAS.197..865W},
  \citenum{Webb20a}:\citet{Webb20a},  
  \citenum{1988MmSAI..59....7W}:\citet{1988MmSAI..59....7W}, 
  \citenum{Whitlock92b}:\citet{Whitlock92b},
  \citenum{Whitlock92}:\citet{Whitlock92},
  \citenum{Winkler03a}:\citet{Winkler03a},
  \citenum{Wood84a}:\citet{Wood84a},
  \citenum{EXSAS}:\citet{EXSAS}}
\label{tab:params}
\resizebox{1\linewidth}{!}{  
\begin{tabular}[hd]{l|l|lll|ll|cccc|lll}
\toprule 
Mission & Energy range & FOV & Spatial res. & Source
radius & Cat. name (search rad.) & Cat. size & UL & Bkg & Exp & P & Instr./Filters & References &
Notes \\ \midrule\midrule

\velaVb & 3--12 (hard) & $6.1\,\deg\times6.1\,\deg$ & 6\,\deg FWHM &  & -- & --
& N & N & N & N & XC & \citenum{Terrell82a,Whitlock92b,Whitlock92} & lightcurves of 99 sources \\ \midrule

\uhuru & 2--6 (hard) & \begin{tabular}{@{}l@{}}$0.52\,\deg\times
                   0.52\,\deg$\\$5.2\,\deg\times 5.2\,\deg$\end{tabular} & &  &
\href{https://heasarc.gsfc.nasa.gov/W3Browse/uhuru/uhuru4.html}{UHURU4} (60\,\amin) & 339 
& N & N & N & N & & \citenum{1978ApJS...38..357F,1971ApJ...165L..27G} & only catalog, 339 sources \\
\midrule

\arielv SSI & 2--18 (hard) & $0.75\,\deg\times 10.6\,\deg$ & & & 
\href{https://heasarc.gsfc.nasa.gov/W3Browse/ariel-v/ariel3a.html}{ARIEL3A} (30\,\amin) & 250
& N & N & N & N & & Instr.: \citenum{Villa76a,Whitlock92b}, Cat.: \citenum{1981MNRAS.197..893M,1981MNRAS.197..865W} & only catalog, 251 sources \\

\arielv ASM & 3--6 (hard) & $0.75\,\deg\times 10.6\,\deg$ & & & -- & -- & N & N & N & N & & \citenum{Holt76a,Kaluzienski77a} & lightcurves of 249 sources \\
\midrule

\heaoI A-1 & 0.5--25 (hard) & \multirow{3}{*}{$1.5\,\deg\times 3\,\deg$} & & & \href{https://heasarc.gsfc.nasa.gov/W3Browse/heao1/a1.html}{A1} (60\,\amin) & 842
& N & N & N & N & & \citenum{Wood84a} & only catalog, 842 sources \\

\heaoI A-2 & \begin{tabular}{@{}l@{}}0.44--2.8 (soft)\\2--10 (hard)\end{tabular} & & & & \begin{tabular}{@{}l@{}}\href{https://heasarc.gsfc.nasa.gov/W3Browse/heao1/a2led.html}{A2LED} (60\,\amin)\\\href{https://heasarc.gsfc.nasa.gov/W3Browse/heao1/a2pic.html}{A2PIC} (60\,\amin)\end{tabular}
& \begin{tabular}{@{}l@{}}114\\ 68\end{tabular} & N & N & N & N & & \begin{tabular}{@{}l@{}}\citenum{1979SSI.....4..269R,Nugent83a}\\\citenum{1982ApJ...253..485P}\end{tabular} & \begin{tabular}{@{}l@{}}only catalog, 114 sources\\only catalog, 68 sources, $|b|>20\,\deg$\end{tabular} \\
\midrule

\multirow{2}{*}{\einstein IPC} & \multirow{2}{*}{0.2--3.5 (soft)} &
\multirow{2}{*}{$75\,\amin\times 75\,\amin$} & \multirow{2}{*}{1\,\amin} &
\multirow{2}{*}{$5.911\,\amin$ @ 0.5 EEF} &
\href{https://heasarc.gsfc.nasa.gov/W3Browse/einstein/ipc.html}{IPC}
(2\,\amin) & 6816 & \multirow{2}{*}{Y} & \multirow{2}{*}{N} & \multirow{2}{*}{N} & \multirow{2}{*}{Y} & & Instr.:
\citenum{1979ApJ...230..540G,1981ITNS...28..869G,1984SAOSR.393.....H}
& \multirow{2}{*}{B1950$\rightarrow$J2000, 3923 images}\\ & &
& & &
\href{https://heasarc.gsfc.nasa.gov/W3Browse/einstein/ipcimage.html}{IPCIMAGE}
(15\,\amin) & 4132 & & & & & & Cat.:
\citenum{1990ApJS...72..567G,1990eoci.book.....H} & \\

\multirow{2}{*}{\einstein HRI} & \multirow{2}{*}{0.15--3.5 (soft)} &
\multirow{2}{*}{25\,\amin diam.} & \multirow{2}{*}{3\,\asec} &
\multirow{2}{*}{18\,\asec @ 0.8 EEF} &
\href{https://heasarc.gsfc.nasa.gov/W3Browse/einstein/hricfa.html}{HRICFA}
(1\,\amin) & 598 & \multirow{2}{*}{Y} & \multirow{2}{*}{N} & \multirow{2}{*}{N} & \multirow{2}{*}{Y} & &
\multirow{2}{*}{Instr.:
  \citenum{RUM,1979ApJ...230..540G,1977SPIE..106..196H}} &
\multirow{2}{*}{B1950$\rightarrow$J2000, 836 images} \\ & & &
& &
\href{https://heasarc.gsfc.nasa.gov/W3Browse/einstein/hriimage.html}{HRIIMAGE}
(15\,\amin) & 870 & & & & \\ \midrule

\exosat LE & 0.05--2 (soft) & 2\,\deg diam. & 18\,\asec & 1\,\amin &
\href{https://heasarc.gsfc.nasa.gov/W3Browse/exosat/le.html}{LE}
(1\,\amin) & 2563 & Y & N & N & Y & CMA & Instr.: \citenum{1981SSRv...30..495D,1988MmSAI..59....7W} &
B1950$\rightarrow$J2000, 3677 imgs. \\

\exosat ME Pointed & 1--15 (hard) & \multirow{2}{*}{$45\,\amin\times 45\,\amin$} & & & \href{https://heasarc.gsfc.nasa.gov/W3Browse/exosat/me.html}{ME} (45\,\amin) & 2291 & \multirow{2}{*}{N} & \multirow{2}{*}{N} & \multirow{2}{*}{N} & \multirow{2}{*}{Y} & & Instr.: \citenum{1981SSRv...30..495D,1981SSRv...30..513T,1988MmSAI..59....7W} & only catalog \\
\exosat ME Slew & 1--8 (hard) & & & & \href{https://heasarc.gsfc.nasa.gov/W3Browse/exosat/exms.html}{EXMS} (60\,\amin) & 1210 & & & & & & Cat.: \citenum{1999AAS..134..287R} & only cat., 1210 srcs., 98\% of sky \\ 
\midrule

\ginga LAC & 2--10 (hard) & $0.8\,\deg\times 1.7\,\deg$ & 6\,\asec & &
\href{https://heasarc.gsfc.nasa.gov/W3Browse/ginga/gingalac.html}{GINGALAC}
(10\,\amin) & 419 & N & N & N & Y & Top & Instr.:
\citenum{1989PASJ...41..373H,1987ApL....25..223M,1989PASJ...41..345T}
& only catalog \\ \midrule

\asca GIS
& \begin{tabular}{@{}l@{}l@{}}0.7--2 (soft)\\2--10 (hard)\\0.7--7 (total)\end{tabular} &
    50\,\amin & 0.5\,\amin & 5\,\amin &
    \href{https://heasarc.gsfc.nasa.gov/W3Browse/asca/ascagis.html}{ASCAGIS}
    (3\,\amin) & 2533 & N & N & N & Y & & \begin{tabular}{@{}l@{}l@{}}Instr.:
      \citenum{AscaTechReport,1996PASJ...48..171M,1996PASJ...48..157O,1995PASJ...47..105S}\\\citenum{Tanaka94A,1995ApOpt..34.4848T}\\Cat.: \citenum{2001ApJS..133....1U,2005ApJS..161..185U}\end{tabular} & only cat., 2533 srcs., 278\,deg$^2$ \\
      
\asca SIS & 0.5--12 (total) & $22\,\amin\times 22\,\amin$ & 30\,\asec & 3.3\,\amin @ 0.5 EEF &
\href{https://heasarc.gsfc.nasa.gov/W3Browse/asca/ascasis.html}{ASCASIS}
(5\,\amin) & 433 & N & N & N & Y & & \begin{tabular}{@{}l@{}}Instr.:
  \citenum{1995PhDT.........7G,1995PASJ...47..105S,Tanaka94A,1995ApOpt..34.4848T}\\Cat.:
  \citenum{1997xisc.conf...31G}\end{tabular} & only catalog \\ \midrule

\rosat PSPC Pointed & 0.1--2.34 (soft) & \multirow{3}{*}{2\,\deg
  diam.} & & \multirow{3}{*}{100\,\asec} &
\href{https://heasarc.gsfc.nasa.gov/W3Browse/rosat/rospspc.html}{ROSPSPC}
(0.5\,\amin) & 54\,133 & Y & N & Y & Y & \rdelim\}{2}{*}[~Open] & Instr.:
\citenum{1988ApOpt..27.1404A,ROSAT_User_Handbook,1987SPIE..733..519P,1982AdSpR...2..241T,EXSAS}
& 5490 imgs., 100\,048 srcs., 17.3\% of sky \\

\rosat PSPC Survey & 0.1--2.4 (soft) & & & &
\href{https://heasarc.gsfc.nasa.gov/W3Browse/rosat/rass2rxs.html}{RASS2RXS} (2\,\amin) & 135\,118
& Y & N & Y & Y & & Cat.: \citenum{2016AA...588A.103B} & 1378 imgs., 135\,000 srcs., all-sky \\
                                                                       
\rosat HRI Pointed & 0.2--2.4 (soft) & 38\,\amin square & 2\,\asec FWHM & 30\,\asec @ 0.94
EEF &
\href{https://heasarc.gsfc.nasa.gov/W3Browse/rosat/roshri.html}{ROSHRI}
(1\,\amin) & 13\,452 & Y & Y & N & Y & & Instr.:
\citenum{1988ApOpt..27.1404A,ROSAT_User_Handbook,1987SPIE..733..519P,1982AdSpR...2..241T,EXSAS}
& 5094 imgs., 56\,401 srcs., 1.94\% of sky \\

\midrule
\xmm Pointed & 0.2--2 (soft)& $27.2\,\amin \times 26.2\,\amin$ & 6\,\asec FWHM & \rdelim\}{3}{*}[~\begin{tabular}{@{}l@{}}30\,\asec @ \\0.876 EEF\end{tabular}] &
\href{http://nxsa.esac.esa.int/nxsa-web/\#tap}{epic\_source\_cat}
(10\,\asec) & 895\,415 & Y & Y & Y & Y & \rdelim\}{3}{*}[~EPIC-pn] &
Instr.:~\citenum{Jansen01a,Strueder01a}, Cat.:~\citenum{Webb20a} & 602\,543 srcs., 12\,210 imgs., 1239\,deg$^2$ \\

\xmm Slew & 2--12 (hard) & undef. & 8\,\asec & &
\href{http://nxsa.esac.esa.int/nxsa-web/\#tap}{slew\_source\_cat}
(20\,\asec) & 72\,352 & Y & Y & Y & Y & & Cat.:~\citenum{Saxton08a} & 23\,252 sources, 85\% of sky \\

\xmm Stacked & 0.2--12 (total)& undef. & & &
\href{http://nxsa.esac.esa.int/nxsa-web/\#tap}{epic\_xmm\_stack\_cat}
(10\,\asec) & 358\,809 & Y & Y & Y & Y & & Cat.:~\citenum{Traulsen19a,Traulsen20a} & 1328 imgs., 358\,809 srcs., 560\,deg$^2$
\\ \midrule

\integral & \begin{tabular}{@{}l@{}l@{}}20--40
  \\ 40--60\\60--100\end{tabular} & $8.3\,\deg \times 8.0\,\deg$ & 12\,\amin
  FWHM & & -- & -- & Y & Y & Y & N & IBIS/ISGRI & Instr.:
  \citenum{Lebrun03a,Ubertini03a,Winkler03a} & no catalog \\ \midrule

\swift XRT & \begin{tabular}{@{}l@{}l@{}}0.2--2 (soft)
  \\ 2--10 (hard)\\0.3--10 (total)\end{tabular} & 12.3\,\amin\ rad. & 18\,\asec\ HEW & 28\,\asec &
  \href{https://www.swift.ac.uk/2SXPS/}{2SXPS} (20\,\asec) & 1\,091\,058 & Y & Y & Y & Y &
  PC & \begin{tabular}{@{}l@{}}Instr.: \citenum{BurrowsXRT,GehrelsSwift}\\Cat.: \citenum{Evans20a}\end{tabular} & 48\,931 imgs., 206\,335 srcs., 3790\,deg$^2$ \\
\bottomrule
\end{tabular}
}
\end{table}
\end{landscape}

\section{Summary}
\label{sec:summary}

The high-energy lightcurve generator (\hiligt) is a simple and fast tool that produces long-term lightcurves by combining archival data and upper limit estimates.
Already by providing a unified access to most of the relevant X-ray catalogs, \hiligt is of excellent use for X-ray and multi-wavelength researchers to quickly get an overview of existing data, and study the past history of the source.
Through the addition of upper limits, valuable information about periods of decreased source activity are added to the lightcurve. This is information that can be used to study, for instance, off-states and transient phenomena. Only through the combination of various missions, variability patterns on decade-long time scales can be resolved, which opens a new discovery space as the data coverage of X-ray sources constantly increases. 

This paper, which describes the back-end of \hiligt, is complementary to Paper~I where we describe the front end together with an overview of the server structure and science use-cases. Here, we described the back-end.
Firstly, we showed how upper limits are calculated in \hiligt and which database implementation we use.
Upper limits from image data are available for the missions \einstein, \exosat, \rosat, \swift, \integral, and \xmm, and we described their footprints, catalog calls, vignetting and encircled energy fraction.
Besides calculating upper limits of these missions, \hiligt provides archival catalog data for formerly mentioned missions, as well as for \uhuru, \arielv, \heaoI, \ginga, and \asca.
With the inclusion of these historical catalogs -- notably the Fourth Uhuru and \heaoI Piccinotti catalog -- \hiligt is able to calculate lightcurves as long as 50 years, covering the greater part of space-based X-ray astronomy.
Additionally, all-sky monitor lightcurves of \velaVb and \arielv are included, which provide the opportunity to directly compare data from some of the earliest X-ray satellites.

Due to the different effective areas of the satellites, count rates cannot be directly compared among the instruments. \hiligt therefore transforms the count rates to a physical flux using a selection of spectral models as an approximation, such as an absorbed power-law or a black body.
We cautioned the user about the importance of the correct choice of the spectral model (see Sect.~\ref{sec:convfactor}), which can otherwise lead to systematic offsets in the derived fluxes. 

Finally, we emphasize that this paper comprises a detailed and universally researched composition of many current and past X-ray satellites. As the extraction of calibration data from old mission papers can be time-consuming, this paper can serve as a look-up reference to easily access the most relevant information.

Future plans on \hiligt are the inclusion of additional missions such as \textsl{Chandra}, the ASM of \textsl{RXTE} and \ginga, and \textsl{NuSTAR}. We also plan to add new spectral models such as APEC and thermal Bremsstrahlung, which are included in PIMMS but not yet in \hiligt. To give the user increased control over spectral models and output energy ranges it is also planned to implement an on-the-fly computation of the flux conversion factors.


\par\addvspace{6pt}
{ \footnotesize
  \begin{spacing}{0.81}
    \noindent\textit{Acknowledgements.}\hskip.5em
    OK thanks the ESAC Trainee Program (\url{https://www.cosmos.esa.int/web/esac-trainees}) and ERASMUS+ stipend for their financial support. IT gratefully acknowledges support by Deutsches Zentrum f\"ur Luft- und Raumfahrt (DLR) through grant 50\,OX\,1901.
  \end{spacing}
}
\addvspace{6pt}

\clearpage
\bibliographystyle{elsarticle-harv}  
\bibliography{references.bib}

\begin{thebibliography}{92}
\expandafter\ifx\csname natexlab\endcsname\relax\def\natexlab#1{#1}\fi
\providecommand{\url}[1]{\texttt{#1}}
\providecommand{\href}[2]{#2}
\providecommand{\path}[1]{#1}
\providecommand{\DOIprefix}{doi:}
\providecommand{\ArXivprefix}{arXiv:}
\providecommand{\URLprefix}{URL: }
\providecommand{\Pubmedprefix}{pmid:}
\providecommand{\doi}[1]{\href{http://dx.doi.org/#1}{\path{#1}}}
\providecommand{\Pubmed}[1]{\href{pmid:#1}{\path{#1}}}
\providecommand{\bibinfo}[2]{#2}
\ifx\xfnm\relax \def\xfnm[#1]{\unskip,\space#1}\fi
\bibitem[{Arida(1998)}]{AscaTechReport}
\bibinfo{author}{Arida, M.}, \bibinfo{year}{1998}.
\newblock \bibinfo{title}{ASCA Technical Description}.
\newblock \bibinfo{type}{Technical Report}. Laboratory of High Energy
  Astrophysics, NASA GSFC, Greenbelt.
\newblock
  \bibinfo{note}{\url{https://heasarc.gsfc.nasa.gov/docs/asca/ao7/appendix_e/ao7_appendix_e.html}}.
\bibitem[{{Arnaud} et~al.(2011){Arnaud}, {Smith} and
  {Siemiginowska}}]{2011hxra.book.....A}
\bibinfo{author}{{Arnaud}, K.}, \bibinfo{author}{{Smith}, R.},
  \bibinfo{author}{{Siemiginowska}, A.}, \bibinfo{year}{2011}.
\newblock \bibinfo{title}{{Handbook of X-ray Astronomy}}.
\newblock \bibinfo{publisher}{{Cambridge University Press}}.
\bibitem[{{Aschenbach}(1988)}]{1988ApOpt..27.1404A}
\bibinfo{author}{{Aschenbach}, B.}, \bibinfo{year}{1988}.
\newblock \bibinfo{title}{{Design, construction, and performance of the ROSAT
  high- resolution x-ray mirror assembly}}.
\newblock \bibinfo{journal}{\ao} \bibinfo{volume}{27},
  \bibinfo{pages}{1404--1413}.
\newblock \DOIprefix\doi{10.1364/AO.27.001404}.
\bibitem[{{Astropy Collaboration} et~al.(2013){Astropy Collaboration},
  {Robitaille}, {Tollerud}, {Greenfield}, {Droettboom}, {Bray}, {Aldcroft},
  {Davis}, {Ginsburg} and {Price-Whelan}}]{2013AA...558A..33A}
\bibinfo{author}{{Astropy Collaboration}}, \bibinfo{author}{{Robitaille},
  T.P.}, \bibinfo{author}{{Tollerud}, E.J.}, \bibinfo{author}{{Greenfield},
  P.}, \bibinfo{author}{{Droettboom}, M.}, \bibinfo{author}{{Bray}, E.},
  \bibinfo{author}{{Aldcroft}, T.}, \bibinfo{author}{{Davis}, M.},
  \bibinfo{author}{{Ginsburg}, A.}, \bibinfo{author}{{Price-Whelan}, A.M.},
  \bibinfo{year}{2013}.
\newblock \bibinfo{title}{{Astropy: A community Python package for astronomy}}.
\newblock \bibinfo{journal}{\aap} \bibinfo{volume}{558}, \bibinfo{pages}{A33}.
\newblock \DOIprefix\doi{10.1051/0004-6361/201322068},
  \href{http://arxiv.org/abs/1307.6212}{{\tt arXiv:1307.6212}}.
\bibitem[{{Belloni} et~al.(2011){Belloni}, {Motta} and
  {Mu{\~n}oz-Darias}}]{Belloni11a}
\bibinfo{author}{{Belloni}, T.M.}, \bibinfo{author}{{Motta}, S.E.},
  \bibinfo{author}{{Mu{\~n}oz-Darias}, T.}, \bibinfo{year}{2011}.
\newblock \bibinfo{title}{{Black hole transients}}.
\newblock \bibinfo{journal}{Bulletin of the Astronomical Society of India}
  \bibinfo{volume}{39}, \bibinfo{pages}{409--428}.
\newblock \href{http://arxiv.org/abs/1109.3388}{{\tt arXiv:1109.3388}}.
\bibitem[{{Boller} et~al.(2016){Boller}, {Freyberg}, {Tr{\"u}mper}, {Haberl},
  {Voges} and {Nandra}}]{2016AA...588A.103B}
\bibinfo{author}{{Boller}, T.}, \bibinfo{author}{{Freyberg}, M.J.},
  \bibinfo{author}{{Tr{\"u}mper}, J.}, \bibinfo{author}{{Haberl}, F.},
  \bibinfo{author}{{Voges}, W.}, \bibinfo{author}{{Nandra}, K.},
  \bibinfo{year}{2016}.
\newblock \bibinfo{title}{{Second ROSAT all-sky survey (2RXS) source
  catalogue}}.
\newblock \bibinfo{journal}{\aap} \bibinfo{volume}{588}, \bibinfo{pages}{A103}.
\newblock \DOIprefix\doi{10.1051/0004-6361/201525648},
  \href{http://arxiv.org/abs/1609.09244}{{\tt arXiv:1609.09244}}.
\bibitem[{Briel et~al.(1996)Briel, Aschenbach, Hasinger, Hippmann, Pfeffermann,
  Predehl, Schmitt, Schwentker, Voges and Zimmermann}]{ROSAT_User_Handbook}
\bibinfo{editor}{Briel, U.}, \bibinfo{editor}{Aschenbach, B.},
  \bibinfo{editor}{Hasinger, G.}, \bibinfo{editor}{Hippmann, H.},
  \bibinfo{editor}{Pfeffermann, E.}, \bibinfo{editor}{Predehl, P.},
  \bibinfo{editor}{Schmitt, J.}, \bibinfo{editor}{Schwentker, O.},
  \bibinfo{editor}{Voges, W.}, \bibinfo{editor}{Zimmermann, U.} (Eds.),
  \bibinfo{year}{1996}.
\newblock \bibinfo{title}{ROSAT User's Handbook},
  \bibinfo{publisher}{Max-Planck-Institut f{\"u}r Extraterrestrische Physik},
  \bibinfo{address}{85740 Garching bei M{\"u}nchen, Germany}.
\newblock
  \bibinfo{note}{\url{https://heasarc.gsfc.nasa.gov/docs/rosat/ruh/handbook/handbook.html}}.
\bibitem[{{Burrows} et~al.(2005){Burrows}, {Hill}, {Nousek}, {Kennea}, {Wells},
  {Osborne}, {Abbey}, {Beardmore}, {Mukerjee}, {Short}, {Chincarini},
  {Campana}, {Citterio}, {Moretti}, {Pagani}, {Tagliaferri}, {Giommi},
  {Capalbi}, {Tamburelli}, {Angelini}, {Cusumano}, {Br{\"a}uninger}, {Burkert}
  and {Hartner}}]{BurrowsXRT}
\bibinfo{author}{{Burrows}, D.N.}, \bibinfo{author}{{Hill}, J.E.},
  \bibinfo{author}{{Nousek}, J.A.}, \bibinfo{author}{{Kennea}, J.A.},
  \bibinfo{author}{{Wells}, A.}, \bibinfo{author}{{Osborne}, J.P.},
  \bibinfo{author}{{Abbey}, A.F.}, \bibinfo{author}{{Beardmore}, A.},
  \bibinfo{author}{{Mukerjee}, K.}, \bibinfo{author}{{Short}, A.D.T.},
  \bibinfo{author}{{Chincarini}, G.}, \bibinfo{author}{{Campana}, S.},
  \bibinfo{author}{{Citterio}, O.}, \bibinfo{author}{{Moretti}, A.},
  \bibinfo{author}{{Pagani}, C.}, \bibinfo{author}{{Tagliaferri}, G.},
  \bibinfo{author}{{Giommi}, P.}, \bibinfo{author}{{Capalbi}, M.},
  \bibinfo{author}{{Tamburelli}, F.}, \bibinfo{author}{{Angelini}, L.},
  \bibinfo{author}{{Cusumano}, G.}, \bibinfo{author}{{Br{\"a}uninger}, H.W.},
  \bibinfo{author}{{Burkert}, W.}, \bibinfo{author}{{Hartner}, G.D.},
  \bibinfo{year}{2005}.
\newblock \bibinfo{title}{{The Swift X-Ray Telescope}}.
\newblock \bibinfo{journal}{\ssr} \bibinfo{volume}{120},
  \bibinfo{pages}{165--195}.
\newblock \DOIprefix\doi{10.1007/s11214-005-5097-2},
  \href{http://arxiv.org/abs/arXiv:astro-ph/0508071}{{\tt
  arXiv:arXiv:astro-ph/0508071}}.
\bibitem[{D.E.~Harris(1984)}]{RUM}
\bibinfo{author}{D.E.~Harris, D.I.}, \bibinfo{year}{1984}.
\newblock \bibinfo{title}{Einstein Observatory Revised User's Manual}.
\newblock \bibinfo{type}{Technical Report} \bibinfo{number}{2}.
  Harvard-Smithonian Center for Astrophysics.
  \bibinfo{address}{\url{https://heasarc.gsfc.nasa.gov/docs/einstein/}}.
\bibitem[{{de Korte} et~al.(1981){de Korte}, {Bleeker}, {den Boggende},
  {Branduardi-Raymont}, {Brinkman}, {Culhane}, {Gronenschild}, {Mason} and
  {McKechnie}}]{1981SSRv...30..495D}
\bibinfo{author}{{de Korte}, P.A.J.}, \bibinfo{author}{{Bleeker}, J.A.M.},
  \bibinfo{author}{{den Boggende}, A.J.F.},
  \bibinfo{author}{{Branduardi-Raymont}, G.}, \bibinfo{author}{{Brinkman},
  A.C.}, \bibinfo{author}{{Culhane}, J.L.}, \bibinfo{author}{{Gronenschild},
  E.H.B.M.}, \bibinfo{author}{{Mason}, I.}, \bibinfo{author}{{McKechnie},
  S.P.}, \bibinfo{year}{1981}.
\newblock \bibinfo{title}{{The X-ray imaging telescopes on EXOSAT}}.
\newblock \bibinfo{journal}{\ssr} \bibinfo{volume}{30},
  \bibinfo{pages}{495--511}.
\newblock \DOIprefix\doi{10.1007/BF01246070}.
\bibitem[{{Evans} et~al.(2020){Evans}, {Page}, {Osborne}, {Beardmore},
  {Willingale}, {Burrows}, {Kennea}, {Perri}, {Capalbi}, {Tagliaferri} and
  {Cenko}}]{Evans20a}
\bibinfo{author}{{Evans}, P.A.}, \bibinfo{author}{{Page}, K.L.},
  \bibinfo{author}{{Osborne}, J.P.}, \bibinfo{author}{{Beardmore}, A.P.},
  \bibinfo{author}{{Willingale}, R.}, \bibinfo{author}{{Burrows}, D.N.},
  \bibinfo{author}{{Kennea}, J.A.}, \bibinfo{author}{{Perri}, M.},
  \bibinfo{author}{{Capalbi}, M.}, \bibinfo{author}{{Tagliaferri}, G.},
  \bibinfo{author}{{Cenko}, S.B.}, \bibinfo{year}{2020}.
\newblock \bibinfo{title}{{2SXPS: An Improved and Expanded Swift X-Ray
  Telescope Point-source Catalog}}.
\newblock \bibinfo{journal}{\apjs} \bibinfo{volume}{247}, \bibinfo{pages}{54}.
\newblock \DOIprefix\doi{10.3847/1538-4365/ab7db9},
  \href{http://arxiv.org/abs/1911.11710}{{\tt arXiv:1911.11710}}.
\bibitem[{{Forman} et~al.(1978){Forman}, {Jones}, {Cominsky}, {Julien},
  {Murray}, {Peters}, {Tananbaum} and {Giacconi}}]{1978ApJS...38..357F}
\bibinfo{author}{{Forman}, W.}, \bibinfo{author}{{Jones}, C.},
  \bibinfo{author}{{Cominsky}, L.}, \bibinfo{author}{{Julien}, P.},
  \bibinfo{author}{{Murray}, S.}, \bibinfo{author}{{Peters}, G.},
  \bibinfo{author}{{Tananbaum}, H.}, \bibinfo{author}{{Giacconi}, R.},
  \bibinfo{year}{1978}.
\newblock \bibinfo{title}{{The fourth Uhuru catalog of X-ray sources.}}
\newblock \bibinfo{journal}{\apjs} \bibinfo{volume}{38},
  \bibinfo{pages}{357--412}.
\newblock \DOIprefix\doi{10.1086/190561}.
\bibitem[{{Gabriel} et~al.(2004){Gabriel}, {Denby}, {Fyfe}, {Hoar}, {Ibarra},
  {Ojero}, {Osborne}, {Saxton}, {Lammers} and {Vacanti}}]{2004ASPC..314..759G}
\bibinfo{author}{{Gabriel}, C.}, \bibinfo{author}{{Denby}, M.},
  \bibinfo{author}{{Fyfe}, D.J.}, \bibinfo{author}{{Hoar}, J.},
  \bibinfo{author}{{Ibarra}, A.}, \bibinfo{author}{{Ojero}, E.},
  \bibinfo{author}{{Osborne}, J.}, \bibinfo{author}{{Saxton}, R.D.},
  \bibinfo{author}{{Lammers}, U.}, \bibinfo{author}{{Vacanti}, G.},
  \bibinfo{year}{2004}.
\newblock \bibinfo{title}{{The XMM-Newton SAS - Distributed Development and
  Maintenance of a Large Science Analysis System: A Critical Analysis}}, in:
  \bibinfo{editor}{{Ochsenbein}, F.}, \bibinfo{editor}{{Allen}, M.G.},
  \bibinfo{editor}{{Egret}, D.} (Eds.), \bibinfo{booktitle}{Astronomical Data
  Analysis Software and Systems (ADASS) XIII}, p. \bibinfo{pages}{759}.
\bibitem[{{Gehrels}(1986)}]{Gehrels86a}
\bibinfo{author}{{Gehrels}, N.}, \bibinfo{year}{1986}.
\newblock \bibinfo{title}{{Confidence Limits for Small Numbers of Events in
  Astrophysical Data}}.
\newblock \bibinfo{journal}{\apj} \bibinfo{volume}{303}, \bibinfo{pages}{336}.
\newblock \DOIprefix\doi{10.1086/164079}.
\bibitem[{{Gehrels} et~al.(2004){Gehrels}, {Chincarini}, {Giommi}, {Mason},
  {Nousek}, {Wells}, {White}, {Barthelmy}, {Burrows}, {Cominsky}, {Hurley},
  {Marshall}, {M{\'e}sz{\'a}ros}, {Roming}, {Angelini}, {Barbier}, {Belloni},
  {Campana}, {Caraveo}, {Chester}, {Citterio}, {Cline}, {Cropper}, {Cummings},
  {Dean}, {Feigelson}, {Fenimore}, {Frail}, {Fruchter}, {Garmire}, {Gendreau},
  {Ghisellini}, {Greiner}, {Hill}, {Hunsberger}, {Krimm}, {Kulkarni}, {Kumar},
  {Lebrun}, {Lloyd-Ronning}, {Markwardt}, {Mattson}, {Mushotzky}, {Norris},
  {Osborne}, {Paczynski}, {Palmer}, {Park}, {Parsons}, {Paul}, {Rees},
  {Reynolds}, {Rhoads}, {Sasseen}, {Schaefer}, {Short}, {Smale}, {Smith},
  {Stella}, {Tagliaferri}, {Takahashi}, {Tashiro}, {Townsley}, {Tueller},
  {Turner}, {Vietri}, {Voges}, {Ward}, {Willingale}, {Zerbi} and
  {Zhang}}]{GehrelsSwift}
\bibinfo{author}{{Gehrels}, N.}, \bibinfo{author}{{Chincarini}, G.},
  \bibinfo{author}{{Giommi}, P.}, \bibinfo{author}{{Mason}, K.O.},
  \bibinfo{author}{{Nousek}, J.A.}, \bibinfo{author}{{Wells}, A.A.},
  \bibinfo{author}{{White}, N.E.}, \bibinfo{author}{{Barthelmy}, S.D.},
  \bibinfo{author}{{Burrows}, D.N.}, \bibinfo{author}{{Cominsky}, L.R.},
  \bibinfo{author}{{Hurley}, K.C.}, \bibinfo{author}{{Marshall}, F.E.},
  \bibinfo{author}{{M{\'e}sz{\'a}ros}, P.}, \bibinfo{author}{{Roming}, P.W.A.},
  \bibinfo{author}{{Angelini}, L.}, \bibinfo{author}{{Barbier}, L.M.},
  \bibinfo{author}{{Belloni}, T.}, \bibinfo{author}{{Campana}, S.},
  \bibinfo{author}{{Caraveo}, P.A.}, \bibinfo{author}{{Chester}, M.M.},
  \bibinfo{author}{{Citterio}, O.}, \bibinfo{author}{{Cline}, T.L.},
  \bibinfo{author}{{Cropper}, M.S.}, \bibinfo{author}{{Cummings}, J.R.},
  \bibinfo{author}{{Dean}, A.J.}, \bibinfo{author}{{Feigelson}, E.D.},
  \bibinfo{author}{{Fenimore}, E.E.}, \bibinfo{author}{{Frail}, D.A.},
  \bibinfo{author}{{Fruchter}, A.S.}, \bibinfo{author}{{Garmire}, G.P.},
  \bibinfo{author}{{Gendreau}, K.}, \bibinfo{author}{{Ghisellini}, G.},
  \bibinfo{author}{{Greiner}, J.}, \bibinfo{author}{{Hill}, J.E.},
  \bibinfo{author}{{Hunsberger}, S.D.}, \bibinfo{author}{{Krimm}, H.A.},
  \bibinfo{author}{{Kulkarni}, S.R.}, \bibinfo{author}{{Kumar}, P.},
  \bibinfo{author}{{Lebrun}, F.}, \bibinfo{author}{{Lloyd-Ronning}, N.M.},
  \bibinfo{author}{{Markwardt}, C.B.}, \bibinfo{author}{{Mattson}, B.J.},
  \bibinfo{author}{{Mushotzky}, R.F.}, \bibinfo{author}{{Norris}, J.P.},
  \bibinfo{author}{{Osborne}, J.}, \bibinfo{author}{{Paczynski}, B.},
  \bibinfo{author}{{Palmer}, D.M.}, \bibinfo{author}{{Park}, H.S.},
  \bibinfo{author}{{Parsons}, A.M.}, \bibinfo{author}{{Paul}, J.},
  \bibinfo{author}{{Rees}, M.J.}, \bibinfo{author}{{Reynolds}, C.S.},
  \bibinfo{author}{{Rhoads}, J.E.}, \bibinfo{author}{{Sasseen}, T.P.},
  \bibinfo{author}{{Schaefer}, B.E.}, \bibinfo{author}{{Short}, A.T.},
  \bibinfo{author}{{Smale}, A.P.}, \bibinfo{author}{{Smith}, I.A.},
  \bibinfo{author}{{Stella}, L.}, \bibinfo{author}{{Tagliaferri}, G.},
  \bibinfo{author}{{Takahashi}, T.}, \bibinfo{author}{{Tashiro}, M.},
  \bibinfo{author}{{Townsley}, L.K.}, \bibinfo{author}{{Tueller}, J.},
  \bibinfo{author}{{Turner}, M.J.L.}, \bibinfo{author}{{Vietri}, M.},
  \bibinfo{author}{{Voges}, W.}, \bibinfo{author}{{Ward}, M.J.},
  \bibinfo{author}{{Willingale}, R.}, \bibinfo{author}{{Zerbi}, F.M.},
  \bibinfo{author}{{Zhang}, W.W.}, \bibinfo{year}{2004}.
\newblock \bibinfo{title}{{The Swift Gamma-Ray Burst Mission}}.
\newblock \bibinfo{journal}{\apj} \bibinfo{volume}{611},
  \bibinfo{pages}{1005--1020}.
\newblock \DOIprefix\doi{10.1086/422091}.
\bibitem[{{Gendreau}(1995)}]{1995PhDT.........7G}
\bibinfo{author}{{Gendreau}, K.C.}, \bibinfo{year}{1995}.
\newblock \bibinfo{title}{{X-Ray Ccds for Space Applications: Calibration,
  Radiation Hardness, and Use for Measuring the Spectrum of the Cosmic X-Ray
  Background}}.
\newblock Ph.D. thesis. MASSACHUSETTS INSTITUTE OF TECHNOLOGY.
\bibitem[{{Giacconi} et~al.(1979){Giacconi}, {Branduardi}, {Briel}, {Epstein},
  {Fabricant}, {Feigelson}, {Forman}, {Gorenstein}, {Grindlay}, {Gursky},
  {Harnden}, {Henry}, {Jones}, {Kellogg}, {Koch}, {Murray}, {Schreier},
  {Seward}, {Tananbaum}, {Topka}, {Van Speybroeck}, {Holt}, {Becker}, {Boldt},
  {Serlemitsos}, {Clark}, {Canizares}, {Markert}, {Novick}, {Helfand} and
  {Long}}]{1979ApJ...230..540G}
\bibinfo{author}{{Giacconi}, R.}, \bibinfo{author}{{Branduardi}, G.},
  \bibinfo{author}{{Briel}, U.}, \bibinfo{author}{{Epstein}, A.},
  \bibinfo{author}{{Fabricant}, D.}, \bibinfo{author}{{Feigelson}, E.},
  \bibinfo{author}{{Forman}, W.}, \bibinfo{author}{{Gorenstein}, P.},
  \bibinfo{author}{{Grindlay}, J.}, \bibinfo{author}{{Gursky}, H.},
  \bibinfo{author}{{Harnden}, F.R.}, \bibinfo{author}{{Henry}, J.P.},
  \bibinfo{author}{{Jones}, C.}, \bibinfo{author}{{Kellogg}, E.},
  \bibinfo{author}{{Koch}, D.}, \bibinfo{author}{{Murray}, S.},
  \bibinfo{author}{{Schreier}, E.}, \bibinfo{author}{{Seward}, F.},
  \bibinfo{author}{{Tananbaum}, H.}, \bibinfo{author}{{Topka}, K.},
  \bibinfo{author}{{Van Speybroeck}, L.}, \bibinfo{author}{{Holt}, S.S.},
  \bibinfo{author}{{Becker}, R.H.}, \bibinfo{author}{{Boldt}, E.A.},
  \bibinfo{author}{{Serlemitsos}, P.J.}, \bibinfo{author}{{Clark}, G.},
  \bibinfo{author}{{Canizares}, C.}, \bibinfo{author}{{Markert}, T.},
  \bibinfo{author}{{Novick}, R.}, \bibinfo{author}{{Helfand}, D.},
  \bibinfo{author}{{Long}, K.}, \bibinfo{year}{1979}.
\newblock \bibinfo{title}{{The Einstein /HEAO 2/ X-ray Observatory}}.
\newblock \bibinfo{journal}{\apj} \bibinfo{volume}{230},
  \bibinfo{pages}{540--550}.
\newblock \DOIprefix\doi{10.1086/157110}.
\bibitem[{{Giacconi} et~al.(1971){Giacconi}, {Kellogg}, {Gorenstein}, {Gursky}
  and {Tananbaum}}]{1971ApJ...165L..27G}
\bibinfo{author}{{Giacconi}, R.}, \bibinfo{author}{{Kellogg}, E.},
  \bibinfo{author}{{Gorenstein}, P.}, \bibinfo{author}{{Gursky}, H.},
  \bibinfo{author}{{Tananbaum}, H.}, \bibinfo{year}{1971}.
\newblock \bibinfo{title}{{An X-Ray Scan of the Galactic Plane from UHURU}}.
\newblock \bibinfo{journal}{\apjl} \bibinfo{volume}{165}, \bibinfo{pages}{L27}.
\newblock \DOIprefix\doi{10.1086/180711}.
\bibitem[{{Gioia} et~al.(1990){Gioia}, {Maccacaro}, {Schild}, {Wolter},
  {Stocke}, {Morris} and {Henry}}]{1990ApJS...72..567G}
\bibinfo{author}{{Gioia}, I.M.}, \bibinfo{author}{{Maccacaro}, T.},
  \bibinfo{author}{{Schild}, R.E.}, \bibinfo{author}{{Wolter}, A.},
  \bibinfo{author}{{Stocke}, J.T.}, \bibinfo{author}{{Morris}, S.L.},
  \bibinfo{author}{{Henry}, J.P.}, \bibinfo{year}{1990}.
\newblock \bibinfo{title}{{The Einstein Observatory Extended Medium-Sensitivity
  Survey. I - X-ray data and analysis}}.
\newblock \bibinfo{journal}{\apjs} \bibinfo{volume}{72},
  \bibinfo{pages}{567--619}.
\newblock \DOIprefix\doi{10.1086/191426}.
\bibitem[{{Gokus} et~al.(2020){Gokus}, {Rau}, {Wilms}, {Ducci}, {Koenig},
  {Weber}, {Boller} and {Malyali}}]{GokusATel20a}
\bibinfo{author}{{Gokus}, A.}, \bibinfo{author}{{Rau}, A.},
  \bibinfo{author}{{Wilms}, J.}, \bibinfo{author}{{Ducci}, L.},
  \bibinfo{author}{{Koenig}, O.}, \bibinfo{author}{{Weber}, P.},
  \bibinfo{author}{{Boller}, T.}, \bibinfo{author}{{Malyali}, A.},
  \bibinfo{year}{2020}.
\newblock \bibinfo{title}{{SRGt J071522.1-191609: SRG/eROSITA discovery of a
  bright transient X-ray source}}.
\newblock \bibinfo{journal}{The Astronomer's Telegram} \bibinfo{volume}{13657},
  \bibinfo{pages}{1}.
\bibitem[{{Goldwurm} et~al.(2003){Goldwurm}, {David}, {Foschini}, {Gros},
  {Laurent}, {Sauvageon}, {Bird}, {Lerusse} and {Produit}}]{Goldwurm03}
\bibinfo{author}{{Goldwurm}, A.}, \bibinfo{author}{{David}, P.},
  \bibinfo{author}{{Foschini}, L.}, \bibinfo{author}{{Gros}, A.},
  \bibinfo{author}{{Laurent}, P.}, \bibinfo{author}{{Sauvageon}, A.},
  \bibinfo{author}{{Bird}, A.J.}, \bibinfo{author}{{Lerusse}, L.},
  \bibinfo{author}{{Produit}, N.}, \bibinfo{year}{2003}.
\newblock \bibinfo{title}{{The INTEGRAL/IBIS scientific data analysis}}.
\newblock \bibinfo{journal}{\aap} \bibinfo{volume}{411},
  \bibinfo{pages}{L223--L229}.
\newblock \DOIprefix\doi{10.1051/0004-6361:20031395},
  \href{http://arxiv.org/abs/astro-ph/0311172}{{\tt arXiv:astro-ph/0311172}}.
\bibitem[{{Gorenstein} et~al.(1981){Gorenstein}, {Harnden} and
  {Fabricant}}]{1981ITNS...28..869G}
\bibinfo{author}{{Gorenstein}, P.}, \bibinfo{author}{{Harnden}, Jr., F.R.},
  \bibinfo{author}{{Fabricant}, D.G.}, \bibinfo{year}{1981}.
\newblock \bibinfo{title}{{In orbit performance of the Einstein
  Observatory/HEAO-2 Imaging Proportional Counter}}.
\newblock \bibinfo{journal}{IEEE Transactions on Nuclear Science}
  \bibinfo{volume}{28}, \bibinfo{pages}{869--874}.
\newblock \DOIprefix\doi{10.1109/TNS.1981.4331295}.
\bibitem[{{Gotthelf} and {White}(1997)}]{1997xisc.conf...31G}
\bibinfo{author}{{Gotthelf}, E.V.}, \bibinfo{author}{{White}, N.E.},
  \bibinfo{year}{1997}.
\newblock \bibinfo{title}{{The ASCA SIS Source Catalog}}, in:
  \bibinfo{editor}{{Makino}, F.}, \bibinfo{editor}{{Mitsuda}, K.} (Eds.),
  \bibinfo{booktitle}{X-Ray Imaging and Spectroscopy of Cosmic Hot Plasmas},
  p.~\bibinfo{pages}{31}.
\bibitem[{{Grindlay} et~al.(1976){Grindlay}, {Gursky}, {Schnopper},
  {Parsignault}, {Heise}, {Brinkman} and {Schrijver}}]{Grindlay76a}
\bibinfo{author}{{Grindlay}, J.}, \bibinfo{author}{{Gursky}, H.},
  \bibinfo{author}{{Schnopper}, H.}, \bibinfo{author}{{Parsignault}, D.R.},
  \bibinfo{author}{{Heise}, J.}, \bibinfo{author}{{Brinkman}, A.C.},
  \bibinfo{author}{{Schrijver}, J.}, \bibinfo{year}{1976}.
\newblock \bibinfo{title}{{Discovery of intense X-ray bursts from the globular
  cluster NGC 6624.}}
\newblock \bibinfo{journal}{\apjl} \bibinfo{volume}{205},
  \bibinfo{pages}{L127--L130}.
\newblock \DOIprefix\doi{10.1086/182105}.
\bibitem[{{Harnden} et~al.(1984){Harnden}, {Fabricant}, {Harris} and
  {Schwarz}}]{1984SAOSR.393.....H}
\bibinfo{author}{{Harnden}, Jr., F.R.}, \bibinfo{author}{{Fabricant}, D.G.},
  \bibinfo{author}{{Harris}, D.E.}, \bibinfo{author}{{Schwarz}, J.},
  \bibinfo{year}{1984}.
\newblock \bibinfo{title}{{Scientific Specification of the Data Analysis System
  for the EINSTEIN Observatory (HEAO-2) Imaging Proportional Counter}}.
\newblock \bibinfo{journal}{SAO Special Report} \bibinfo{volume}{393}.
\bibitem[{{Harris}(1990)}]{1990eoci.book.....H}
\bibinfo{author}{{Harris}, D.E.}, \bibinfo{year}{1990}.
\newblock \bibinfo{title}{{The Einstein Observatory Catalog of IPC X-ray
  Sources}}.
\newblock \bibinfo{publisher}{Smithsonian Institution, Astrophysical
  Observatory}.
\bibitem[{{Hartman} et~al.(2008){Hartman}, {Patruno}, {Chakrabarty}, {Kaplan},
  {Markwardt}, {Morgan}, {Ray}, {van der Klis} and {Wijnands}}]{Hartman08a}
\bibinfo{author}{{Hartman}, J.M.}, \bibinfo{author}{{Patruno}, A.},
  \bibinfo{author}{{Chakrabarty}, D.}, \bibinfo{author}{{Kaplan}, D.L.},
  \bibinfo{author}{{Markwardt}, C.B.}, \bibinfo{author}{{Morgan}, E.H.},
  \bibinfo{author}{{Ray}, P.S.}, \bibinfo{author}{{van der Klis}, M.},
  \bibinfo{author}{{Wijnands}, R.}, \bibinfo{year}{2008}.
\newblock \bibinfo{title}{{The Long-Term Evolution of the Spin, Pulse Shape,
  and Orbit of the Accretion-powered Millisecond Pulsar SAX J1808.4-3658}}.
\newblock \bibinfo{journal}{\apj} \bibinfo{volume}{675},
  \bibinfo{pages}{1468--1486}.
\newblock \DOIprefix\doi{10.1086/527461},
  \href{http://arxiv.org/abs/0708.0211}{{\tt arXiv:0708.0211}}.
\bibitem[{{Hasinger} and {van der Klis}(1989)}]{Hasinger89a}
\bibinfo{author}{{Hasinger}, G.}, \bibinfo{author}{{van der Klis}, M.},
  \bibinfo{year}{1989}.
\newblock \bibinfo{title}{{Two patterns of correlated X-ray timing and spectral
  behaviour in low-mass X-ray binaries.}}
\newblock \bibinfo{journal}{\aap} \bibinfo{volume}{225},
  \bibinfo{pages}{79--96}.
\bibitem[{{Hayashida} et~al.(1989){Hayashida}, {Inoue}, {Koyama}, {Awaki} and
  {Takano}}]{1989PASJ...41..373H}
\bibinfo{author}{{Hayashida}, K.}, \bibinfo{author}{{Inoue}, H.},
  \bibinfo{author}{{Koyama}, K.}, \bibinfo{author}{{Awaki}, H.},
  \bibinfo{author}{{Takano}, S.}, \bibinfo{year}{1989}.
\newblock \bibinfo{title}{{The origin and behavior of the background in the
  large area counters on GINGA and its effect on the sensitivity}}.
\newblock \bibinfo{journal}{\pasj} \bibinfo{volume}{41},
  \bibinfo{pages}{373--389}.
\bibitem[{{Henry} et~al.(1977){Henry}, {Kellogg}, {Murray}, {van Speybroeck},
  {Bjorkholm} and {Briel}}]{1977SPIE..106..196H}
\bibinfo{author}{{Henry}, J.P.}, \bibinfo{author}{{Kellogg}, E.M.},
  \bibinfo{author}{{Murray}, S.S.}, \bibinfo{author}{{van Speybroeck}, L.P.},
  \bibinfo{author}{{Bjorkholm}, P.J.}, \bibinfo{author}{{Briel}, U.G.},
  \bibinfo{year}{1977}.
\newblock \bibinfo{title}{{High resolution imaging X-ray detector for
  astronomical measurements}}, in: \bibinfo{editor}{{Chase}, R.C.},
  \bibinfo{editor}{{Kuswa}, G.W.} (Eds.), \bibinfo{booktitle}{X-ray imaging},
  pp. \bibinfo{pages}{196--205}.
\newblock \DOIprefix\doi{10.1117/12.955472}.
\bibitem[{{Holt}(1976)}]{Holt76a}
\bibinfo{author}{{Holt}, S.S.}, \bibinfo{year}{1976}.
\newblock \bibinfo{title}{{Temporal X-Ray Astronomy with a Pinhole Camera}}.
\newblock \bibinfo{journal}{\apss} \bibinfo{volume}{42},
  \bibinfo{pages}{123--141}.
\newblock \DOIprefix\doi{10.1007/BF00645534}.
\bibitem[{{Houck}(2002)}]{2002hrxs.confE..17H}
\bibinfo{author}{{Houck}, J.C.}, \bibinfo{year}{2002}.
\newblock \bibinfo{title}{{ISIS: The Interactive Spectral Interpretation
  System}}, in: \bibinfo{editor}{{Branduardi-Raymont}, G.} (Ed.),
  \bibinfo{booktitle}{High Resolution X-ray Spectroscopy with XMM-Newton and
  Chandra}, p.~\bibinfo{pages}{17}.
\bibitem[{{Jansen} et~al.(2001){Jansen}, {Lumb}, {Altieri}, {Clavel}, {Ehle},
  {Erd}, {Gabriel}, {Guainazzi}, {Gondoin}, {Much}, {Munoz}, {Santos},
  {Schartel}, {Texier} and {Vacanti}}]{Jansen01a}
\bibinfo{author}{{Jansen}, F.}, \bibinfo{author}{{Lumb}, D.},
  \bibinfo{author}{{Altieri}, B.}, \bibinfo{author}{{Clavel}, J.},
  \bibinfo{author}{{Ehle}, M.}, \bibinfo{author}{{Erd}, C.},
  \bibinfo{author}{{Gabriel}, C.}, \bibinfo{author}{{Guainazzi}, M.},
  \bibinfo{author}{{Gondoin}, P.}, \bibinfo{author}{{Much}, R.},
  \bibinfo{author}{{Munoz}, R.}, \bibinfo{author}{{Santos}, M.},
  \bibinfo{author}{{Schartel}, N.}, \bibinfo{author}{{Texier}, D.},
  \bibinfo{author}{{Vacanti}, G.}, \bibinfo{year}{2001}.
\newblock \bibinfo{title}{{XMM-Newton observatory. I. The spacecraft and
  operations}}.
\newblock \bibinfo{journal}{\aap} \bibinfo{volume}{365},
  \bibinfo{pages}{L1--L6}.
\newblock \DOIprefix\doi{10.1051/0004-6361:20000036}.
\bibitem[{{Joye} and {Mandel}(2003)}]{ds9Joye03}
\bibinfo{author}{{Joye}, W.A.}, \bibinfo{author}{{Mandel}, E.},
  \bibinfo{year}{2003}.
\newblock \bibinfo{title}{{New Features of SAOImage DS9}}, in:
  \bibinfo{editor}{{Payne}, H.E.}, \bibinfo{editor}{{Jedrzejewski}, R.I.},
  \bibinfo{editor}{{Hook}, R.N.} (Eds.), \bibinfo{booktitle}{Astronomical Data
  Analysis Software and Systems XII}, p. \bibinfo{pages}{489}.
\bibitem[{{Kaluzienski}(1977)}]{Kaluzienski77a}
\bibinfo{author}{{Kaluzienski}, L.J.}, \bibinfo{year}{1977}.
\newblock \bibinfo{title}{{Studies of Transient X-Ray Sources with the Ariel 5
  All-Sky Monitor.}}
\newblock Ph.D. thesis. National Aeronautics and Space Administration. Goddard
  Space Flight Center, Greenbelt, MD.
\bibitem[{{Komossa} and {Bade}(1999)}]{Komossa99a}
\bibinfo{author}{{Komossa}, S.}, \bibinfo{author}{{Bade}, N.},
  \bibinfo{year}{1999}.
\newblock \bibinfo{title}{{The giant X-ray outbursts in NGC 5905 and IC 3599:()
  hfill Follow-up observations and outburst scenarios}}.
\newblock \bibinfo{journal}{\aap} \bibinfo{volume}{343},
  \bibinfo{pages}{775--787}.
\newblock \href{http://arxiv.org/abs/astro-ph/9901141}{{\tt
  arXiv:astro-ph/9901141}}.
\bibitem[{K\"onig(2019)}]{KoenigMastersThesis19}
\bibinfo{author}{K\"onig, O.}, \bibinfo{year}{2019}.
\newblock \bibinfo{title}{{Extension of Upper Limit Servers and spectral
  analysis of the X-ray binary GRO J1744-28}}.
\newblock Master's thesis. Friedrich-Alexander-Universit\"at
  Erlangen-N\"urnberg. \bibinfo{address}{Germany}.
\newblock
  \bibinfo{note}{\url{https://www.sternwarte.uni-erlangen.de/docs/theses/2019-07_Koenig.pdf}}.
\bibitem[{{Kraft} et~al.(1991){Kraft}, {Burrows} and
  {Nousek}}]{1991ApJ...374..344K}
\bibinfo{author}{{Kraft}, R.P.}, \bibinfo{author}{{Burrows}, D.N.},
  \bibinfo{author}{{Nousek}, J.A.}, \bibinfo{year}{1991}.
\newblock \bibinfo{title}{{Determination of confidence limits for experiments
  with low numbers of counts}}.
\newblock \bibinfo{journal}{\apj} \bibinfo{volume}{374},
  \bibinfo{pages}{344--355}.
\newblock \DOIprefix\doi{10.1086/170124}.
\bibitem[{{Krivonos} et~al.(2007){Krivonos}, {Revnivtsev}, {Lutovinov},
  {Sazonov}, {Churazov} and {Sunyaev}}]{Krivonos07a}
\bibinfo{author}{{Krivonos}, R.}, \bibinfo{author}{{Revnivtsev}, M.},
  \bibinfo{author}{{Lutovinov}, A.}, \bibinfo{author}{{Sazonov}, S.},
  \bibinfo{author}{{Churazov}, E.}, \bibinfo{author}{{Sunyaev}, R.},
  \bibinfo{year}{2007}.
\newblock \bibinfo{title}{{INTEGRAL/IBIS all-sky survey in hard X-rays}}.
\newblock \bibinfo{journal}{\aap} \bibinfo{volume}{475},
  \bibinfo{pages}{775--784}.
\newblock \DOIprefix\doi{10.1051/0004-6361:20077191},
  \href{http://arxiv.org/abs/astro-ph/0701836}{{\tt arXiv:astro-ph/0701836}}.
\bibitem[{{Kuulkers} et~al.(2009){Kuulkers}, {in't Zand} and
  {Lasota}}]{Kuulkers09a}
\bibinfo{author}{{Kuulkers}, E.}, \bibinfo{author}{{in't Zand}, J.J.M.},
  \bibinfo{author}{{Lasota}, J.P.}, \bibinfo{year}{2009}.
\newblock \bibinfo{title}{{Restless quiescence: thermonuclear flashes between
  transient X-ray outbursts}}.
\newblock \bibinfo{journal}{\aap} \bibinfo{volume}{503},
  \bibinfo{pages}{889--897}.
\newblock \DOIprefix\doi{10.1051/0004-6361/200810981},
  \href{http://arxiv.org/abs/0809.3323}{{\tt arXiv:0809.3323}}.
\bibitem[{{Lampton} et~al.(1976){Lampton}, {Margon} and {Bowyer}}]{Lampton76a}
\bibinfo{author}{{Lampton}, M.}, \bibinfo{author}{{Margon}, B.},
  \bibinfo{author}{{Bowyer}, S.}, \bibinfo{year}{1976}.
\newblock \bibinfo{title}{{Parameter estimation in X-ray astronomy.}}
\newblock \bibinfo{journal}{\apj} \bibinfo{volume}{208},
  \bibinfo{pages}{177--190}.
\newblock \DOIprefix\doi{10.1086/154592}.
\bibitem[{{Lebrun} et~al.(2003){Lebrun}, {Leray}, {Lavocat}, {Cr{\'e}tolle},
  {Arqu{\`e}s}, {Blondel}, {Bonnin}, {Bou{\`e}re}, {Cara}, {Chaleil}, {Daly},
  {Desages}, {Dzitko}, {Horeau}, {Laurent}, {Limousin}, {Mathy}, {Mauguen},
  {Meignier}, {Molini{\'e}}, {Poindron}, {Rouger}, {Sauvageon} and
  {Tourrette}}]{Lebrun03a}
\bibinfo{author}{{Lebrun}, F.}, \bibinfo{author}{{Leray}, J.P.},
  \bibinfo{author}{{Lavocat}, P.}, \bibinfo{author}{{Cr{\'e}tolle}, J.},
  \bibinfo{author}{{Arqu{\`e}s}, M.}, \bibinfo{author}{{Blondel}, C.},
  \bibinfo{author}{{Bonnin}, C.}, \bibinfo{author}{{Bou{\`e}re}, A.},
  \bibinfo{author}{{Cara}, C.}, \bibinfo{author}{{Chaleil}, T.},
  \bibinfo{author}{{Daly}, F.}, \bibinfo{author}{{Desages}, F.},
  \bibinfo{author}{{Dzitko}, H.}, \bibinfo{author}{{Horeau}, B.},
  \bibinfo{author}{{Laurent}, P.}, \bibinfo{author}{{Limousin}, O.},
  \bibinfo{author}{{Mathy}, F.}, \bibinfo{author}{{Mauguen}, V.},
  \bibinfo{author}{{Meignier}, F.}, \bibinfo{author}{{Molini{\'e}}, F.},
  \bibinfo{author}{{Poindron}, E.}, \bibinfo{author}{{Rouger}, M.},
  \bibinfo{author}{{Sauvageon}, A.}, \bibinfo{author}{{Tourrette}, T.},
  \bibinfo{year}{2003}.
\newblock \bibinfo{title}{{ISGRI: The INTEGRAL Soft Gamma-Ray Imager}}.
\newblock \bibinfo{journal}{\aap} \bibinfo{volume}{411},
  \bibinfo{pages}{L141--L148}.
\newblock \DOIprefix\doi{10.1051/0004-6361:20031367},
  \href{http://arxiv.org/abs/astro-ph/0310362}{{\tt arXiv:astro-ph/0310362}}.
\bibitem[{{Levine} et~al.(1984){Levine}, {Lang}, {Lewin}, {Primini}, {Dobson},
  {Doty}, {Hoffman}, {Howe}, {Scheepmaker}, {Wheaton}, {Matteson}, {Baity},
  {Gruber}, {Knight}, {Nolan}, {Pelling}, {Rothschild} and
  {Peterson}}]{Levine84a}
\bibinfo{author}{{Levine}, A.M.}, \bibinfo{author}{{Lang}, F.L.},
  \bibinfo{author}{{Lewin}, W.H.G.}, \bibinfo{author}{{Primini}, F.A.},
  \bibinfo{author}{{Dobson}, C.A.}, \bibinfo{author}{{Doty}, J.P.},
  \bibinfo{author}{{Hoffman}, J.A.}, \bibinfo{author}{{Howe}, S.K.},
  \bibinfo{author}{{Scheepmaker}, A.}, \bibinfo{author}{{Wheaton}, W.A.},
  \bibinfo{author}{{Matteson}, J.L.}, \bibinfo{author}{{Baity}, W.A.},
  \bibinfo{author}{{Gruber}, D.E.}, \bibinfo{author}{{Knight}, F.K.},
  \bibinfo{author}{{Nolan}, P.L.}, \bibinfo{author}{{Pelling}, R.M.},
  \bibinfo{author}{{Rothschild}, R.E.}, \bibinfo{author}{{Peterson}, L.E.},
  \bibinfo{year}{1984}.
\newblock \bibinfo{title}{{The HEAO1 A-4 catalog of high-energy X-ray
  sources.}}
\newblock \bibinfo{journal}{\apjs} \bibinfo{volume}{54},
  \bibinfo{pages}{581--617}.
\newblock \DOIprefix\doi{10.1086/190944}.
\bibitem[{{Lin} et~al.(2017){Lin}, {Guillochon}, {Komossa}, {Ramirez-Ruiz},
  {Irwin}, {Maksym}, {Grupe}, {Godet}, {Webb}, {Barret}, {Zauderer}, {Duc},
  {Carrasco} and {Gwyn}}]{Lin17a}
\bibinfo{author}{{Lin}, D.}, \bibinfo{author}{{Guillochon}, J.},
  \bibinfo{author}{{Komossa}, S.}, \bibinfo{author}{{Ramirez-Ruiz}, E.},
  \bibinfo{author}{{Irwin}, J.A.}, \bibinfo{author}{{Maksym}, W.P.},
  \bibinfo{author}{{Grupe}, D.}, \bibinfo{author}{{Godet}, O.},
  \bibinfo{author}{{Webb}, N.A.}, \bibinfo{author}{{Barret}, D.},
  \bibinfo{author}{{Zauderer}, B.A.}, \bibinfo{author}{{Duc}, P.A.},
  \bibinfo{author}{{Carrasco}, E.R.}, \bibinfo{author}{{Gwyn}, S.D.J.},
  \bibinfo{year}{2017}.
\newblock \bibinfo{title}{{A likely decade-long sustained tidal disruption
  event}}.
\newblock \bibinfo{journal}{Nature Astronomy} \bibinfo{volume}{1},
  \bibinfo{pages}{0033}.
\newblock \DOIprefix\doi{10.1038/s41550-016-0033},
  \href{http://arxiv.org/abs/1702.00792}{{\tt arXiv:1702.00792}}.
\bibitem[{{Loiseau} et~al.(2017){Loiseau}, {Baines}, {Colomo}, {Giordano},
  {Mer{\'\i}n}, {Racero}, {Rodr{\'\i}guez}, {Salgado} and
  {Sarmiento}}]{2017RMxAC..49..146L}
\bibinfo{author}{{Loiseau}, N.}, \bibinfo{author}{{Baines}, D.},
  \bibinfo{author}{{Colomo}, E.}, \bibinfo{author}{{Giordano}, F.},
  \bibinfo{author}{{Mer{\'\i}n}, B.}, \bibinfo{author}{{Racero}, E.},
  \bibinfo{author}{{Rodr{\'\i}guez}, P.}, \bibinfo{author}{{Salgado}, J.},
  \bibinfo{author}{{Sarmiento}, M.}, \bibinfo{year}{2017}.
\newblock \bibinfo{title}{{The XMM-Newton Science Archive and its integration
  into ESASky}}, in: \bibinfo{booktitle}{Revista Mexicana de Astronomia y
  Astrofisica Conference Series}, pp. \bibinfo{pages}{146--146}.
\newblock \bibinfo{note}{\url{https://www.cosmos.esa.int/web/xmm-newton/xsa}}.
\bibitem[{{Makino} and {ASTRO-C Team}(1987)}]{1987ApL....25..223M}
\bibinfo{author}{{Makino}, F.}, \bibinfo{author}{{ASTRO-C Team}},
  \bibinfo{year}{1987}.
\newblock \bibinfo{title}{{The X-ray astronomy satellite Astro-C}}.
\newblock \bibinfo{journal}{\aplett} \bibinfo{volume}{25},
  \bibinfo{pages}{223--233}.
\bibitem[{{Makishima} et~al.(1996){Makishima}, {Tashiro}, {Ebisawa}, {Ezawa},
  {Fukazawa}, {Gunji}, {Hirayama}, {Idesawa}, {Ikebe} and
  {Ishida}}]{1996PASJ...48..171M}
\bibinfo{author}{{Makishima}, K.}, \bibinfo{author}{{Tashiro}, M.},
  \bibinfo{author}{{Ebisawa}, K.}, \bibinfo{author}{{Ezawa}, H.},
  \bibinfo{author}{{Fukazawa}, Y.}, \bibinfo{author}{{Gunji}, S.},
  \bibinfo{author}{{Hirayama}, M.}, \bibinfo{author}{{Idesawa}, E.},
  \bibinfo{author}{{Ikebe}, Y.}, \bibinfo{author}{{Ishida}, M.},
  \bibinfo{year}{1996}.
\newblock \bibinfo{title}{{In-Orbit Performance of the Gas Imaging Spectrometer
  onboard ASCA}}.
\newblock \bibinfo{journal}{\pasj} \bibinfo{volume}{48},
  \bibinfo{pages}{171--189}.
\newblock \DOIprefix\doi{10.1093/pasj/48.2.171}.
\bibitem[{{Malyali} et~al.(2020){Malyali}, {Rau}, {Arcodia}, {Boller},
  {Carpano}, {Haberl}, {Merloni}, {Nandra}, {Buchner}, {Liu}, {Salvato},
  {Wilms}, {Koenig}, {Krumpe} and {Lamer}}]{MalyaliATel20a}
\bibinfo{author}{{Malyali}, A.}, \bibinfo{author}{{Rau}, A.},
  \bibinfo{author}{{Arcodia}, R.}, \bibinfo{author}{{Boller}, T.},
  \bibinfo{author}{{Carpano}, S.}, \bibinfo{author}{{Haberl}, F.},
  \bibinfo{author}{{Merloni}, A.}, \bibinfo{author}{{Nandra}, K.},
  \bibinfo{author}{{Buchner}, J.}, \bibinfo{author}{{Liu}, T.},
  \bibinfo{author}{{Salvato}, M.}, \bibinfo{author}{{Wilms}, J.},
  \bibinfo{author}{{Koenig}, O.}, \bibinfo{author}{{Krumpe}, M.},
  \bibinfo{author}{{Lamer}, G.}, \bibinfo{year}{2020}.
\newblock \bibinfo{title}{{eRASSt J082337+042303: A bright, ultra-soft,
  high-amplitude transient in the direction of 2MASX J08233674+042300}}.
\newblock \bibinfo{journal}{The Astronomer's Telegram} \bibinfo{volume}{13712},
  \bibinfo{pages}{1}.
\bibitem[{{McHardy} et~al.(1981){McHardy}, {Lawrence}, {Pye} and
  {Pounds}}]{1981MNRAS.197..893M}
\bibinfo{author}{{McHardy}, I.M.}, \bibinfo{author}{{Lawrence}, A.},
  \bibinfo{author}{{Pye}, J.P.}, \bibinfo{author}{{Pounds}, K.A.},
  \bibinfo{year}{1981}.
\newblock \bibinfo{title}{{The Ariel V (3A) catalogue of X-ray sources. II.}}
\newblock \bibinfo{journal}{\mnras} \bibinfo{volume}{197},
  \bibinfo{pages}{893--919}.
\newblock \DOIprefix\doi{10.1093/mnras/197.4.893}.
\bibitem[{{Menou} et~al.(1999){Menou}, {Esin}, {Narayan}, {Garcia}, {Lasota}
  and {McClintock}}]{Menou99a}
\bibinfo{author}{{Menou}, K.}, \bibinfo{author}{{Esin}, A.A.},
  \bibinfo{author}{{Narayan}, R.}, \bibinfo{author}{{Garcia}, M.R.},
  \bibinfo{author}{{Lasota}, J.P.}, \bibinfo{author}{{McClintock}, J.E.},
  \bibinfo{year}{1999}.
\newblock \bibinfo{title}{{Black Hole and Neutron Star Transients in
  Quiescence}}.
\newblock \bibinfo{journal}{\apj} \bibinfo{volume}{520},
  \bibinfo{pages}{276--291}.
\newblock \DOIprefix\doi{10.1086/307443},
  \href{http://arxiv.org/abs/astro-ph/9810323}{{\tt arXiv:astro-ph/9810323}}.
\bibitem[{{Mukai}(1993)}]{Mukai93PIMMS}
\bibinfo{author}{{Mukai}, K.}, \bibinfo{year}{1993}.
\newblock \bibinfo{title}{{PIMMS and Viewing: proposal preparation tools}}.
\newblock \bibinfo{journal}{Legacy} \bibinfo{volume}{3}.
\newblock \URLprefix
  \url{https://heasarc.gsfc.nasa.gov/docs/journal/pimms3.html}.
\bibitem[{{Nugent} et~al.(1983){Nugent}, {Jensen}, {Nousek}, {Garmire},
  {Mason}, {Walter}, {Bowyer}, {Stern} and {Riegler}}]{Nugent83a}
\bibinfo{author}{{Nugent}, J.J.}, \bibinfo{author}{{Jensen}, K.A.},
  \bibinfo{author}{{Nousek}, J.A.}, \bibinfo{author}{{Garmire}, G.P.},
  \bibinfo{author}{{Mason}, K.O.}, \bibinfo{author}{{Walter}, F.M.},
  \bibinfo{author}{{Bowyer}, C.S.}, \bibinfo{author}{{Stern}, R.A.},
  \bibinfo{author}{{Riegler}, G.R.}, \bibinfo{year}{1983}.
\newblock \bibinfo{title}{{HEAO A-2 soft X-ray source catalog.}}
\newblock \bibinfo{journal}{\apjs} \bibinfo{volume}{51},
  \bibinfo{pages}{1--28}.
\newblock \DOIprefix\doi{10.1086/190838}.
\bibitem[{{Ohashi} et~al.(1996){Ohashi}, {Ebisawa}, {Fukazawa}, {Hiyoshi},
  {Horii}, {Ikebe}, {Ikeda}, {Inoue}, {Ishida} and
  {Ishisaki}}]{1996PASJ...48..157O}
\bibinfo{author}{{Ohashi}, T.}, \bibinfo{author}{{Ebisawa}, K.},
  \bibinfo{author}{{Fukazawa}, Y.}, \bibinfo{author}{{Hiyoshi}, K.},
  \bibinfo{author}{{Horii}, M.}, \bibinfo{author}{{Ikebe}, Y.},
  \bibinfo{author}{{Ikeda}, H.}, \bibinfo{author}{{Inoue}, H.},
  \bibinfo{author}{{Ishida}, M.}, \bibinfo{author}{{Ishisaki}, Y.},
  \bibinfo{year}{1996}.
\newblock \bibinfo{title}{{The Gas Imaging Spectrometer on Board ASCA}}.
\newblock \bibinfo{journal}{\pasj} \bibinfo{volume}{48},
  \bibinfo{pages}{157--170}.
\newblock \DOIprefix\doi{10.1093/pasj/48.2.157}.
\bibitem[{{Peacock} et~al.(1981){Peacock}, {Andresen}, {Manzo}, {Taylor},
  {Villa}, {Re}, {Ives} and {Kellock}}]{1981SSRv...30..525P}
\bibinfo{author}{{Peacock}, A.}, \bibinfo{author}{{Andresen}, R.D.},
  \bibinfo{author}{{Manzo}, G.}, \bibinfo{author}{{Taylor}, B.G.},
  \bibinfo{author}{{Villa}, G.}, \bibinfo{author}{{Re}, S.},
  \bibinfo{author}{{Ives}, J.C.}, \bibinfo{author}{{Kellock}, S.},
  \bibinfo{year}{1981}.
\newblock \bibinfo{title}{{The gas scintillation proportional counter on
  EXOSAT.}}
\newblock \bibinfo{journal}{\ssr} \bibinfo{volume}{30},
  \bibinfo{pages}{525--534}.
\newblock \DOIprefix\doi{10.1007/BF01246072}.
\bibitem[{{Pfeffermann} et~al.(1987){Pfeffermann}, {Briel}, {Hippmann},
  {Kettenring}, {Metzner}, {Predehl}, {Reger}, {Stephan}, {Zombeck}, {Chappell}
  and {Murray}}]{1987SPIE..733..519P}
\bibinfo{author}{{Pfeffermann}, E.}, \bibinfo{author}{{Briel}, U.G.},
  \bibinfo{author}{{Hippmann}, H.}, \bibinfo{author}{{Kettenring}, G.},
  \bibinfo{author}{{Metzner}, G.}, \bibinfo{author}{{Predehl}, P.},
  \bibinfo{author}{{Reger}, G.}, \bibinfo{author}{{Stephan}, K.H.},
  \bibinfo{author}{{Zombeck}, M.}, \bibinfo{author}{{Chappell}, J.},
  \bibinfo{author}{{Murray}, S.S.}, \bibinfo{year}{1987}.
\newblock \bibinfo{title}{{The focal plane instrumentation of the ROSAT
  Telescope}}, in: \bibinfo{booktitle}{Soft X-ray optics and technology}, p.
  \bibinfo{pages}{519}.
\bibitem[{{Piccinotti} et~al.(1982){Piccinotti}, {Mushotzky}, {Boldt}, {Holt},
  {Marshall}, {Serlemitsos} and {Shafer}}]{1982ApJ...253..485P}
\bibinfo{author}{{Piccinotti}, G.}, \bibinfo{author}{{Mushotzky}, R.F.},
  \bibinfo{author}{{Boldt}, E.A.}, \bibinfo{author}{{Holt}, S.S.},
  \bibinfo{author}{{Marshall}, F.E.}, \bibinfo{author}{{Serlemitsos}, P.J.},
  \bibinfo{author}{{Shafer}, R.A.}, \bibinfo{year}{1982}.
\newblock \bibinfo{title}{{A complete X-ray sample of the high-latitude
  /absolute value of B greater than 20 deg/ sky from HEAO 1 A-2 - Log N-log S
  and luminosity functions}}.
\newblock \bibinfo{journal}{\apj} \bibinfo{volume}{253},
  \bibinfo{pages}{485--503}.
\newblock \DOIprefix\doi{10.1086/159651}.
\bibitem[{{Pierre} et~al.(2007){Pierre}, {Chiappetti}, {Pacaud}, {Gueguen},
  {Libbrecht}, {Altieri}, {Aussel}, {Gandhi}, {Garcet}, {Gosset}, {Paioro},
  {Ponman}, {Read}, {Refregier}, {Starck}, {Surdej}, {Valtchanov}, {Adami},
  {Alloin}, {Alshino}, {Andreon}, {Birkinshaw}, {Bremer}, {Detal}, {Duc},
  {Galaz}, {Jones}, {Le F{\`e}vre}, {Le F{\`e}vre}, {Maccagni}, {Mazure},
  {Quintana}, {R{\"o}ttgering}, {Sprimont}, {Tasse}, {Trinchieri} and
  {Willis}}]{Pierre07a}
\bibinfo{author}{{Pierre}, M.}, \bibinfo{author}{{Chiappetti}, L.},
  \bibinfo{author}{{Pacaud}, F.}, \bibinfo{author}{{Gueguen}, A.},
  \bibinfo{author}{{Libbrecht}, C.}, \bibinfo{author}{{Altieri}, B.},
  \bibinfo{author}{{Aussel}, H.}, \bibinfo{author}{{Gandhi}, P.},
  \bibinfo{author}{{Garcet}, O.}, \bibinfo{author}{{Gosset}, E.},
  \bibinfo{author}{{Paioro}, L.}, \bibinfo{author}{{Ponman}, T.J.},
  \bibinfo{author}{{Read}, A.M.}, \bibinfo{author}{{Refregier}, A.},
  \bibinfo{author}{{Starck}, J.L.}, \bibinfo{author}{{Surdej}, J.},
  \bibinfo{author}{{Valtchanov}, I.}, \bibinfo{author}{{Adami}, C.},
  \bibinfo{author}{{Alloin}, D.}, \bibinfo{author}{{Alshino}, A.},
  \bibinfo{author}{{Andreon}, S.}, \bibinfo{author}{{Birkinshaw}, M.},
  \bibinfo{author}{{Bremer}, M.}, \bibinfo{author}{{Detal}, A.},
  \bibinfo{author}{{Duc}, P.A.}, \bibinfo{author}{{Galaz}, G.},
  \bibinfo{author}{{Jones}, L.}, \bibinfo{author}{{Le F{\`e}vre}, J.P.},
  \bibinfo{author}{{Le F{\`e}vre}, O.}, \bibinfo{author}{{Maccagni}, D.},
  \bibinfo{author}{{Mazure}, A.}, \bibinfo{author}{{Quintana}, H.},
  \bibinfo{author}{{R{\"o}ttgering}, H.J.A.}, \bibinfo{author}{{Sprimont},
  P.G.}, \bibinfo{author}{{Tasse}, C.}, \bibinfo{author}{{Trinchieri}, G.},
  \bibinfo{author}{{Willis}, J.P.}, \bibinfo{year}{2007}.
\newblock \bibinfo{title}{{The XMM-Large Scale Structure catalogue: X-ray
  sources and associated optical data. Version I}}.
\newblock \bibinfo{journal}{\mnras} \bibinfo{volume}{382},
  \bibinfo{pages}{279--290}.
\newblock \DOIprefix\doi{10.1111/j.1365-2966.2007.12354.x},
  \href{http://arxiv.org/abs/0708.3299}{{\tt arXiv:0708.3299}}.
\bibitem[{{Pounds} et~al.(1993){Pounds}, {Allan}, {Barber}, {Barstow},
  {Bertram}, {Branduardi-Raymont}, {Brebner}, {Buckley}, {Bromage}, {Cole},
  {Courtier}, {Cruise}, {Culhane}, {Denby}, {Donoghue}, {Dunford},
  {Georgantopoulos}, {Goodall}, {Gondhalekar}, {Gourlay}, {Harris}, {Hassall},
  {Hellier}, {Hodgkin}, {Jeffries}, {Kellett}, {Kent}, {Lieu}, {Lloyd},
  {McGale}, {Mason}, {Matthews}, {Mittaz}, {Page}, {Pankiewicz}, {Pike},
  {Ponman}, {Puchnarewicz}, {Pye}, {Quenby}, {Ricketts}, {Rosen}, {Sansom},
  {Sembay}, {Sidher}, {Sims}, {Stewart}, {Sumner}, {Vallance}, {Watson},
  {Warwick}, {Wells}, {Willingale}, {Willmore}, {Willoughby} and
  {Wonnacott}}]{1993MNRAS.260...77P}
\bibinfo{author}{{Pounds}, K.A.}, \bibinfo{author}{{Allan}, D.J.},
  \bibinfo{author}{{Barber}, C.}, \bibinfo{author}{{Barstow}, M.A.},
  \bibinfo{author}{{Bertram}, D.}, \bibinfo{author}{{Branduardi-Raymont}, G.},
  \bibinfo{author}{{Brebner}, G.E.C.}, \bibinfo{author}{{Buckley}, D.},
  \bibinfo{author}{{Bromage}, G.E.}, \bibinfo{author}{{Cole}, R.E.},
  \bibinfo{author}{{Courtier}, M.}, \bibinfo{author}{{Cruise}, A.M.},
  \bibinfo{author}{{Culhane}, J.L.}, \bibinfo{author}{{Denby}, M.},
  \bibinfo{author}{{Donoghue}, D.O.}, \bibinfo{author}{{Dunford}, E.},
  \bibinfo{author}{{Georgantopoulos}, I.}, \bibinfo{author}{{Goodall}, C.V.},
  \bibinfo{author}{{Gondhalekar}, P.M.}, \bibinfo{author}{{Gourlay}, J.A.},
  \bibinfo{author}{{Harris}, A.W.}, \bibinfo{author}{{Hassall}, B.J.M.},
  \bibinfo{author}{{Hellier}, C.}, \bibinfo{author}{{Hodgkin}, S.},
  \bibinfo{author}{{Jeffries}, R.D.}, \bibinfo{author}{{Kellett}, B.J.},
  \bibinfo{author}{{Kent}, B.J.}, \bibinfo{author}{{Lieu}, R.},
  \bibinfo{author}{{Lloyd}, C.}, \bibinfo{author}{{McGale}, P.},
  \bibinfo{author}{{Mason}, K.O.}, \bibinfo{author}{{Matthews}, L.},
  \bibinfo{author}{{Mittaz}, J.P.D.}, \bibinfo{author}{{Page}, C.G.},
  \bibinfo{author}{{Pankiewicz}, G.S.}, \bibinfo{author}{{Pike}, C.D.},
  \bibinfo{author}{{Ponman}, T.J.}, \bibinfo{author}{{Puchnarewicz}, E.M.},
  \bibinfo{author}{{Pye}, J.P.}, \bibinfo{author}{{Quenby}, J.J.},
  \bibinfo{author}{{Ricketts}, M.J.}, \bibinfo{author}{{Rosen}, S.R.},
  \bibinfo{author}{{Sansom}, A.E.}, \bibinfo{author}{{Sembay}, S.},
  \bibinfo{author}{{Sidher}, S.}, \bibinfo{author}{{Sims}, M.R.},
  \bibinfo{author}{{Stewart}, B.C.}, \bibinfo{author}{{Sumner}, T.J.},
  \bibinfo{author}{{Vallance}, R.J.}, \bibinfo{author}{{Watson}, M.G.},
  \bibinfo{author}{{Warwick}, R.S.}, \bibinfo{author}{{Wells}, A.A.},
  \bibinfo{author}{{Willingale}, R.}, \bibinfo{author}{{Willmore}, A.P.},
  \bibinfo{author}{{Willoughby}, G.A.}, \bibinfo{author}{{Wonnacott}, D.},
  \bibinfo{year}{1993}.
\newblock \bibinfo{title}{{The ROSAT Wide Field Camera all-sky survey of
  extreme-ultraviolet sources. I. The bright source catalogue.}}
\newblock \bibinfo{journal}{\mnras} \bibinfo{volume}{260},
  \bibinfo{pages}{77--102}.
\newblock \DOIprefix\doi{10.1093/mnras/260.1.77}.
\bibitem[{{Reeves} et~al.(2001){Reeves}, {Turner}, {Bennie}, {Pounds}, {Short},
  {O'Brien}, {Boller}, {Kuster} and {Tiengo}}]{Reeves01a}
\bibinfo{author}{{Reeves}, J.N.}, \bibinfo{author}{{Turner}, M.J.L.},
  \bibinfo{author}{{Bennie}, P.J.}, \bibinfo{author}{{Pounds}, K.A.},
  \bibinfo{author}{{Short}, A.}, \bibinfo{author}{{O'Brien}, P.T.},
  \bibinfo{author}{{Boller}, T.}, \bibinfo{author}{{Kuster}, M.},
  \bibinfo{author}{{Tiengo}, A.}, \bibinfo{year}{2001}.
\newblock \bibinfo{title}{{The first XMM-Newton spectrum of a high redshift
  quasar - PKS 0537-286}}.
\newblock \bibinfo{journal}{\aap} \bibinfo{volume}{365},
  \bibinfo{pages}{L116--L121}.
\newblock \DOIprefix\doi{10.1051/0004-6361:20000423},
  \href{http://arxiv.org/abs/astro-ph/0010367}{{\tt arXiv:astro-ph/0010367}}.
\bibitem[{{Remillard} and {McClintock}(2006)}]{Remillard06a}
\bibinfo{author}{{Remillard}, R.A.}, \bibinfo{author}{{McClintock}, J.E.},
  \bibinfo{year}{2006}.
\newblock \bibinfo{title}{{X-Ray Properties of Black-Hole Binaries}}.
\newblock \bibinfo{journal}{\araa} \bibinfo{volume}{44},
  \bibinfo{pages}{49--92}.
\newblock \DOIprefix\doi{10.1146/annurev.astro.44.051905.092532},
  \href{http://arxiv.org/abs/astro-ph/0606352}{{\tt arXiv:astro-ph/0606352}}.
\bibitem[{{Reynolds} et~al.(1999){Reynolds}, {Parmar}, {Hakala}, {Pollock},
  {Williams}, {Peacock} and {Taylor}}]{1999AAS..134..287R}
\bibinfo{author}{{Reynolds}, A.P.}, \bibinfo{author}{{Parmar}, A.N.},
  \bibinfo{author}{{Hakala}, P.J.}, \bibinfo{author}{{Pollock}, A.M.T.},
  \bibinfo{author}{{Williams}, O.R.}, \bibinfo{author}{{Peacock}, A.},
  \bibinfo{author}{{Taylor}, B.G.}, \bibinfo{year}{1999}.
\newblock \bibinfo{title}{{The EXOSAT medium-energy slew survey catalog}}.
\newblock \bibinfo{journal}{\aaps} \bibinfo{volume}{134},
  \bibinfo{pages}{287--300}.
\newblock \DOIprefix\doi{10.1051/aas:1999140},
  \href{http://arxiv.org/abs/astro-ph/9807318}{{\tt arXiv:astro-ph/9807318}}.
\bibitem[{{Rosenberg} et~al.(1975){Rosenberg}, {Eyles}, {Skinner} and
  {Willmore}}]{Rosenberg75a}
\bibinfo{author}{{Rosenberg}, F.D.}, \bibinfo{author}{{Eyles}, C.J.},
  \bibinfo{author}{{Skinner}, G.K.}, \bibinfo{author}{{Willmore}, A.P.},
  \bibinfo{year}{1975}.
\newblock \bibinfo{title}{{Observations of a transient X-ray source with a
  period of 104 S}}.
\newblock \bibinfo{journal}{\nat} \bibinfo{volume}{256},
  \bibinfo{pages}{628--630}.
\newblock \DOIprefix\doi{10.1038/256628a0}.
\bibitem[{{Rothschild} et~al.(1979){Rothschild}, {Boldt}, {Holt},
  {Serlemitsos}, {Garmire}, {Agrawal}, {Riegler}, {Bowyer} and
  {Lampton}}]{1979SSI.....4..269R}
\bibinfo{author}{{Rothschild}, R.}, \bibinfo{author}{{Boldt}, E.},
  \bibinfo{author}{{Holt}, S.}, \bibinfo{author}{{Serlemitsos}, P.},
  \bibinfo{author}{{Garmire}, G.}, \bibinfo{author}{{Agrawal}, P.},
  \bibinfo{author}{{Riegler}, G.}, \bibinfo{author}{{Bowyer}, S.},
  \bibinfo{author}{{Lampton}, M.}, \bibinfo{year}{1979}.
\newblock \bibinfo{title}{{The cosmic X-ray experiment aboard HEAO-1}}.
\newblock \bibinfo{journal}{Space Science Instrumentation} \bibinfo{volume}{4},
  \bibinfo{pages}{269--301}.
\bibitem[{{Saxton} et~al.(2008){Saxton}, {Read}, {Esquej}, {Freyberg},
  {Altieri} and {Bermejo}}]{Saxton08a}
\bibinfo{author}{{Saxton}, R.D.}, \bibinfo{author}{{Read}, A.M.},
  \bibinfo{author}{{Esquej}, P.}, \bibinfo{author}{{Freyberg}, M.J.},
  \bibinfo{author}{{Altieri}, B.}, \bibinfo{author}{{Bermejo}, D.},
  \bibinfo{year}{2008}.
\newblock \bibinfo{title}{{The first XMM-Newton slew survey catalogue:
  XMMSL1}}.
\newblock \bibinfo{journal}{\aap} \bibinfo{volume}{480},
  \bibinfo{pages}{611--622}.
\newblock \DOIprefix\doi{10.1051/0004-6361:20079193},
  \href{http://arxiv.org/abs/0801.3732}{{\tt arXiv:0801.3732}}.
\bibitem[{{Serlemitsos} et~al.(1995){Serlemitsos}, {Jalota}, {Soong},
  {Kunieda}, {Tawara}, {Tsusaka}, {Suzuki}, {Sakima}, {Yamazaki}, {Yoshioka},
  {Furuzawa}, {Yamashita}, {Awaki}, {Itoh}, {Ogasaka}, {Honda} and
  {Uchibori}}]{1995PASJ...47..105S}
\bibinfo{author}{{Serlemitsos}, P.J.}, \bibinfo{author}{{Jalota}, L.},
  \bibinfo{author}{{Soong}, Y.}, \bibinfo{author}{{Kunieda}, H.},
  \bibinfo{author}{{Tawara}, Y.}, \bibinfo{author}{{Tsusaka}, Y.},
  \bibinfo{author}{{Suzuki}, H.}, \bibinfo{author}{{Sakima}, Y.},
  \bibinfo{author}{{Yamazaki}, T.}, \bibinfo{author}{{Yoshioka}, H.},
  \bibinfo{author}{{Furuzawa}, A.}, \bibinfo{author}{{Yamashita}, K.},
  \bibinfo{author}{{Awaki}, H.}, \bibinfo{author}{{Itoh}, M.},
  \bibinfo{author}{{Ogasaka}, Y.}, \bibinfo{author}{{Honda}, H.},
  \bibinfo{author}{{Uchibori}, Y.}, \bibinfo{year}{1995}.
\newblock \bibinfo{title}{{The X-ray telescope on board ASCA}}.
\newblock \bibinfo{journal}{\pasj} \bibinfo{volume}{47},
  \bibinfo{pages}{105--114}.
\bibitem[{SOC(2020)}]{XMM_UHB}
\bibinfo{author}{SOC, E.X.N.}, \bibinfo{year}{2020}.
\newblock \bibinfo{title}{Xmm-newton users handbook}.
\newblock \URLprefix
  \url{https://xmm-tools.cosmos.esa.int/external/xmm_user_support/documentation/uhb/}.
  \bibinfo{note}{is. 2.18}.
\bibitem[{{Sokolovsky} et~al.(2021){Sokolovsky}, {Li}, {Lopes de Oliveira},
  {Ness}, {Mukai}, {Chomiuk}, {Aydi}, {Steinberg}, {Vurm}, {Metzger}, {Babul},
  {Kawash}, {Linford}, {Nelson}, {Page}, {Rupen}, {Sokoloski}, {Strader} and
  {Kilkenny}}]{Sokolovsky21a}
\bibinfo{author}{{Sokolovsky}, K.V.}, \bibinfo{author}{{Li}, K.L.},
  \bibinfo{author}{{Lopes de Oliveira}, R.}, \bibinfo{author}{{Ness}, J.U.},
  \bibinfo{author}{{Mukai}, K.}, \bibinfo{author}{{Chomiuk}, L.},
  \bibinfo{author}{{Aydi}, E.}, \bibinfo{author}{{Steinberg}, E.},
  \bibinfo{author}{{Vurm}, I.}, \bibinfo{author}{{Metzger}, B.D.},
  \bibinfo{author}{{Babul}, A.N.}, \bibinfo{author}{{Kawash}, A.},
  \bibinfo{author}{{Linford}, J.D.}, \bibinfo{author}{{Nelson}, T.},
  \bibinfo{author}{{Page}, K.L.}, \bibinfo{author}{{Rupen}, M.P.},
  \bibinfo{author}{{Sokoloski}, J.L.}, \bibinfo{author}{{Strader}, J.},
  \bibinfo{author}{{Kilkenny}, D.}, \bibinfo{year}{2021}.
\newblock \bibinfo{title}{{The first nova eruption in a novalike variable: YZ
  Ret as seen in X-rays and gamma-rays}}.
\newblock \bibinfo{journal}{Submitted to MNRAS} ,
  \bibinfo{pages}{arXiv:2108.03241}\href{http://arxiv.org/abs/2108.03241}{{\tt
  arXiv:2108.03241}}.
\bibitem[{{Strohmayer}(2001)}]{Strohmayer01a}
\bibinfo{author}{{Strohmayer}, T.E.}, \bibinfo{year}{2001}.
\newblock \bibinfo{title}{{Discovery of a 450 HZ Quasi-periodic Oscillation
  from the Microquasar GRO J1655-40 with the Rossi X-Ray Timing Explorer}}.
\newblock \bibinfo{journal}{\apjl} \bibinfo{volume}{552},
  \bibinfo{pages}{L49--L53}.
\newblock \DOIprefix\doi{10.1086/320258}.
\bibitem[{{Str{\"u}der} et~al.(2001){Str{\"u}der}, {Briel}, {Dennerl},
  {Hartmann}, {Kendziorra}, {Meidinger}, {Pfeffermann}, {Reppin}, {Aschenbach},
  {Bornemann}, {Br{\"a}uninger}, {Burkert}, {Elender}, {Freyberg}, {Haberl},
  {Hartner}, {Heuschmann}, {Hippmann}, {Kastelic}, {Kemmer}, {Kettenring},
  {Kink}, {Krause}, {M{\"u}ller}, {Oppitz}, {Pietsch}, {Popp}, {Predehl},
  {Read}, {Stephan}, {St{\"o}tter}, {Tr{\"u}mper}, {Holl}, {Kemmer}, {Soltau},
  {St{\"o}tter}, {Weber}, {Weichert}, {von Zanthier}, {Carathanassis}, {Lutz},
  {Richter}, {Solc}, {B{\"o}ttcher}, {Kuster}, {Staubert}, {Abbey}, {Holland},
  {Turner}, {Balasini}, {Bignami}, {La Palombara}, {Villa}, {Buttler},
  {Gianini}, {Lain{\'e}}, {Lumb} and {Dhez}}]{Strueder01a}
\bibinfo{author}{{Str{\"u}der}, L.}, \bibinfo{author}{{Briel}, U.},
  \bibinfo{author}{{Dennerl}, K.}, \bibinfo{author}{{Hartmann}, R.},
  \bibinfo{author}{{Kendziorra}, E.}, \bibinfo{author}{{Meidinger}, N.},
  \bibinfo{author}{{Pfeffermann}, E.}, \bibinfo{author}{{Reppin}, C.},
  \bibinfo{author}{{Aschenbach}, B.}, \bibinfo{author}{{Bornemann}, W.},
  \bibinfo{author}{{Br{\"a}uninger}, H.}, \bibinfo{author}{{Burkert}, W.},
  \bibinfo{author}{{Elender}, M.}, \bibinfo{author}{{Freyberg}, M.},
  \bibinfo{author}{{Haberl}, F.}, \bibinfo{author}{{Hartner}, G.},
  \bibinfo{author}{{Heuschmann}, F.}, \bibinfo{author}{{Hippmann}, H.},
  \bibinfo{author}{{Kastelic}, E.}, \bibinfo{author}{{Kemmer}, S.},
  \bibinfo{author}{{Kettenring}, G.}, \bibinfo{author}{{Kink}, W.},
  \bibinfo{author}{{Krause}, N.}, \bibinfo{author}{{M{\"u}ller}, S.},
  \bibinfo{author}{{Oppitz}, A.}, \bibinfo{author}{{Pietsch}, W.},
  \bibinfo{author}{{Popp}, M.}, \bibinfo{author}{{Predehl}, P.},
  \bibinfo{author}{{Read}, A.}, \bibinfo{author}{{Stephan}, K.H.},
  \bibinfo{author}{{St{\"o}tter}, D.}, \bibinfo{author}{{Tr{\"u}mper}, J.},
  \bibinfo{author}{{Holl}, P.}, \bibinfo{author}{{Kemmer}, J.},
  \bibinfo{author}{{Soltau}, H.}, \bibinfo{author}{{St{\"o}tter}, R.},
  \bibinfo{author}{{Weber}, U.}, \bibinfo{author}{{Weichert}, U.},
  \bibinfo{author}{{von Zanthier}, C.}, \bibinfo{author}{{Carathanassis}, D.},
  \bibinfo{author}{{Lutz}, G.}, \bibinfo{author}{{Richter}, R.H.},
  \bibinfo{author}{{Solc}, P.}, \bibinfo{author}{{B{\"o}ttcher}, H.},
  \bibinfo{author}{{Kuster}, M.}, \bibinfo{author}{{Staubert}, R.},
  \bibinfo{author}{{Abbey}, A.}, \bibinfo{author}{{Holland}, A.},
  \bibinfo{author}{{Turner}, M.}, \bibinfo{author}{{Balasini}, M.},
  \bibinfo{author}{{Bignami}, G.F.}, \bibinfo{author}{{La Palombara}, N.},
  \bibinfo{author}{{Villa}, G.}, \bibinfo{author}{{Buttler}, W.},
  \bibinfo{author}{{Gianini}, F.}, \bibinfo{author}{{Lain{\'e}}, R.},
  \bibinfo{author}{{Lumb}, D.}, \bibinfo{author}{{Dhez}, P.},
  \bibinfo{year}{2001}.
\newblock \bibinfo{title}{{The European Photon Imaging Camera on XMM-Newton:
  The pn-CCD camera}}.
\newblock \bibinfo{journal}{\aap} \bibinfo{volume}{365},
  \bibinfo{pages}{L18--L26}.
\newblock \DOIprefix\doi{10.1051/0004-6361:20000066}.
\bibitem[{{Tanaka} et~al.(1994){Tanaka}, {Inoue} and {Holt}}]{Tanaka94A}
\bibinfo{author}{{Tanaka}, Y.}, \bibinfo{author}{{Inoue}, H.},
  \bibinfo{author}{{Holt}, S.S.}, \bibinfo{year}{1994}.
\newblock \bibinfo{title}{{The X-Ray Astronomy Satellite ASCA}}.
\newblock \bibinfo{journal}{\pasj} \bibinfo{volume}{46},
  \bibinfo{pages}{L37--L41}.
\bibitem[{{Terrell} et~al.(1982){Terrell}, {Fenimore}, {Klebesadel} and
  {Desai}}]{Terrell82a}
\bibinfo{author}{{Terrell}, J.}, \bibinfo{author}{{Fenimore}, E.E.},
  \bibinfo{author}{{Klebesadel}, R.W.}, \bibinfo{author}{{Desai}, U.D.},
  \bibinfo{year}{1982}.
\newblock \bibinfo{title}{{Observation of two gamma-ray bursts by VELA X-ray
  detectors.}}
\newblock \bibinfo{journal}{\apj} \bibinfo{volume}{254},
  \bibinfo{pages}{279--286}.
\newblock \DOIprefix\doi{10.1086/159731}.
\bibitem[{{Traulsen} et~al.(2019){Traulsen}, {Schwope}, {Lamer}, {Ballet},
  {Carrera}, {Coriat}, {Freyberg}, {Michel}, {Motch}, {Rosen}, {Webb},
  {Ceballos}, {Koliopanos}, {Kurpas}, {Page} and {Watson}}]{Traulsen19a}
\bibinfo{author}{{Traulsen}, I.}, \bibinfo{author}{{Schwope}, A.D.},
  \bibinfo{author}{{Lamer}, G.}, \bibinfo{author}{{Ballet}, J.},
  \bibinfo{author}{{Carrera}, F.}, \bibinfo{author}{{Coriat}, M.},
  \bibinfo{author}{{Freyberg}, M.J.}, \bibinfo{author}{{Michel}, L.},
  \bibinfo{author}{{Motch}, C.}, \bibinfo{author}{{Rosen}, S.R.},
  \bibinfo{author}{{Webb}, N.}, \bibinfo{author}{{Ceballos}, M.T.},
  \bibinfo{author}{{Koliopanos}, F.}, \bibinfo{author}{{Kurpas}, J.},
  \bibinfo{author}{{Page}, M.J.}, \bibinfo{author}{{Watson}, M.G.},
  \bibinfo{year}{2019}.
\newblock \bibinfo{title}{{The XMM-Newton serendipitous survey. VIII. The first
  XMM-Newton serendipitous source catalogue from overlapping observations}}.
\newblock \bibinfo{journal}{\aap} \bibinfo{volume}{624}, \bibinfo{pages}{A77}.
\newblock \DOIprefix\doi{10.1051/0004-6361/201833938},
  \href{http://arxiv.org/abs/1807.09178}{{\tt arXiv:1807.09178}}.
\bibitem[{{Traulsen} et~al.(2020){Traulsen}, {Schwope}, {Lamer}, {Ballet},
  {Carrera}, {Ceballos}, {Coriat}, {Freyberg}, {Koliopanos}, {Kurpas},
  {Michel}, {Motch}, {Page}, {Watson} and {Webb}}]{Traulsen20a}
\bibinfo{author}{{Traulsen}, I.}, \bibinfo{author}{{Schwope}, A.D.},
  \bibinfo{author}{{Lamer}, G.}, \bibinfo{author}{{Ballet}, J.},
  \bibinfo{author}{{Carrera}, F.J.}, \bibinfo{author}{{Ceballos}, M.T.},
  \bibinfo{author}{{Coriat}, M.}, \bibinfo{author}{{Freyberg}, M.J.},
  \bibinfo{author}{{Koliopanos}, F.}, \bibinfo{author}{{Kurpas}, J.},
  \bibinfo{author}{{Michel}, L.}, \bibinfo{author}{{Motch}, C.},
  \bibinfo{author}{{Page}, M.J.}, \bibinfo{author}{{Watson}, M.G.},
  \bibinfo{author}{{Webb}, N.A.}, \bibinfo{year}{2020}.
\newblock \bibinfo{title}{{The XMM-Newton serendipitous survey. X. The second
  source catalogue from overlapping XMM-Newton observations and its long-term
  variable content}}.
\newblock \bibinfo{journal}{\aap} \bibinfo{volume}{641}, \bibinfo{pages}{A137}.
\newblock \DOIprefix\doi{10.1051/0004-6361/202037706},
  \href{http://arxiv.org/abs/2007.02932}{{\tt arXiv:2007.02932}}.
\bibitem[{{Tr{\"u}mper}(1982)}]{1982AdSpR...2..241T}
\bibinfo{author}{{Tr{\"u}mper}, J.}, \bibinfo{year}{1982}.
\newblock \bibinfo{title}{{The ROSAT mission}}.
\newblock \bibinfo{journal}{Advances in Space Research} \bibinfo{volume}{2},
  \bibinfo{pages}{241--249}.
\newblock \DOIprefix\doi{10.1016/0273-1177(82)90070-9}.
\bibitem[{{Tsusaka} et~al.(1995){Tsusaka}, {Suzuki}, {Yamashita}, {Kunieda},
  {Tawara}, {Ogasaka}, {Uchibori}, {Honda}, {Itoh}, {Awaki}, {Tsunemi},
  {Hayashida}, {Nomoto}, {Wada}, {Miyata}, {Serlemitsos}, {Jalota} and
  {Soong}}]{1995ApOpt..34.4848T}
\bibinfo{author}{{Tsusaka}, Y.}, \bibinfo{author}{{Suzuki}, H.},
  \bibinfo{author}{{Yamashita}, K.}, \bibinfo{author}{{Kunieda}, H.},
  \bibinfo{author}{{Tawara}, Y.}, \bibinfo{author}{{Ogasaka}, Y.},
  \bibinfo{author}{{Uchibori}, Y.}, \bibinfo{author}{{Honda}, H.},
  \bibinfo{author}{{Itoh}, M.}, \bibinfo{author}{{Awaki}, H.},
  \bibinfo{author}{{Tsunemi}, H.}, \bibinfo{author}{{Hayashida}, K.},
  \bibinfo{author}{{Nomoto}, S.}, \bibinfo{author}{{Wada}, M.},
  \bibinfo{author}{{Miyata}, E.}, \bibinfo{author}{{Serlemitsos}, P.J.},
  \bibinfo{author}{{Jalota}, L.}, \bibinfo{author}{{Soong}, Y.},
  \bibinfo{year}{1995}.
\newblock \bibinfo{title}{{Characterization of the Advanced Satellite for
  Cosmology and Astrophysics x-ray telescope: preflight calibration and ray
  tracing}}.
\newblock \bibinfo{journal}{\ao} \bibinfo{volume}{34},
  \bibinfo{pages}{4848--4856}.
\newblock \DOIprefix\doi{10.1364/AO.34.004848}.
\bibitem[{{Turner} et~al.(2001){Turner}, {Abbey}, {Arnaud}, {Balasini},
  {Barbera}, {Belsole}, {Bennie}, {Bernard}, {Bignami}, {Boer}, {Briel},
  {Butler}, {Cara}, {Chabaud}, {Cole}, {Collura}, {Conte}, {Cros}, {Denby},
  {Dhez}, {Di Coco}, {Dowson}, {Ferrando}, {Ghizzardi}, {Gianotti}, {Goodall},
  {Gretton}, {Griffiths}, {Hainaut}, {Hochedez}, {Holland}, {Jourdain},
  {Kendziorra}, {Lagostina}, {Laine}, {La Palombara}, {Lortholary}, {Lumb},
  {Marty}, {Molendi}, {Pigot}, {Poindron}, {Pounds}, {Reeves}, {Reppin},
  {Rothenflug}, {Salvetat}, {Sauvageot}, {Schmitt}, {Sembay}, {Short},
  {Spragg}, {Stephen}, {Str{\"u}der}, {Tiengo}, {Trifoglio}, {Tr{\"u}mper},
  {Vercellone}, {Vigroux}, {Villa}, {Ward}, {Whitehead} and
  {Zonca}}]{Turner01a}
\bibinfo{author}{{Turner}, M.J.L.}, \bibinfo{author}{{Abbey}, A.},
  \bibinfo{author}{{Arnaud}, M.}, \bibinfo{author}{{Balasini}, M.},
  \bibinfo{author}{{Barbera}, M.}, \bibinfo{author}{{Belsole}, E.},
  \bibinfo{author}{{Bennie}, P.J.}, \bibinfo{author}{{Bernard}, J.P.},
  \bibinfo{author}{{Bignami}, G.F.}, \bibinfo{author}{{Boer}, M.},
  \bibinfo{author}{{Briel}, U.}, \bibinfo{author}{{Butler}, I.},
  \bibinfo{author}{{Cara}, C.}, \bibinfo{author}{{Chabaud}, C.},
  \bibinfo{author}{{Cole}, R.}, \bibinfo{author}{{Collura}, A.},
  \bibinfo{author}{{Conte}, M.}, \bibinfo{author}{{Cros}, A.},
  \bibinfo{author}{{Denby}, M.}, \bibinfo{author}{{Dhez}, P.},
  \bibinfo{author}{{Di Coco}, G.}, \bibinfo{author}{{Dowson}, J.},
  \bibinfo{author}{{Ferrando}, P.}, \bibinfo{author}{{Ghizzardi}, S.},
  \bibinfo{author}{{Gianotti}, F.}, \bibinfo{author}{{Goodall}, C.V.},
  \bibinfo{author}{{Gretton}, L.}, \bibinfo{author}{{Griffiths}, R.G.},
  \bibinfo{author}{{Hainaut}, O.}, \bibinfo{author}{{Hochedez}, J.F.},
  \bibinfo{author}{{Holland}, A.D.}, \bibinfo{author}{{Jourdain}, E.},
  \bibinfo{author}{{Kendziorra}, E.}, \bibinfo{author}{{Lagostina}, A.},
  \bibinfo{author}{{Laine}, R.}, \bibinfo{author}{{La Palombara}, N.},
  \bibinfo{author}{{Lortholary}, M.}, \bibinfo{author}{{Lumb}, D.},
  \bibinfo{author}{{Marty}, P.}, \bibinfo{author}{{Molendi}, S.},
  \bibinfo{author}{{Pigot}, C.}, \bibinfo{author}{{Poindron}, E.},
  \bibinfo{author}{{Pounds}, K.A.}, \bibinfo{author}{{Reeves}, J.N.},
  \bibinfo{author}{{Reppin}, C.}, \bibinfo{author}{{Rothenflug}, R.},
  \bibinfo{author}{{Salvetat}, P.}, \bibinfo{author}{{Sauvageot}, J.L.},
  \bibinfo{author}{{Schmitt}, D.}, \bibinfo{author}{{Sembay}, S.},
  \bibinfo{author}{{Short}, A.D.T.}, \bibinfo{author}{{Spragg}, J.},
  \bibinfo{author}{{Stephen}, J.}, \bibinfo{author}{{Str{\"u}der}, L.},
  \bibinfo{author}{{Tiengo}, A.}, \bibinfo{author}{{Trifoglio}, M.},
  \bibinfo{author}{{Tr{\"u}mper}, J.}, \bibinfo{author}{{Vercellone}, S.},
  \bibinfo{author}{{Vigroux}, L.}, \bibinfo{author}{{Villa}, G.},
  \bibinfo{author}{{Ward}, M.J.}, \bibinfo{author}{{Whitehead}, S.},
  \bibinfo{author}{{Zonca}, E.}, \bibinfo{year}{2001}.
\newblock \bibinfo{title}{{The European Photon Imaging Camera on XMM-Newton:
  The MOS cameras}}.
\newblock \bibinfo{journal}{\aap} \bibinfo{volume}{365},
  \bibinfo{pages}{L27--L35}.
\newblock \DOIprefix\doi{10.1051/0004-6361:20000087},
  \href{http://arxiv.org/abs/astro-ph/0011498}{{\tt arXiv:astro-ph/0011498}}.
\bibitem[{{Turner} et~al.(1981){Turner}, {Smith} and
  {Zimmermann}}]{1981SSRv...30..513T}
\bibinfo{author}{{Turner}, M.J.L.}, \bibinfo{author}{{Smith}, A.},
  \bibinfo{author}{{Zimmermann}, H.U.}, \bibinfo{year}{1981}.
\newblock \bibinfo{title}{{The medium energy instrument on EXOSAT}}.
\newblock \bibinfo{journal}{\ssr} \bibinfo{volume}{30},
  \bibinfo{pages}{513--524}.
\newblock \DOIprefix\doi{10.1007/BF01246071}.
\bibitem[{{Turner} et~al.(1989){Turner}, {Thomas}, {Patchett}, {Reading},
  {Makishima}, {Ohashi}, {Dotani}, {Hayashida}, {Inoue}, {Kondo}, {Koyama},
  {Mitsusa}, {Ogawara}, {Takano}, {Awaki}, {Tawara} and
  {Nakamura}}]{1989PASJ...41..345T}
\bibinfo{author}{{Turner}, M.J.L.}, \bibinfo{author}{{Thomas}, H.D.},
  \bibinfo{author}{{Patchett}, B.E.}, \bibinfo{author}{{Reading}, D.H.},
  \bibinfo{author}{{Makishima}, K.}, \bibinfo{author}{{Ohashi}, T.},
  \bibinfo{author}{{Dotani}, T.}, \bibinfo{author}{{Hayashida}, K.},
  \bibinfo{author}{{Inoue}, H.}, \bibinfo{author}{{Kondo}, H.},
  \bibinfo{author}{{Koyama}, K.}, \bibinfo{author}{{Mitsusa}, K.},
  \bibinfo{author}{{Ogawara}, Y.}, \bibinfo{author}{{Takano}, S.},
  \bibinfo{author}{{Awaki}, H.}, \bibinfo{author}{{Tawara}, Y.},
  \bibinfo{author}{{Nakamura}, N.}, \bibinfo{year}{1989}.
\newblock \bibinfo{title}{{The large area counter on GINGA}}.
\newblock \bibinfo{journal}{\pasj} \bibinfo{volume}{41},
  \bibinfo{pages}{345--372}.
\bibitem[{{Ubertini} et~al.(2003){Ubertini}, {Lebrun}, {Di Cocco}, {Bazzano},
  {Bird}, {Broenstad}, {Goldwurm}, {La Rosa}, {Labanti}, {Laurent}, {Mirabel},
  {Quadrini}, {Ramsey}, {Reglero}, {Sabau}, {Sacco}, {Staubert}, {Vigroux},
  {Weisskopf} and {Zdziarski}}]{Ubertini03a}
\bibinfo{author}{{Ubertini}, P.}, \bibinfo{author}{{Lebrun}, F.},
  \bibinfo{author}{{Di Cocco}, G.}, \bibinfo{author}{{Bazzano}, A.},
  \bibinfo{author}{{Bird}, A.J.}, \bibinfo{author}{{Broenstad}, K.},
  \bibinfo{author}{{Goldwurm}, A.}, \bibinfo{author}{{La Rosa}, G.},
  \bibinfo{author}{{Labanti}, C.}, \bibinfo{author}{{Laurent}, P.},
  \bibinfo{author}{{Mirabel}, I.F.}, \bibinfo{author}{{Quadrini}, E.M.},
  \bibinfo{author}{{Ramsey}, B.}, \bibinfo{author}{{Reglero}, V.},
  \bibinfo{author}{{Sabau}, L.}, \bibinfo{author}{{Sacco}, B.},
  \bibinfo{author}{{Staubert}, R.}, \bibinfo{author}{{Vigroux}, L.},
  \bibinfo{author}{{Weisskopf}, M.C.}, \bibinfo{author}{{Zdziarski}, A.A.},
  \bibinfo{year}{2003}.
\newblock \bibinfo{title}{{IBIS: The Imager on-board INTEGRAL}}.
\newblock \bibinfo{journal}{\aap} \bibinfo{volume}{411},
  \bibinfo{pages}{L131--L139}.
\newblock \DOIprefix\doi{10.1051/0004-6361:20031224}.
\bibitem[{{Ueda} et~al.(2001){Ueda}, {Ishisaki}, {Takahashi}, {Makishima} and
  {Ohashi}}]{2001ApJS..133....1U}
\bibinfo{author}{{Ueda}, Y.}, \bibinfo{author}{{Ishisaki}, Y.},
  \bibinfo{author}{{Takahashi}, T.}, \bibinfo{author}{{Makishima}, K.},
  \bibinfo{author}{{Ohashi}, T.}, \bibinfo{year}{2001}.
\newblock \bibinfo{title}{{The ASCA Medium Sensitivity Survey (the GIS Catalog
  Project): Source Catalog}}.
\newblock \bibinfo{journal}{\apjs} \bibinfo{volume}{133},
  \bibinfo{pages}{1--52}.
\newblock \DOIprefix\doi{10.1086/319189},
  \href{http://arxiv.org/abs/astro-ph/9908128}{{\tt arXiv:astro-ph/9908128}}.
\bibitem[{{Ueda} et~al.(2005){Ueda}, {Ishisaki}, {Takahashi}, {Makishima} and
  {Ohashi}}]{2005ApJS..161..185U}
\bibinfo{author}{{Ueda}, Y.}, \bibinfo{author}{{Ishisaki}, Y.},
  \bibinfo{author}{{Takahashi}, T.}, \bibinfo{author}{{Makishima}, K.},
  \bibinfo{author}{{Ohashi}, T.}, \bibinfo{year}{2005}.
\newblock \bibinfo{title}{{The ASCA Medium Sensitivity Survey (The GIS Catalog
  Project): Source Catalog II.}}
\newblock \bibinfo{journal}{\apjs} \bibinfo{volume}{161},
  \bibinfo{pages}{185--223}.
\newblock \DOIprefix\doi{10.1086/468187}.
\bibitem[{{Villa} et~al.(1976){Villa}, {Page}, {Turner}, {Cooke}, {Ricketts},
  {Pounds} and {Adams}}]{Villa76a}
\bibinfo{author}{{Villa}, G.}, \bibinfo{author}{{Page}, C.G.},
  \bibinfo{author}{{Turner}, M.J.L.}, \bibinfo{author}{{Cooke}, B.A.},
  \bibinfo{author}{{Ricketts}, M.J.}, \bibinfo{author}{{Pounds}, K.A.},
  \bibinfo{author}{{Adams}, D.J.}, \bibinfo{year}{1976}.
\newblock \bibinfo{title}{{The Ariel V Sky Survey Instrument and new
  observations of the Milky Way.}}
\newblock \bibinfo{journal}{\mnras} \bibinfo{volume}{176},
  \bibinfo{pages}{609--620}.
\newblock \DOIprefix\doi{10.1093/mnras/176.3.609}.
\bibitem[{{Warwick} et~al.(1981){Warwick}, {Marshall}, {Fraser}, {Watson},
  {Lawrence}, {Page}, {Pounds}, {Ricketts}, {Sims} and
  {Smith}}]{1981MNRAS.197..865W}
\bibinfo{author}{{Warwick}, R.S.}, \bibinfo{author}{{Marshall}, N.},
  \bibinfo{author}{{Fraser}, G.W.}, \bibinfo{author}{{Watson}, M.G.},
  \bibinfo{author}{{Lawrence}, A.}, \bibinfo{author}{{Page}, C.G.},
  \bibinfo{author}{{Pounds}, K.A.}, \bibinfo{author}{{Ricketts}, M.J.},
  \bibinfo{author}{{Sims}, M.R.}, \bibinfo{author}{{Smith}, A.},
  \bibinfo{year}{1981}.
\newblock \bibinfo{title}{{The Ariel V /3 A/ catalogue of X-ray sources. I -
  Sources at low galactic latitude /absolute value of B less than 10 deg/}}.
\newblock \bibinfo{journal}{\mnras} \bibinfo{volume}{197},
  \bibinfo{pages}{865--891}.
\newblock \DOIprefix\doi{10.1093/mnras/197.4.865}.
\bibitem[{{Webb} et~al.(2020){Webb}, {Coriat}, {Traulsen}, {Ballet}, {Motch},
  {Carrera}, {Koliopanos}, {Authier}, {de la Calle}, {Ceballos}, {Colomo},
  {Chuard}, {Freyberg}, {Garcia}, {Kolehmainen}, {Lamer}, {Lin}, {Maggi},
  {Michel}, {Page}, {Page}, {Perea-Calderon}, {Pineau}, {Rodriguez}, {Rosen},
  {Santos Lleo}, {Saxton}, {Schwope}, {Tom{\'a}s}, {Watson} and
  {Zakardjian}}]{Webb20a}
\bibinfo{author}{{Webb}, N.A.}, \bibinfo{author}{{Coriat}, M.},
  \bibinfo{author}{{Traulsen}, I.}, \bibinfo{author}{{Ballet}, J.},
  \bibinfo{author}{{Motch}, C.}, \bibinfo{author}{{Carrera}, F.J.},
  \bibinfo{author}{{Koliopanos}, F.}, \bibinfo{author}{{Authier}, J.},
  \bibinfo{author}{{de la Calle}, I.}, \bibinfo{author}{{Ceballos}, M.T.},
  \bibinfo{author}{{Colomo}, E.}, \bibinfo{author}{{Chuard}, D.},
  \bibinfo{author}{{Freyberg}, M.}, \bibinfo{author}{{Garcia}, T.},
  \bibinfo{author}{{Kolehmainen}, M.}, \bibinfo{author}{{Lamer}, G.},
  \bibinfo{author}{{Lin}, D.}, \bibinfo{author}{{Maggi}, P.},
  \bibinfo{author}{{Michel}, L.}, \bibinfo{author}{{Page}, C.G.},
  \bibinfo{author}{{Page}, M.J.}, \bibinfo{author}{{Perea-Calderon}, J.V.},
  \bibinfo{author}{{Pineau}, F.X.}, \bibinfo{author}{{Rodriguez}, P.},
  \bibinfo{author}{{Rosen}, S.R.}, \bibinfo{author}{{Santos Lleo}, M.},
  \bibinfo{author}{{Saxton}, R.D.}, \bibinfo{author}{{Schwope}, A.},
  \bibinfo{author}{{Tom{\'a}s}, L.}, \bibinfo{author}{{Watson}, M.G.},
  \bibinfo{author}{{Zakardjian}, A.}, \bibinfo{year}{2020}.
\newblock \bibinfo{title}{{The XMM-Newton serendipitous survey. IX. The fourth
  XMM-Newton serendipitous source catalogue}}.
\newblock \bibinfo{journal}{\aap} \bibinfo{volume}{641}, \bibinfo{pages}{A136}.
\newblock \DOIprefix\doi{10.1051/0004-6361/201937353},
  \href{http://arxiv.org/abs/2007.02899}{{\tt arXiv:2007.02899}}.
\bibitem[{{Wells} et~al.(1981){Wells}, {Greisen} and {Harten}}]{Wells81a}
\bibinfo{author}{{Wells}, D.C.}, \bibinfo{author}{{Greisen}, E.W.},
  \bibinfo{author}{{Harten}, R.H.}, \bibinfo{year}{1981}.
\newblock \bibinfo{title}{{FITS - a Flexible Image Transport System}}.
\newblock \bibinfo{journal}{\aaps} \bibinfo{volume}{44}, \bibinfo{pages}{363}.
\bibitem[{{White} and {Peacock}(1988)}]{1988MmSAI..59....7W}
\bibinfo{author}{{White}, N.E.}, \bibinfo{author}{{Peacock}, A.},
  \bibinfo{year}{1988}.
\newblock \bibinfo{title}{{The EXOSAT observatory}}.
\newblock \bibinfo{journal}{{Mem. S.A.It.}} \bibinfo{volume}{59},
  \bibinfo{pages}{7--31}.
\bibitem[{{Whitlock} et~al.(1992a){Whitlock}, {Lochner} and
  {Rhode}}]{Whitlock92b}
\bibinfo{author}{{Whitlock}, L.}, \bibinfo{author}{{Lochner}, J.},
  \bibinfo{author}{{Rhode}, K.}, \bibinfo{year}{1992}a.
\newblock \bibinfo{title}{{The Ariel 5 and Vela 5B All-Sky Monitor Databases}}.
\newblock \bibinfo{journal}{Legacy} \bibinfo{volume}{2}.
\newblock \URLprefix
  \url{https://heasarc.gsfc.nasa.gov/docs/journal/ariel2.html}.
\bibitem[{{Whitlock} et~al.(1992b){Whitlock}, {Lochner} and
  {Rhode}}]{Whitlock92}
\bibinfo{author}{{Whitlock}, L.}, \bibinfo{author}{{Lochner}, J.},
  \bibinfo{author}{{Rhode}, K.}, \bibinfo{year}{1992}b.
\newblock \bibinfo{title}{Vela 5B ASM Calibration Guide}.
\newblock \bibinfo{type}{Technical Report}. Laboratory of High Energy
  Astrophysics, Office of Guest Observer Programmes, NASA GSFC, Greenbelt.
\newblock
  \bibinfo{note}{\url{https://heasarc.gsfc.nasa.gov/FTP/vela5b/doc/doctex/}}.
\bibitem[{{Wilson} et~al.(2002){Wilson}, {Finger}, {Coe}, {Laycock} and
  {Fabregat}}]{Wilson02a}
\bibinfo{author}{{Wilson}, C.A.}, \bibinfo{author}{{Finger}, M.H.},
  \bibinfo{author}{{Coe}, M.J.}, \bibinfo{author}{{Laycock}, S.},
  \bibinfo{author}{{Fabregat}, J.}, \bibinfo{year}{2002}.
\newblock \bibinfo{title}{{A Decade in the Life of EXO 2030+375: A
  Multiwavelength Study of an Accreting X-Ray Pulsar}}.
\newblock \bibinfo{journal}{\apj} \bibinfo{volume}{570},
  \bibinfo{pages}{287--302}.
\newblock \DOIprefix\doi{10.1086/339739},
  \href{http://arxiv.org/abs/astro-ph/0201227}{{\tt arXiv:astro-ph/0201227}}.
\bibitem[{{Winkler} et~al.(2003){Winkler}, {Courvoisier}, {Di Cocco},
  {Gehrels}, {Gim{\'e}nez}, {Grebenev}, {Hermsen}, {Mas-Hesse}, {Lebrun},
  {Lund}, {Palumbo}, {Paul}, {Roques}, {Schnopper}, {Sch{\"o}nfelder},
  {Sunyaev}, {Teegarden}, {Ubertini}, {Vedrenne} and {Dean}}]{Winkler03a}
\bibinfo{author}{{Winkler}, C.}, \bibinfo{author}{{Courvoisier}, T.J.L.},
  \bibinfo{author}{{Di Cocco}, G.}, \bibinfo{author}{{Gehrels}, N.},
  \bibinfo{author}{{Gim{\'e}nez}, A.}, \bibinfo{author}{{Grebenev}, S.},
  \bibinfo{author}{{Hermsen}, W.}, \bibinfo{author}{{Mas-Hesse}, J.M.},
  \bibinfo{author}{{Lebrun}, F.}, \bibinfo{author}{{Lund}, N.},
  \bibinfo{author}{{Palumbo}, G.G.C.}, \bibinfo{author}{{Paul}, J.},
  \bibinfo{author}{{Roques}, J.P.}, \bibinfo{author}{{Schnopper}, H.},
  \bibinfo{author}{{Sch{\"o}nfelder}, V.}, \bibinfo{author}{{Sunyaev}, R.},
  \bibinfo{author}{{Teegarden}, B.}, \bibinfo{author}{{Ubertini}, P.},
  \bibinfo{author}{{Vedrenne}, G.}, \bibinfo{author}{{Dean}, A.J.},
  \bibinfo{year}{2003}.
\newblock \bibinfo{title}{{The INTEGRAL mission}}.
\newblock \bibinfo{journal}{\aap} \bibinfo{volume}{411},
  \bibinfo{pages}{L1--L6}.
\newblock \DOIprefix\doi{10.1051/0004-6361:20031288}.
\bibitem[{{Wood} et~al.(1984){Wood}, {Meekins}, {Yentis}, {Smathers}, {McNutt},
  {Bleach}, {Byram}, {Chupp}, {Friedman} and {Meidav}}]{Wood84a}
\bibinfo{author}{{Wood}, K.S.}, \bibinfo{author}{{Meekins}, J.F.},
  \bibinfo{author}{{Yentis}, D.J.}, \bibinfo{author}{{Smathers}, H.W.},
  \bibinfo{author}{{McNutt}, D.P.}, \bibinfo{author}{{Bleach}, R.D.},
  \bibinfo{author}{{Byram}, E.T.}, \bibinfo{author}{{Chupp}, T.A.},
  \bibinfo{author}{{Friedman}, H.}, \bibinfo{author}{{Meidav}, M.},
  \bibinfo{year}{1984}.
\newblock \bibinfo{title}{{The HEAO A-1 X-ray source catalog.}}
\newblock \bibinfo{journal}{\apjs} \bibinfo{volume}{56},
  \bibinfo{pages}{507--649}.
\newblock \DOIprefix\doi{10.1086/190992}.
\bibitem[{{Zimmermann} et~al.(1998){Zimmermann}, {Boese}, {Becker}, {Belloni},
  {D{\"o}bereiner}, {Izzo}, {Kahabka} and {Schwentker}}]{EXSAS}
\bibinfo{author}{{Zimmermann}, U.}, \bibinfo{author}{{Boese}, G.},
  \bibinfo{author}{{Becker}, W.}, \bibinfo{author}{{Belloni}, T.},
  \bibinfo{author}{{D{\"o}bereiner}, S.}, \bibinfo{author}{{Izzo}, C.},
  \bibinfo{author}{{Kahabka}, P.}, \bibinfo{author}{{Schwentker}, O.},
  \bibinfo{year}{1998}.
\newblock \bibinfo{title}{EXSAS User's Guide}.
\newblock \bibinfo{type}{Technical Report} \bibinfo{number}{1}.
  Max-Planck-Institut f{\"u}r Extraterrestrische Physik.
  \bibinfo{address}{85740 Garching bei M{\"u}nchen, Germany}.
\newblock
  \bibinfo{note}{\url{ftp://ftp.xray.mpe.mpg.de/people/mjf/2RXS/exsas_guide-980708.pdf}}.

\end{thebibliography}

\appendix
\onecolumn
\section{Appendix}
\begin{table}[h] 
\centering
\caption{Pixel positions for the footprint definition of the \exosat LE CMA1 and CMA2 images that have an octagonal shape. The images have a size of $2048 \times 2048$ pixels.}
\begin{tabular}{llllllc}
\hline\hline
Shape & \multicolumn{5}{c}{Pixel position of the footprint (\textsc{rawx}/\textsc{rawy})} & Number of images\\
\hline
CMA1 & 124/902  & 149/1278  & 921/2048  & 1181/2013 & 1479/2037 & \multirow{2}{*}{81.1\%}\\
(e.g., Fig.~\ref{fig:exosat_le_footprint}\subref{fig:exosat_le_footprint:a}) & 2048/1464 & 2048/773 & 1435/226 & 886/112 & \\
CMA2 & 1/512  & 1/1427  & 590/2048 & 1452/2048 & & \multirow{2}{*}{10.9\%} \\
(e.g., Fig.~\ref{fig:exosat_le_footprint}\subref{fig:exosat_le_footprint:b})
      & 2048/1520 & 2048/547 & 1530/1 & 537/1 & \\
\hline
\end{tabular}
\label{tab:exosat_pixelpos}
\end{table}

{\footnotesize{  

\begin{longtable}{l|llll|ccc} 
\caption{Conversion factors (CFs) in $\text{erg\,cm}^{-2}\text{cnt}^{-1}$, computed with PIMMS version 4.11a \citep{Mukai93PIMMS} or, in case of \velaVb, \uhuru, \arielv and \heaoI, inferred from the lightcurve FITS files (Sect.~\ref{subsec:vela}), \citet[][see also Sect.~\ref{subsec:uhuru}]{1978ApJS...38..357F}, \citet[][Sect.~\ref{subsec:arielv}]{Kaluzienski77a}, and \citet[][Sect.~\ref{subsec:heaoI}]{Wood84a,Nugent83a}, respectively. These conversion factors were extrapolated to the \hiligt energy bands using Eq.~\ref{eq:CV_old}. Note that for the hard band of \heaoI A2 the conversion factors do not have a unit as \hiligt receives already converted fluxes \citep{1982ApJ...253..485P} and only extrapolated to the 2--12\,keV band. In the case of \arielv ASM the error is only the systematic error of \citet[p.~106]{Kaluzienski77a}, while an additional statistical error on each lightcurve point is added in quadrature. ``plaw'' denotes a powerlaw with a slope specified in the index column. ``bbody'' denotes a black body model with a temperature defined in units of keV in the index column.}\\
\label{tab:cf}
\centering
Mission & Filter & Model & Index & $\mathrm{N}_\mathrm{H}$ [cm$^{-2}$] & CF for 0.2--2\,keV & CF for 2--12\,keV & CF for 0.2--12\,keV \\
\hline
ROSAT-PSPC & Open & plaw & 0.5 & $1\cdot 10^{20}$ & $1.166\cdot 10^{-11}$ & $-$ & $-$ \\
ROSAT-PSPC & Open & plaw & 1.0 & $1\cdot 10^{20}$ & $1.044\cdot 10^{-11}$ & $-$ & $-$ \\
ROSAT-PSPC & Open & plaw & 1.5 & $1\cdot 10^{20}$ & $8.863\cdot 10^{-12}$ & $-$ & $-$ \\
ROSAT-PSPC & Open & plaw & 1.7 & $1\cdot 10^{20}$ & $8.158\cdot 10^{-12}$ & $-$ & $-$ \\
ROSAT-PSPC & Open & plaw & 2.0 & $1\cdot 10^{20}$ & $7.076\cdot 10^{-12}$ & $-$ & $-$ \\
ROSAT-PSPC & Open & plaw & 2.5 & $1\cdot 10^{20}$ & $5.391\cdot 10^{-12}$ & $-$ & $-$ \\
\end{longtable}

}}  

\begin{longtable}{l|llll|ccc} 
\caption{Conversion factors (CFs) for \textsl{INTEGRAL} computed for an unabsorbed powerlaw (plaw) using the OSA~10 software.}\\
\label{tab:cf_integral}
\centering
Mission & Filter & Model & Index & $\mathrm{N}_\mathrm{H}$ [cm$^{-2}$] & CF 20--40\,keV & CF 40--60\,keV & CF 60--100\,keV \\
\hline
INTEGRAL & $-$ & plaw & 1.5 & $-$ & $5.959\cdot 10^{-11}$ & $1.411\cdot 10^{-11}$ & $2.237\cdot 10^{-11}$ \\
INTEGRAL & $-$ & plaw & 2.0 & $-$ & $1.844\cdot 10^{-11}$ & $1.29\cdot 10^{-11}$ & $1.624\cdot 10^{-11}$ \\
INTEGRAL & $-$ & plaw & 2.5 & $-$ & $2.583\cdot 10^{-11}$ & $1.068\cdot 10^{-11}$ & $1.116\cdot 10^{-11}$ \\
INTEGRAL & $-$ & plaw & 3.0 & $-$ & $2.024\cdot 10^{-11}$ & $8.36\cdot 10^{-12}$ & $6.34\cdot 10^{-12}$ \\
INTEGRAL & $-$ & plaw & 3.5 & $-$ & $1.115\cdot 10^{-11}$ & $3.73\cdot 10^{-12}$ & $3.61\cdot 10^{-12}$ \\
\end{longtable}

\end{document}